\documentclass[aps,prd,preprint,superscriptaddress,tightenlines,%
nofootinbib]{revtex4}
\newcommand{\PRE}[1]{{#1}}   
\usepackage{amsmath,amssymb,latexsym}
\usepackage{bm}
\usepackage{epsfig}

\usepackage{lineno}

\newcommand{\MET}{\mbox{${\hbox{$E$\kern-0.6em\lower-.1ex\hbox{/}}}_T$}} 

\newcommand{\En}{E_n}
\newcommand{\Fn}{F_n}
\newcommand{\Enu}{E_{\overline{\nu}}}

\newcommand{\Enubar}{\epsilon_{\overline{\nu}}}
\newcommand{\thnu}{\overline{\theta}_{\overline{\nu}}}
\newcommand{\cth}{\cos\overline{\theta}_{\overline{\nu}}}
\newcommand{\Fnu}{F_{\overline{\nu}}}
\newcommand{\tbar}{\overline{\tau}_n}

\newcommand{\bwide}{\begin{widetext}}
\newcommand{\ewide}{\end{widetext}}

\newcommand{\gev}{\rm{GeV}}
\newcommand{\tev}{\rm{TeV}}

\newcommand{\cm}{\rm{cm}}
\newcommand{\km}{\rm{km}}

\newcommand{\s}{\rm{s}}
\newcommand{\yr}{\rm{yr}}
\newcommand{\sr}{\rm{sr}}

\newcommand{\Dmq}{\delta m_{ij}^2}

\newcommand{\phinu}{\phi^{\nu}}
\newcommand{\phinumax}{\phi^{\nu}_{\rm{max}}}
\newcommand{\phiWB}{\phi^{\nu}_{\rm{WB}}}

\newcommand{\sigmanuN}{\sigma_{\nu N}}
\newcommand{\sigmaSM}{\sigma_{\rm{SM}}}
\newcommand{\Ndown}{{\cal N}_{\rm{down}}}
\newcommand{\Nup}{{\cal N}_{\rm{up}}}
\newcommand{\Ndownobs}{\Ndown^{\rm{obs}}}
\newcommand{\Nupobs}{\Nup^{\rm{obs}}}

\newcommand{\postscript}[2]{\setlength{\epsfxsize}{#2\hsize}
   \centerline{\epsfbox{#1}}}

\begin{document}


\rightline{\tt hep-ph/0510389}

\title{
\PRE{\vspace*{1.in}}
IceHEP\\ High Energy Physics at the South Pole
\PRE{\vspace*{0.2in}}
}

\author{Luis Anchordoqui}
\affiliation{
Department of Physics,
Northeastern University,\\ Boston, MA 02115, USA
\PRE{\vspace*{.1in}}
}

\author{Francis  Halzen}
\affiliation{Department of Physics,
University of Wisconsin,\\ Madison, WI 53706, USA
\PRE{\vspace*{.1in}}
}

\setcounter{footnote}{0}

\date{October 2005}


\begin{abstract}

\PRE{\vspace*{.5in}} 

\noindent With the solar and SN87a neutrino observations as proofs of
concepts, the kilometer-scale neutrino experiment IceCube will
scrutinize its data for new particle physics. In this paper we review
the prospects for the realization of such a program. We begin with a
short overview of the detector response and discuss the reach of
``beam'' luminosity. After that we discuss the potential of IceCube to
probe deviations of neutrino-nucleon cross sections from the Standard
Model predictions at center-of-mass energies well beyond those
accessible in man-made accelerators. Then we review the prospects for
extremely long-baseline analyses and discuss the sensitivity to
measure tiny deviations of the flavor mixing angle, expected to be
induced by quantum gravity effects. Finally we discuss the potential
to uncover annihilation of dark matter particles gravitationally
trapped at the center of the Sun, as well as processes occurring in
the early Universe at energies close to the Grand Unification scale.
\end{abstract}



\maketitle


\tableofcontents

\newpage
\section{Introduction}

Conventional astronomy spans 60 octaves in photon frequency, from
$10^4$~cm radio-waves to $10^{-14}$~cm $\gamma$-rays of GeV energy.
This is an amazing expansion of the power of our eyes which scan the
sky over less than a single octave just above $10^{-5}$~cm wavelength.
In recent years, detection and data handling techniques of particle
physics have been reborn in instrumentation to probe the Universe at
new wavelengths, smaller than $10^{-14}$~cm. In addition to the
traditional photon signals, neutrinos and very high-energy protons
(that are only weakly deflected by the magnetic field of our galaxy)
have become astronomical messengers from the Universe. As exemplified
time and again, the development of novel ways of looking into the
cosmos invariably results in the discovery of unanticipated phenomena.
For particle physicists, the sexiest astrophysical problem is
undoubtedly how Nature manages to impart an energy of more than one
Joule to a single elementary particle.

Although cosmic rays were discovered almost a century ago, we do not
know how and where they are accelerated. This may be the oldest
mystery in astronomy and solving it is challenging as can be seen by
the following argument.  It is reasonable to assume that, in order to
accelerate a proton to energy $E$ in a magnetic field $B$, the size
$R$ of the accelerator must encompass the gyro radius of the particle:
$R > R_{\rm gyro} = E/B,$ i.e. the accelerating magnetic field must
contain the particle's orbit. By dimensional analysis, this condition
yields a maximum energy $E = \Gamma BR.$ The $\Gamma$-factor has been
included to allow for the possibility that we may not be at rest in
the frame of the cosmic accelerator, resulting in the observation of
boosted particle energies. Opportunity for particle acceleration to
the highest energies is limited to dense regions where exceptional
gravitational forces create relativistic particle flows. All
speculations involve collapsed objects and we can therefore replace
$R$ by the Schwarzschild radius $R \sim GM/c^2$ to obtain $E < \Gamma
BM.$

Cosmic accelerators are also cosmic beam dumps producing secondary
photons and neutrino beams. Particles accelerated near black holes
pass through intense radiation fields or dense clouds of gas leading
to production of secondary photons and neutrinos that accompany the
primary cosmic ray beam. The target material, whether a gas or photons,
is likely to be sufficiently tenuous so that the primary beam and the
photon beam are only partially attenuated.

At this point a reality check is in order. Such a dimensional analysis
applies to the Fermilab accelerator: 10 kilogauss fields over several
kilometers (covered with a repetition rate of $10^5$ revolutions per
second) yield 1~TeV. The argument holds because, with optimized design
and perfect alignment of magnets, the accelerator reaches efficiencies
matching the dimensional limit. It is highly questionable that nature
can achieve this feat. Theorists can imagine acceleration in shocks
with an efficiency of perhaps $1-10\%.$

Given the microgauss magnetic field of our galaxy, no structures are
large or massive enough to reach the energies of the highest energy
cosmic rays. Dimensional analysis therefore limits their sources to
extragalactic objects. A common speculation is that they may be
relatively nearby active galactic nuclei powered by a billion solar
mass black holes. With kilo-Gauss fields we reach $10^{11}$~GeV. The
jets (blazars) emitted by the central black hole could reach similar
energies in accelerating sub-structures boosted in our direction by a
$\Gamma$-factor of 10, possible higher. The neutron star or black hole
remnant of a collapsing supermassive star could support magnetic
fields of $10^{12}$~G, possible larger. Shocks emanating from the
collapse black hole could be the origin of gamma ray bursts and,
possibly, the source of the highest energy cosmic rays.

The astrophysics problem is so daunting that many believe that cosmic
rays are not the beam of cosmic accelerators but the decay products of
remnants from the early Universe, for instance topological defects
associated with phase transitions (near $10^{16}$~GeV) in Grand
Unified Theories (GUT's). A topological defect will suffer a chain
decay into GUT particles $X$ that subsequently decay to familiar weak
bosons, leptons, quarks- or gluon-jets.  Cosmic rays are the
fragmentation products of these jets.

All in all, where the highest energy cosmic rays are concerned, both
the accelerator mechanism and the particle physics are enigmatic.
There is a realistic hope that the oldest problem in astronomy will be
resolved soon by ambitious experimentation. One such experiment is the
IceCube neutrino telescope, which is required to be sensitive to the
best estimates of potential cosmic ray neutrino fluxes. Though this
telescope is primarily motivated by these astronomical goals, it has
also appeared in the U.S. Roadmap to Particle Physics, and as we will
argue in this review, deservedly so.

As the lightest of fermions and the most weakly
interacting of particles, neutrinos occupy a fragile corner of the
Standard Model and one can realistically hope that they will
reveal the first and most dramatic signatures of new physics.
IceCube's opportunities for particle physics are only limited by
imagination; they include
\begin{itemize}
  \item The search for theories where particle interactions, including
    gravity, unify at the TeV scale.  Neutrinos with energies
    approaching this scale will interact gravitationally with large
    cross sections, similar to those of quarks and leptons, and this
    increase should yield dramatic signatures in a neutrino telescope
    including, possibly, the production of black holes.
  \item The search for deviations from the neutrino's established
    oscillatory behavior that result from non-standard interactions,
    for instance neutrino decay or quantum decoherence.
  \item The search for a breakdown of the equivalence principle as a
    result of non-universal interactions with the gravitational field
    of neutrinos with different flavors. Similarly, the search for
    breakdown of Lorentz invariance resulting from different limiting
    velocities of neutrinos of different flavors.
  \item The search for neutrinos from the annihilation of dark matter
    particles gravitationally trapped at the center of the Sun.
  \item The search from particle emission from cosmic strings or other
    topological defects and heavy cosmological remnants created in the
    early Universe.
  \item The search for magnetic monopoles.
\end{itemize}
Alternatively, it is possible that we may be guessing the future while
holding too small deck of cards and IceCube will open a new world that
we did not anticipate.

The case for doing particle physics with cosmic neutrinos is
compelling, the challenge has been to deliver the technology to build
the instrumentation for a neutrino detector with the largest possible
effective telescope area to overcome the small neutrino cross section
with matter, and the best possible angular and energy resolution to
address the wide diversity of possible signals. We discuss this next.

\section{Overall Detector Performance}
\label{ODP}

In deep ice neutrinos are detected by observation of the \v {C}erenkov
light emitted by charged particles produced in charged current (CC)
and neutral current (NC) interactions.  To a first approximation, a
neutrino (of energy $E_\nu$) incident on a side of area $L^2$ will be
detected provided it interacts within the lattice of photomultiplier
tubes (PMT's) constituting the sensitive volume $\sim L^3$ of the
detector.  That probability is
\begin{equation}
P (E_\nu) = 1 - \exp[-L/l_{\nu} (E_\nu)] \simeq L/l_{\nu}(E_\nu) \,,
\end{equation}
where $l_{\nu} (E_\nu) = [\rho_{\rm ice} \ N_A \ 
\sigma_{\nu N} (E_\nu) ]^{-1}$ is the 
mean free path. Here $\rho_{\rm ice} = 0.9~{\rm g}\ {\rm cm}^{-3}$ is
the density of the ice, $N_A = 6.022 \times 10^{23}$ 
is Avogadro's number and
$\sigma_{\nu N}(E_\nu)$ is the neutrino nucleon cross section.
A neutrino flux $dF/dE_\nu$ (neutrinos per GeV per cm$^2$ per s) crossing a
detector with energy threshold $E_\nu^{\rm th}$ and cross sectional area 
$A\ (\simeq L^2)$ facing the
incident beam 
will produce
\begin{equation}
{\cal N} = T \, \int_{E_\nu^{\rm th}} 
A(E_\nu) \,\, \, P (E_\nu) \, \frac{dF}{dE_\nu}\,\, dE_\nu
\end{equation}
events after a time $T$. In practice, the "effective" detector area
$A$ is not strictly equal to the geometric cross section of the
instrumented volume facing the incoming neutrino because even
neutrinos interacting outside the instrumented volume, may produce a
sufficient amount of light inside the detector to be detected.
Therefore, $A$ is determined as a function of the incident neutrino
direction by simulation of the full detector, including the trigger.
To be realistic such a simulation must account for all of the detector
properties that we discuss next.

The Antarctic Muon And Neutrino Detector Array
(AMANDA)~\cite{Andres:1999hm} is located below the surface of the
Antarctic ice sheet at the geographic South pole.  During 1993 and
1994, in an exploratory phase project, the four-string AMANDA-A array
was deployed and instrumented with 80 PMT's spaced at 10~m intervals
from 810 to 1000~m. A deeper array of 10 strings, referred to as
AMANDA-B10, was deployed during the austral summers between 1995 and
1997, to depths between 1500 and 2000~m. The instrument volume of
AMANDA-B10 forms a cylinder with diameter 120~m, overlooked by 302
PMT's. During December 1997 and January 2000, the detector was expanded
by adding nine strings of PMT's. The composite array of 19 strings and
677 PMT's forms the AMANDA-II array.

Overall, AMANDA represents a proof of concept for the kilometer-scale
neutrino observatory, IceCube~\cite{Ahrens:2002dv}, now under
construction. IceCube will consist of 80 kilometer-length strings,
each instrumented with 60 10-inch photomultipliers spaced by 17~m.
The deepest module is 2.4~km below the surface. The strings are
arranged at the apexes of equilateral triangles 125\,m on a side. The 
instrumented (not effective!)
detector volume is a cubic kilometer. A surface air shower detector,
IceTop, consisting of 160 Auger-style~\cite{Abraham:2004dt} \v
{C}erenkov detectors deployed over 1\,km$^{2}$ above IceCube, augments
the deep-ice component by providing a tool for calibration, background
rejection and air-shower physics.  The expected energy resolution is
$\pm 0.1$ on a log$_{10}$ scale.  Construction of the detector started
in the Austral summer of 2004/2005 and will continue for 6 years,
possibly less. At the time of writing, data collection by the first
string has begun.

The event signatures are grouped as tracks, showers, or a combination
of the two. Tracks include muons resulting from both cosmic ray
showers and from CC interaction of muon neutrinos. Tracks can also be
produced by $\tau$ leptons arising in ultra-high energy $\nu_\tau$ CC
interactions. Showers are generated by neutrino collisions $(\nu_e$ or
low energy $\nu_\tau$ CC interactions, and all NC interactions) inside
or near the detector, and by muon bremsstrahlung radiation near the
detector. 

\subsection{Muon Tracks}
\label{MT}

In a CC event a $\nu_\mu$ produces a muon traveling in nearly the
same direction as the neutrino. Secondary muons range out over
kilometers at $E_\mu \sim 10^3$~GeV, to tens of kilometers at $E_\mu
\sim 10^9$~GeV, generating showers along their track by
bremsstrahlung, pair production and photonuclear interactions. All of
these are sources of blue \v {C}erenkov light. As the energy of the
muon degrades along its track, the energy of the secondary showers
diminishes and the distance from the track over which the associated
\v{C}erenkov light can trigger a PMT becomes smaller. The geometry of
the lightpool surrounding the muon track over which single
photo-electron are produced is therefore a kilometer-long cone with
gradually decreasing radius. The orientation of the \v {C}erenkov cone
reveals the neutrino direction, with an angular resolution $\approx
0.7^\circ$~\cite{Ahrens:2002dv}. Muons created by cosmic ray
interactions in the atmosphere constitute the main background. Full
event reconstruction, with quality cuts applied for the rejection of
cosmic ray muon background, is referred to as ``level
2''~\cite{Ahrens:2002dv}.  Down-going atmospheric neutrinos at level 2
are highly suppressed because low energy muons need to be rejected with
energy cuts.

High energy muons lose energy catastrophically according to
\begin{equation}
\frac{dE_\mu}{dl}=-\alpha - \beta_\mu E_\mu \, ,
\end{equation}
where $\alpha=2.0 \times 10^{-6}~{\rm TeV}\, {\rm cm}^2 \ {\rm
  g}^{-1}$ (characterizes the ionization process) and $\beta_\mu = 4.2
\times 10^{-6}~{\rm cm}^2 \ {\rm g}^{-1}$ (takes into account
bremsstrahlung, $e^+e^-$ pair production and nuclear
interactions)~\cite{Eidelman:2004wy}. The distance a muon travels
before its energy drops below some energy threshold $E^{\rm th}_\mu$,
called the muon range, is then given by
\begin{equation}  
l_\mu^{{\rm min}} = \frac{1}{\beta_\mu} \ln \left[ 
\frac{\alpha + \beta_\mu\, E_\mu}{\alpha + \beta_\mu\, 
E^{\rm th}_\mu} \right] \,.
\label{murange}
\end{equation}

In the first kilometer a high energy muon typically loses energy in a
couple of showers of one tenth its energy. So the initial size of the
cone is the radius of a shower with 10\% of the muon energy, e.g.\
130\,m for a 100\,TeV muon. Near the end of its range the muon becomes
minimum ionizing emitting light that creates single photoelectron
signals at a distance of just over 10\,m from the track. For 0.3
photoelectrons, the standard PMT threshold setting, this distance
reaches 45~m.

\begin{figure}[ht]
\includegraphics[width=4.5in]{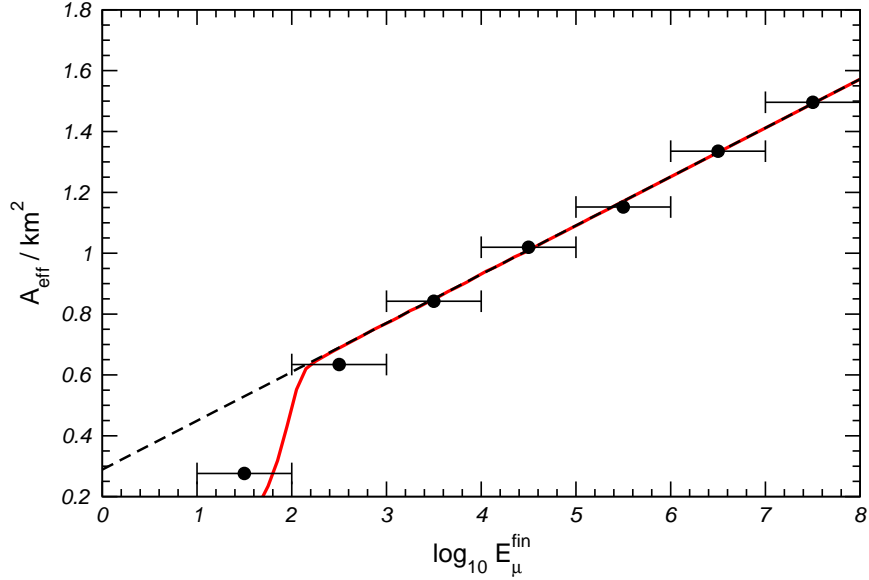}
\caption{\label{fig:aeff} Effective area as a function of the final
  muon energy (in GeV) after level 2 cuts. The full line indicates the result
  from the semi-analytical calculation~\cite{Gonzalez-Garcia:2005xw},
  whereas the data points are from Monte Carlo
  simulation~\cite{Ahrens:2003ix}. For comparison, also shown by a
  dashed line is $A_0(E_\mu^{\rm fin})$.}
\end{figure}

\begin{figure}[ht]
\includegraphics[width=5in]{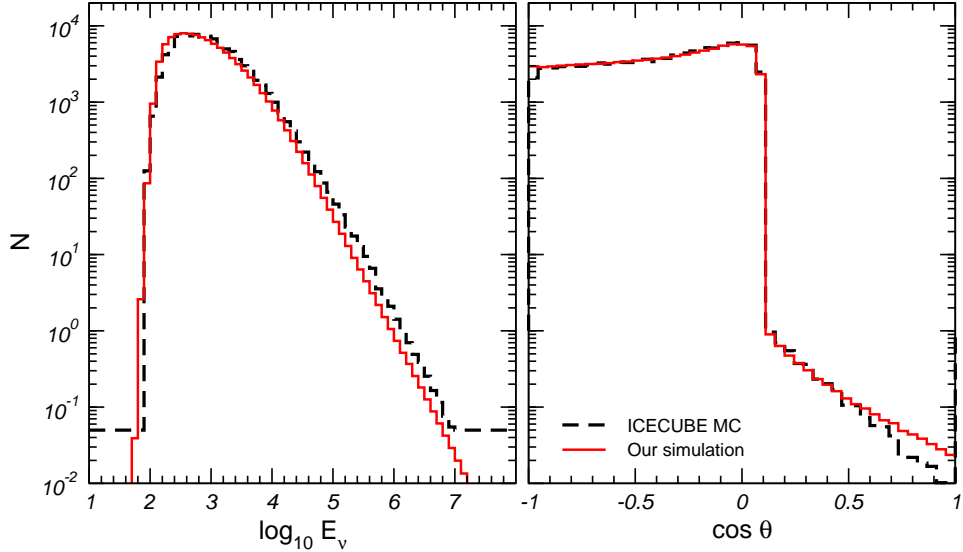}
\caption{\label{fig:nudist} Expected ${\cal N}_{\nu_\mu}$ event rate after
  level 2 cuts for one year exposure obtained with the semianalytical
  calculation (full lines)~\cite{Gonzalez-Garcia:2005xw} and with Monte Carlo
  simualtions (dashed lines)~\cite{Ahrens:2003ix}.  The left panel shows the
  energy spectrum ($E_\nu$ in GeV) and the right panel the zenith angular distribution.}
\label{fluxcv}
\end{figure}

The expected  $\nu_\mu$ event rate at IceCube can be estimated through
semi-analytical calculations~\cite{Gonzalez-Garcia:2005xw},
\begin{eqnarray}
{\cal N}_{\nu_\mu}
&=& T \int^{1}_{-1} d\cos\theta\,  
\int^\infty_{l^{\prime \rm min}_\mu} dl_\mu\,
\int_{m_\mu}^\infty dE_\mu^{\rm fin}\,
\int_{E_\mu^{\rm fin}}^\infty dE_\mu^0\, 
\int_{E_\mu^0}^\infty dE_\nu \\ \nonumber
&&\frac{dF^{\nu_\mu}}{dE_\nu\, d\cos\theta}(E_\nu,\cos\theta)
\frac{d\sigma_{\rm CC}}{dE_\mu^0}(E_\nu,E_\mu^0)\, n_T\, 
F(E^0_\mu,E_\mu^{\rm fin},l_\mu)\, A^0_{\rm eff}\, \, ,
\label{eq:numuevents}
\end{eqnarray}
where $dF^{\nu_\mu}/dE_\nu\, d\cos\theta$ is the differential muon
neutrino neutrino flux, $d\sigma_{\rm CC}/dE_\mu^0$ is the
differential CC interaction cross section producing a muon of energy
$E_\mu^0$ and $n_T$ is the number density of nucleons in the matter
surrounding the detector.  After production with energy $E_\mu^0$, the
muon ranges out in the rock and in the ice surrounding the detector
and looses energy. Here, $F(E^0_\mu,E_\mu^{\rm fin},l_\mu)$ is the
function that describes the energy spectrum of the muons arriving at
the detector.  Hence, $F(E^0_\mu,E_\mu^{\rm fin},l_\mu)$ represents
the probability that a muon produced with energy $E_\mu^0$ arrives at
the detector with energy $E_\mu^{\rm fin}$ after traveling a distance
$l_\mu$.  The function $F(E^0_\mu,E_\mu^{\rm fin},l_\mu)$ is computed
by propagating the muons to the detector taking into account energy
losses as given in Eq.~(\ref{murange}). The possibility of
fluctuations around the average muon energy loss (using the average
energy loss would equalize $l_\mu$ to the average muon range distance)
is included in $F(E^0_\mu,E_\mu^{\rm fin},l_\mu)$.  Thus, $E^0_\mu$,
$E_\mu^{\rm fin}$, and, $l_\mu$ are kept as independent variables.
For simplicity, $n_T$ and $F(E^0_\mu,E_\mu^{\rm fin},l_\mu)$ in ice
are used and the effect of the rock bed below the ice is accounted for
in the form of an additional angular dependence of the effective area
for upward going events.  The details of the detector response are
encoded in the effective area $A^0_{\rm eff}$.  A phenomenological
parametrization has been used to simulate the response of the IceCube
detector after events that are not neutrinos have been rejected (i.e.,
level 2 cuts~\cite{Ahrens:2003ix})
\begin{equation}
A^0_{\rm eff} =  A_0(E_\mu^{\rm fin})\times 
R(\cos\theta,E_\mu^{\rm fin})\times R(l_\mu^{\rm min})  \,,
\label{eq:aeff}
\end{equation}
where $A_0(E_\mu^{\rm fin})$ includes the energy dependence of 
the effective area due to trigger requirements~\cite{Ahrens:2003ix}.   
Here $R(l_\mu^{\rm min})$ represents the smearing in the minimum 
track length cut, 
$l_\mu^{\rm min}=300$~m, due to the uncertainty in the track length
reconstruction which can be parametrized by a Gaussian
\begin{equation}
R(l_\mu^{\rm min})=\frac{1}{\sqrt{2\pi} \sigma_l} \int_0^\infty 
dl_\mu^{\prime \rm min}
\exp{-\frac{(l_\mu^{\rm min}-l_\mu^{\prime \rm min})^2}{2\sigma_l^2}} \; ,
\label{eq:almin}
\end{equation}
with $\sigma_l=50$ m.
Introduction of a simple straight line dependence 
on $\log_{10}(E_\mu^{\rm fin})$
\begin{equation}
A_0(E_\mu^{\rm fin}) = {\cal A}_0 \left[1+0.55 \log_{10}
\left(\frac{E_\mu^{\rm fin}}{\rm GeV}\right)\right]    \; ,
\label{eq:a0}
\end{equation}
leads to a good agreement with the results of Monte Carlo
simulations~\cite{Ahrens:2003ix}.  Here ${\cal A}_0$ is an overall
normalization constant, which is calibrated using the flux of
atmospheric neutrinos as seen by AMANDA~\cite{Ahrens:2004qq}. After
level 2 cuts are applied the atmospheric neutrino flux yields 91000
events/yr at IceCube. Next, one has to ``simulate" the cuts in the
muon tracklength $l_\mu^{\rm min}$ and the number of PMT's reporting
signals in an event $N_{\rm CH}^{\rm min}$~\cite{Ahrens:2003ix}. The
angular dependence of the effective area for downgoing events
($\theta<80^\circ$) is determined by the level 2 cut on the minimum
number of channels $N_{\rm CH}>N_{\rm CH}^{\rm min}(\cos\theta)=150
+250\, \cos\theta$.  This requirement leads to an $E_\mu^{\rm
  fin}$-dependent angular constraint
\begin{eqnarray}
R(\cos\theta,E_\mu^{\rm fin})&=&\frac{1}{\sqrt{2\pi} \sigma_{N_{\rm CH}}}
\int_{N_{\rm CH}^{\rm min}}
^\infty d N_{\rm CH}\exp{-\frac{(N_{\rm CH}-\langle N_{\rm CH}
\rangle_{E_\mu^{\rm fin}})^2}{2\sigma^2_{N_{\rm CH}}} }\; , 
\label{eq:athetadown}
\end{eqnarray}
where $\langle N_{\rm CH} \rangle _{E_\mu^{\rm fin}}$ 
is the average channel multiplicity produced by a muon which 
reaches the detector with energy 
$E_\mu^{\rm fin}$ and $\sigma^2_{N_{\rm CH}}$ is the spread on
the distribution. The parametrization
\begin{equation}
\log_{10}\left(\langle N_{\rm CH}
 \rangle_{E_\mu^{\rm fin}}\right)=2.0+0.88\frac{X}{\sqrt{1+X^2}} \ ,
\end{equation}
with
\begin{equation}
X=0.47\left(\log_{10}\left(\frac{E_\mu^{\rm fin}}{\rm GeV}\right)-4.6\right) 
\end{equation}
and
\begin{equation}
\sigma_{N_{\rm CH}}=0.4 \, \langle N_{\rm CH}
\rangle_{E_\mu^{\rm fin}} 
\end{equation}
shows good agreement with the charged multiplicity obtained through Monte 
Carlo simulations~\cite{Ahrens:2003ix}.
Finally, to account for the presence of the rock bed below the
detector, one can introduce a phenomenological angular dependence 
of the effective area for upward going muons  
\begin{equation}
R(\cos\theta)= 0.70-0.48\,\cos\theta \hspace*{1cm} {\rm for}
\, \theta>85^\circ \; ,
\label{eq:athetaup}
\end{equation} 
independent of the muon energy.  Figure~\ref{fig:aeff} shows the
effective area $A_{\rm eff}$, defined as the ratio of the number of
upgoing muon events, with/without the inclusion of $A_0(E_\mu^{\rm
  fin})\times R(l_\mu^{\rm min})$ and the level 2 cuts on $l_\mu^{\rm
  min}$, and compare the results of the semi-analytical calculation
with Monte Carlo simulations of the detector
response~\cite{Ahrens:2003ix}.  Figure~\ref{fig:nudist} compares the
energy spectrum and the zenith angular distribution of the events
(after level 2 cuts are applied) obtained with the semi-analytical
calculation and with Monte Carlo simulations.

\subsection{Electromagnetic Showers}

Showers are generated by neutrino collisions --- $\nu_e\ \mbox{or}\,\,
\nu_\tau$ CC interactions, and all NC interactions
--- inside of or near the detector, and by muon bremsstrahlung
radiation near the detector. Normally, a reduction of the muon bremsstrahlung
background is effected by placing a cut of $4 \times 10^4$~GeV on the minimum
reconstructed energy~\cite{Ackermann:2004zw}.

Electron neutrinos deposit 0.5-0.8\% of their energy into an
electromagnetic shower initiated by the leading final state electron.
The rest of the energy goes into the fragments of the target that
produce a second subdominant shower. Because the size of the shower,
of order meters in ice, is small compared to the spacing of the PMT's,
it represents, to a good approximation, a point source of \v{C}erenkov
photons radiated by the shower particles. These trigger the PMT at the
single photoelectron level over a spherical volume whose radius scales
linearly with the shower energy. For ice, the radius is 130~m at
$10^{4}$~GeV and 460~m at $10^{10}$~GeV, i.e. the shower radius grows
by just over 50\,m per decade in energy.

\begin{figure}[!htb]
\postscript{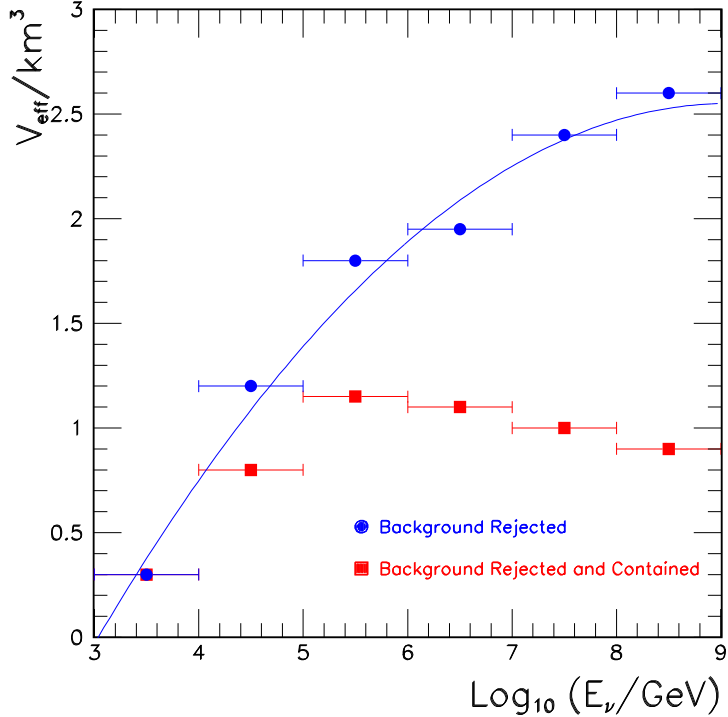}{0.6}
\caption{Effective volume of IceCube for showers initiated by neutrinos. 
The parametrization shown by a solid curve corresponds to  
${\cal V}_{\rm eff} (E_\nu) \approx - 0.07 \, {\rm log}_{10}^2 E_\nu + 1.27 
\,{\rm log}_{10} E_\nu - 3.21.$}
\label{ice3v}
\end{figure}

The measurement of the radius of the sphere in the lattice of PMT
determines the energy and turns neutrino-detection-experiments into
total energy calorimeters. Note that a contained ``direct hit" by a 
$10^{10}$~GeV neutrino will not saturate a km$^3$ detector volume. 
Because the shower and its accompanying \v{C}erenkov lightpool are not
totally symmetric but elongated in the direction of the leading
electron (and incident neutrino), its direction can be reconstructed.
Pointing is however far inferior to what can be achieved for muon
neutrinos and estimated to be precise to ${\sim}10^\circ$ only. 

The expected shower event rate at IceCube can also be estimated through
semi-analytical calculations~\cite{Anchordoqui:2005gj},
\begin{equation}
{\cal N}_{\rm sh}  = 
{\cal N}_{\rm sh,CC}+{\cal N}_{\rm sh,NC} \,\,,
\label{FTjpg}
\end{equation}
where
\begin{widetext}
\begin{equation}
{\cal N}_{\rm sh,CC} =  T\, n_{\rm T}\, 
\int_{E^{\rm min}_{\rm sh}}^\infty {\rm d}E_{\nu} \, \sum_{\alpha=e,\tau}\frac
{dF^{\nu_\alpha}}{dE_{\nu}}(E_{\nu})
\sigma_{\rm CC}(E_{\nu})\, {\cal V}_{\rm eff} (E_\nu) \,\,,
\label{eq:shsourcecc}
\end{equation}
and
\begin{equation}
{\cal N}_{\rm sh,NC} =  T\, n_{\rm T}\,
\int_{E_{\nu}-E^{\rm min}_{\rm sh}}^\infty {\rm d}E'_{\nu}
\int_{E^{\rm min}_{\rm sh}}^\infty {\rm d}E_{\nu} \, \sum_{\alpha=e,\mu,\tau}
\frac{d F^{\nu_\alpha}}{dE_{\nu}}(E_{\nu})
\frac{d\sigma_{\rm NC}}{dE'_{\nu}}
(E_{\nu}, E'_{\nu})\, {\cal V}_{\rm eff} (E_\nu) \,\,.
\label{eq:shsourcenc}
\end{equation}
\end{widetext}
Here, $d\sigma_{\rm NC}/dE'_{\nu}$ is the differential NC
interaction cross section producing a secondary neutrino of
energy $E'_{\nu}$, and  ${\cal V}_{\rm eff} (E_\nu)$ is the effective volume shown in Fig.~\ref{ice3v}. In writing Eqs.~({\ref{eq:shsourcecc}})
and (\ref{eq:shsourcenc}) we are assuming that for contained events
the shower energy corresponds with the interacting $\nu_e$ or
$\nu_\tau$ neutrino energy ($E_{\rm
sh}=E_{\nu}$) in a CC interaction, while for NC the shower
energy corresponds to the energy in the hadronic shower $E_{\rm
sh}=E_{\nu}-E'_{\nu}\equiv E_{\nu}\, y,$
where $y$ is the usual inelasticity parameter in deep inelastic 
scattering~\cite{Halzen:1984mc}.

\subsection{Shower-Track Combo}

For $\nu_\tau$'s, CC current interactions have different signals
depending on the energy.  The $\tau$ has a decay length 
\begin{equation}
l_\tau \approx 50~{\rm m} \times (E_\tau/10^6~{\rm GeV})\,,
\label{taudecay}
\end{equation}
and loses energy in the Earth according to
\begin{equation}
\frac{dE_\tau}{dl} = - \alpha - 
\beta_\tau E_\tau\, ,
\label{tauloss}
\end{equation}
where $\beta_\tau = 8 \times 10^{-7}~{\rm cm^2} \ {\rm
  g}^{-1}$~\cite{Eidelman:2004wy}. From Eqs.~(\ref{taudecay}) and
(\ref{tauloss}) it is easily seen that for energies above about
$10^{8}$ GeV the $\tau$ is more likely to interact electromagnetically
than to decay, whereas for $E_\tau < 10^{7.5}$~GeV the energy loss
before decay is negligible.

For $\tau$ leptons less energetic than $10^6$~GeV, the shower (hadronic or
electromagnetic) from the tau decay cannot be separated from the
hadronic shower of the initial $\nu_\tau$ interaction. At a few times
$10^6$~GeV, the range of the tau becomes a few hundred meters and the
two showers produced may be easily separated and be identify as a
double bang event. At energies between $10^7 - 10^{7.5}$~GeV, the tau
decay length is comparable to the instrumented volume. In such cases,
one may observe a $\tau$ track followed by the $\tau$-decay shower
(``lollipop topology''), or a hadronic shower followed by a $\tau$
track which leaves the detector (``popillol topology'').  At energies
$> 10^{7.5}$~GeV, $l_\tau \gg 1$~km and $\tau$'s leave only a track
like muons. However, a $\tau$ going through the detector at high
energies without decaying will not deposit as much energy in the
detector as a comparable-energy muon, due to the mass difference.
(The direct-pair production process scales inversely with mass, so it
dominates tau-lepton energy loss~\cite{Becattini:2000fj} resulting in
1/20$^{\rm th}$ the light produced by a muon.) Such a tau might then
be indistinguishable from a low energy muon track, and thus might not
be flagged as an interesting event. In summary, the energy range from
$10^6 - 10^{7.5}$~GeV is the ``sweet spot'' for $\tau$ detection in
IceCube, since here one can observe all the distinctive topologies.

\vspace{1cm}

Although neutrino "telescopes" are designed as discovery instruments, be it for
particle or astrophysics, their conceptual design is very much anchored to the
observational fact that Nature produces protons and photons with energies in
excess of $10^{11}$~GeV and $10^{4}$~GeV, respectively. The cosmic ray connection
sets the scale of cosmic neutrino fluxes. We discuss this next.

\section{Luminosity}

IceCube is unique among high energy neutrino experiments in being
sensitive to ten decades in energy. Unlike terrestrial experiments
with a man-made beam, the luminosity varies significantly with the
center-of-mass energy~\cite{Gaisser:1994yf}.  In what follows we
discuss the different energy regimes accesible to IceCube.

\subsection{Atmospheric Neutrinos}

A guaranteed beam of high energy neutrinos originates in the
atmospheric cascades initiated by cosmic rays~\cite{Gaisser:2002jj}.
When protons and nuclei enter the atmosphere, they collide with the
air molecules and produce all kind of secondary particles, which in
turn interact or decay or propagate to the ground, depending on their
intrinsic propeties and energies. In the GeV range, the most abundant
particles are neutrinos.  The production mechanism is the decay chains
of mesons created in these cascades.

\begin{figure}[!thb]
\postscript{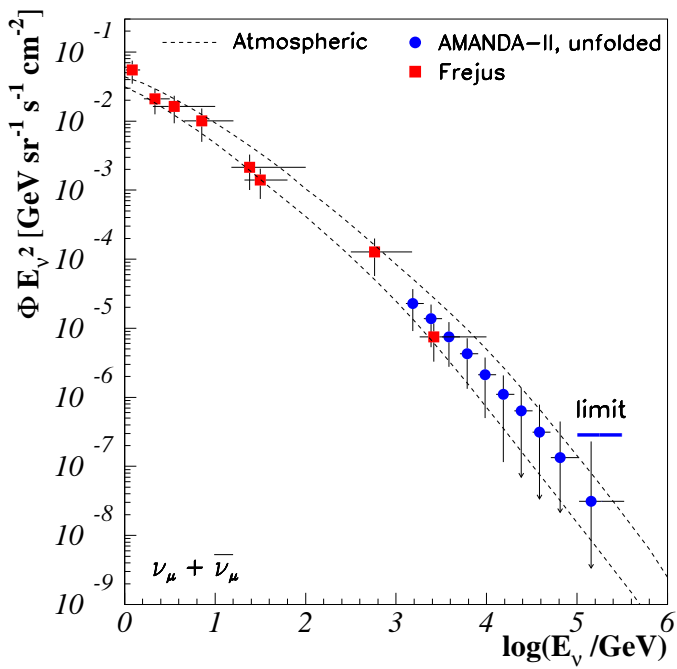}{0.6}
\caption{Energy spectrum of atmospheric neutrinos as seen by the
  AMANDA II~\cite{Ribordy:2005fi} and Frejus~\cite{Daum:1994bf}
  experiments. The two dotted curves indicates model
  predictions~\cite{Volkova:1980sw} for the horizontal (upper) and
  vertical (lower) flux.}
\label{atmflux}
\end{figure}

Pion decay dominates the atmospheric neutrino production: $\pi^+ \to
\mu^+ \nu_\mu \to e^+ \nu_e \nu_\mu \overline \nu_\mu$ and the
conjugate process. This decay chain determines the neutrino energy
spectra up to about 100 GeV above which they become increasingly 
 modified by the kaon contribution, which asymptotically reaches 90\%. In the 
atmosphere mesons encounter the interaction--decay
competition.  Therefore, neutrinos from meson decay will have a spectrum one
power of energy steeper than the primary cosmic ray spectrum. The muon
daughter neutrinos will have a spectrum steeper by two powers of
energy, because the muon spectrum itself is steeper by $1/E$.
Electron neutrinos have a differential spectrum (approximately)
$\propto E_\nu^{-4.7}.$ The muon neutrino spectrum is
flatter,
\begin{equation}
dF_{\rm atm}^{\nu_\mu}/dE_\nu \equiv \phi_{\rm atm}^{\nu_\mu} \approx 
\left\{\begin{array}{c} 
1.6 \times 10^{-16}\, E_\nu^{-3.7}~{\rm GeV}\, {\rm cm}^{-2}\, 
{\rm s}^{-1}\, {\rm sr}^{-1}\,\,\,\,\,\,\,\,E_\nu < 10^5~{\rm GeV} \\
5.0 \times 10^{-18}\, E_\nu^{-4.0}~{\rm GeV} \, {\rm cm}^{-2} \, {\rm s}^{-1}\, {\rm sr}^{-1} \,\,\,\,\,\,\,\,E_\nu > 10^5~{\rm GeV}
\end{array} \right. \,\,.
\end{equation}
In this energy window, the flavor ratios are $w_e : w_\mu :
w_\tau \approx 1/20 : 19/20 : 0$~\cite{Lipari:1993hd} and, 
as can be seen in Fig.~\ref{atmflux}, the energy 
spectra are a
function of the zenith angle of the atmospheric cascades. This is
because mesons in inclined showers spend more time in tenuous
atmosphere where they are more likely to decay rather than interact.
For this reason the spectra of highly inclined neutrinos are flatter
than those of almost vertical neutrinos. In the sub-GeV
energy range the spectra are significantly flatter (parallel to the
primary cosmic ray spectrum) as all mesons and muons decay. Thus, one
can immediately predict that the flavor ratios at production are 
$1/3 :2/3: 0$. Above about $10^5$~GeV, kaons are also significantly
attenuated before decaying and the ``prompt'' component, arising
mainly from very short-live charmed mesons ($D^\pm,$ $D^0,$ $D_s$ and
$\Lambda_c$) dominates the spectrum~\cite{Zas:1992ci}. Such a ``prompt'' 
neutrino flux is isotropic with flavor ratios $12/25 : 12/25 : 1/25.$

The neutrino flux arising from pion and kaon decay is reasonable well
understood, with an uncertainty in the range $10\% -
20\%$~\cite{Gaisser:2002jj}.  The prompt atmospheric neutrino flux,
however, is much less understood, because of uncertainty about cosmic
ray composition and relatively poor knowledge of small-$x$ QCD
processes~\cite{Beacom:2004jb}. For numerical estimates, throughout
this review we have chosen to use the 3-dimensional estimates of
conventional atmospheric neutrino flux given in~\cite{Honda:2004yz},
which we extrapolate to match at higher energies the 1-dimensional
calculations of Volkova~\cite{Volkova:1980sw}. We also incorporate
``prompt'' neutrinos from charm decay as calculated
in~\cite{Bugaev:1998bi}.

In 10~yr of operation IceCube will collect more than $7 \times 10^{5}$
atmospheric neutrino events with $E_\mu^{\rm fin} > 100$~GeV. These
events are generated by neutrinos with large enough energy for
standard mass-induced oscillations to be very much suppressed so they
should behave as flavor eigenstates. As we discuss in Sec.~\ref{FM},
this high-statistics high-energy event sample offers a unique
opportunity to probe new physics mechanisms for leptonic flavor
mixing.

\subsection{Extraterrestrial Neutrinos}

Interactions of ultra-high energy cosmic ray protons propagating over
cosmological distances with the cosmic microwave background (CMB) generates
a nearly guaranteed cosmogenic flux of
neutrinos~\cite{Berezinsky:1969} through the decay of charged pions
produced in $p \gamma$ interactions, which should also result in a
suppression of the cosmic ray spectrum above the ``GZK
cutoff'': $E_{\rm GZK} \sim 10^{10.7}$~GeV~\cite{Greisen:1966jv}. 
The intermediate state of the reaction $p \gamma_{\rm CMB} \to N
\pi$ is dominated by the $\Delta^+$ resonance, because the $n$ decay
length is smaller than the nucleon mean free path on the relic
photons.  Hence, there is roughly an equal number of $\pi^+$ and
$\pi^0$. Gamma rays, produced via $\pi^0$ decay, subsequently cascade
electromagnetically on the cosmic radiation fields through $e^+ e^-$
pair production followed by inverse Compton scattering.  The net
result is a pile up of $\gamma$ rays at GeV energies, just below the
threshold for further pair production.  On the other hand, each
$\pi^+$ decays to 3 neutrinos and a positron.  The $e^+$ readily loses
its energy through synchrotron radiation in the cosmic magnetic
fields.  The neutrinos carry away about 3/4 of the $\pi^+$ energy, and
therefore the energy in cosmogenic neutrinos is about 3/4 of the one
produced in $\gamma$-rays.

The normalization of the neutrino flux depends critically on the
cosmological evolution of the cosmic ray sources and on their proton
injection spectra~\cite{Yoshida:pt,engel}. It also depends on the
assumed spatial distribution of sources; for example, relatively local
objects, such as sources in the Virgo cluster~\cite{Hill:1985mk},
would dominate the high energy tail of the neutrino spectrum.  Another
source of uncertainty in the cosmogenic neutrino flux is the energy at
which there is a transition from Galactic to extragalactic cosmic rays
as inferred from a change in the spectral slope~\cite{Fodor:2003ph}.  
A fourth source of uncertainty in the
cosmogenic flux is the chemical composition --- if ultra-high energy
cosmic rays are heavy nuclei rather than protons the corresponding
cosmogenic neutrino flux may be somewhat
reduced~\cite{Hooper:2004jc}.

In addition to being produced in the propagation of ultra-high energy
cosmic rays, neutrinos are also expected to be generated in their
sources.  How many neutrinos are produced in association with the
cosmic ray beam? The answer to this question, among many others,
provides the rationale for building kilometer-scale neutrino
detectors. We first consider a neutrino beam produced at an
accelerator laboratory. The accelerator beam is ``dumped'' into a large
target, for instance a hundred meters of steel. The target absorbs all
parent protons as well as the secondary electromagnetic and hadronic
showers. Only neutrinos, the decay products of charged pions, exit the
dump. If Nature constructed such a "hidden source" in the heavens,
conventional astronomy will not reveal it. It cannot be the source of
the cosmic rays, however, because in this case the dump must be
transparent to protons.

A more generic "transparent" source can be imagined as follows:
protons are accelerated in a region of high magnetic fields near a
black hole or neutron star. They inevitably interact with the
radiation surrounding collapsed objects via the processes
%
$p + \gamma \rightarrow \Delta \rightarrow \pi^0 + p$
and
$p + \gamma \rightarrow \Delta \rightarrow \pi^+ + n$.
%
While the secondary protons may remain trapped in the acceleration
region, equal numbers of neutrons, neutral and charged pions escape/decay.
The energies escaping the source in the form of cosmic rays, gamma
rays and neutrinos produced by the decay of neutrons and neutral and
charged pions, are related by the physics of photoproduction. The
neutrino flux from a generic transparent cosmic ray source is often
referred to as the Waxman-Bahcall (WB) flux~\cite{Waxman:1998yy}. It is easy to
calculate and the derivation is revealing.

We will concentrate on the population of extragalactic cosmic rays
with energies in excess of  $10^{9.9}$~GeV, i.e., beyond the
``ankle'' in the spectrum. The flux above the ankle is often
summarized as "one $10^{10}$\,GeV particle per kilometer square per
year per steradian". This can be translated into an energy flux
\begin{eqnarray}
E \left\{ E{dN_{\rm CR}\over dE} \right\} & = & {10^{10}\,{\rm GeV} 
\over \rm (10^{10}\,cm^2)(3\times 10^7\,s) \, sr} \nonumber \\
 & = & 3\times 10^{-8}\rm\, GeV\ cm^{-2} \, s^{-1} \, sr^{-1} \,.
\end{eqnarray}
From this we can derive the energy density $\rho_E$ in cosmic rays using  
flux${}={}$velocity${}\times{}$density, or
\begin{equation}
4\pi \int  dE \left\{ E{dN_{\rm CR}\over dE} \right\} =  c\rho_E\,.
\end{equation}
We obtain
\begin{equation}
\rho_E = {4\pi\over c} \int_{E_{\rm min}}^{E_{\rm max}} {3\times 10^{-8}\over E} 
dE \, {\rm {GeV\over cm^3}} \simeq 10^{-19} \, {\rm {TeV\over cm^3}} \,,
\end{equation}
taking the extreme energies of the accelerator(s) to be $E_{\rm max} /
E_{\rm min} \simeq 10^3$.

This energy density derived more carefully by integrating the spectrum
beyond the ``ankle" assuming an $E^{-2}$ energy spectrum with a GZK
cutoff is closer to $\sim 3 \times 10^{-19}\rm\,erg\ 
cm^{-3}$~\cite{Waxman:1995dg}.  The
power required for a population of sources to generate this energy
density over the Hubble time of $10^{10}$\,years is $\sim 3 \times
10^{37}\rm\,erg\ s^{-1}$ per (Mpc)$^3$ or, as often quoted in the
literature, $\sim 5\times10^{44}\rm\,TeV$ per year per (Mpc)$^3$. This
works out to~\cite{Gaisser:1997aw}
\begin{itemize}
\item $\sim 3 \times 10^{39}\rm\,erg\ s^{-1}$ per galaxy,
\item $\sim 3 \times 10^{42}\rm\,erg\ s^{-1}$ per cluster of galaxies,
\item $\sim 2 \times 10^{44}\rm\,erg\ s^{-1}$ per active galaxy, or
\item $\sim 2 \times 10^{52}$\,erg per cosmological gamma ray burst.
\end{itemize}
The coincidence between these numbers and the observed output in
electromagnetic energy of these sources explains why they have emerged
as the leading candidates for the cosmic ray accelerators. The
coincidence is consistent with the relationship between cosmic rays
and photons built into the "transparent" source. In the
photoproduction processes roughly equal energy goes into the secondary
neutrons, neutral and charged pions whose energy ends up in cosmic
rays, gamma rays and neutrinos, respectively. For $E_{\rm max} =
10^{12}$\,GeV, the generic source of the highest energy cosmic rays
produces a flux of $ {E_\nu}^2 dN_\nu / dE_{\nu} \sim 4 \times
10^{-8}\rm\, GeV \,cm^{-2}\, s^{-1}\, sr^{-1}$, close to the
back-of-the-envelope calculation performed earlier.

There are several ways to modify this simple prediction:
\begin{itemize} 
\item The derivation fails to take into account the fact that there are more
  cosmic rays in the Universe producing neutrinos than observed at Earth because
  of the GZK-effect and neglects evolution of the sources with redshift. This
  increases the neutrino flux by a factor $d_H/d_{\rm CMB}$, the ratio of the Hubble
  to the attenuation length radius over which cosmic rays above the
  ankle penetrate the cosmic microwave background.
\item For proton-$\gamma$ interactions muon neutrinos (and muon
  antineutrinos receive only 1/2 of the energy of the charged pion in
  the decay chain $\pi^+\rightarrow \mu^+ +\nu_{\mu}\rightarrow e^+
  +\nu_e +\bar{\nu}_{\mu} +\nu_{\mu}$ assuming that the energy is
  equally shared between the 4 leptons. Furthermore half the muon
  neutrinos oscillate into tau neutrinos over cosmic distances.
\end{itemize}

In summary,
\begin{equation}
E_\nu{dN_\nu\over dE_\nu} = {1\over2} \times {1\over2} 
\times E{dN_{\rm CR}\over dE} \times {d_H\over d_{\rm CMB}} 
\simeq E{dN_{\rm CR}\over dE}
\end{equation}
The corrections approximately cancel. The transition from galactic to
extragalactic sources is debated; a transition at lower energy
significantly increases the energy in the extragalactic component.
This raises the possibility of an increase in the associated neutrino
flux~\cite{Ahlers:2005sn}. Waxman and Bahcall referred to this as a
bound because in reality more energy is transferred to the neutron
than to the charged pion in the source; in the case of the
photoproduction reaction $p + \gamma \rightarrow \Delta \rightarrow
\pi^+ + n$ it is four times more.  Therefore
\begin{equation}
E_\nu{dN_\nu\over dE_\nu} = {1\over4} E{dN_{\rm CR}\over dE} \, .
\end{equation}
This is, for instance, the diffuse flux expected from all gamma ray bursts.

In the end we estimate that for each neutrino flavor the flux
associated with the sources of the highest energy cosmic rays is
loosely confined to the range,
\begin{equation}
E_\nu^2 \phi_{\rm WB}^\nu \simeq 1.3 \times 10^{-8}~{\rm GeV} \, 
{\rm cm}^{-2} \, {\rm s}^{-1}\, {\rm sr}^{-1} \, ,
\label{WB}
\end{equation}
yielding $\sim 10 - 50$ detected
muon neutrinos per km$^2$ per year. This number depends weakly on
$E_{\rm max}$ and the spectral slope. The observed event rate
is obtained by folding the predicted flux with the probability that
the neutrino is actually detected in a high energy neutrino telescope;
only one in a million is at TeV energy. As we discussed in Sec.~\ref{MT}, this 
probability is given by
the ratio of the muon and neutrino interaction lengths in the detector
medium, $l_\mu / l_\nu$.

This flux has to be compared with the sensitivity of
${\sim}10^{-7}\rm\, GeV\ cm^{-2}\, s^{-1}\,sr^{-1}$ reached during the
first 4 years of operation of the completed AMANDA detector in
2000--2003~\cite{Ahrens:2003qd}. The analysis of the data has not been
completed, but a preliminary limit of $2.9 \times 10^{-7}\rm\,GeV\
cm^{-2}\,s^{-1}\,sr^{-1}$ has been obtained with a single year of
data~\cite{Ahrens:2003ee}. On the other hand, after three years of
operation IceCube will reach a diffuse flux limit of $E_{\nu}^2 \phi^{\nu_\mu}_{\rm IceCube} \sim 
2\, - \, 7 \times 10^{-9}\rm\,GeV \,cm^{-2}\, s^{-1}\,
sr^{-1}$. The exact value depends on the magnitude of the dominant
high energy atmospheric neutrino background from the prompt decay of
atmospheric charmed particles~\cite{Ahrens:2003ix}. As noted in the 
previous section, the level of this background
is difficult to anticipate. A cosmic flux at the "Waxman-Bahcall"
level will result in the observation of several hundred neutrinos in
IceCube~\cite{Ahrens:2003ix}.

\begin{figure} [t]
\postscript{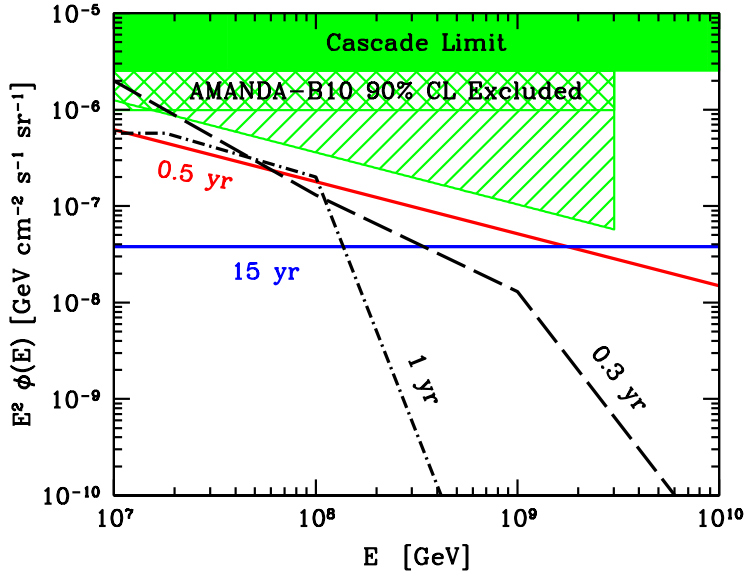}{0.6}
\caption{Neutrino fluxes from transparent and hidden
  sources. The horizontal solid line indicates the WB prediction
  (summed over all flavors) which corresponds to a
  Galactic/extra-galactic crossover energy $\sim
  10^{10}$~GeV~\cite{Waxman:1998yy}.  The falling solid line indicates
  the expected neutrino flux from transparent sources, if one
  assumes the onset of dominance by the extra-galactic component is at
  $10^{8.6}$~GeV~\cite{Ahlers:2005sn}.  The dash and dashed-dotted
  lines indicate neutrino flux estimates from hidden
  sources~\cite{Stecker:1991vm}.  The label on each
  curve indicates the time required for IceCube to achieve an
  integrated luminosity ${\cal L} \approx 30~{\rm nb}^{-1}$ at
  $\sqrt{s} \simeq 6$~TeV. The cross-hatched region excludes an
  $E_\nu^{-2}$ spectrum at the 90\% CL by measurements of
  AMANDA-B10~\cite{Ackermann:2005sb}.  The single hatched region,
  obtained by rescaling the AMANDA integrated bolometric flux limit to
  an $E_\nu^{-2.54}$ power law, is the exclusion region for the low
  crossover model. The shaded region indicates the cascade
  limit.}
\label{reachlum}
\end{figure}

In Fig.~\ref{reachlum}, we plot the expected spectrum of ultra-high
energy cosmic neutrinos in the models discussed above. For comparison
we show the upper limits on neutrino fluxes reported by the AMANDA
Collaboration~\cite{Ackermann:2005sb} and the cascade
limit.\footnote{The cascade limit arises from the requirement that the
  diffuse $\gamma$-ray fluxes initiated by photons and electrons from
  pion decays should not exceed measurements.  The cascade limit from
  Ref.~\cite{Mannheim:1998wp} exploits the measurement of the diffuse
  $\gamma$-ray background by EGRET~\cite{Sreekumar:1997un}.  A lower
  extragalactic contribution to the $\gamma$-ray background than that
  inferred in above reference, by roughly a factor of two, has been
  proposed recently~\cite{Strong:2003ex}.  The cascade limit may
  therefore be stronger by a corresponding factor. Note that
  logarithmic corrections to the spectrum $\propto E_\nu^{-2}$ could
  result in sizeable deviations of the cascade limit at very high
  energy~\cite{Anchordoqui:2004bd}.} Hidden sources must have a
steeper spectrum than transparent sources in order to be consistent
with data. In particular, the flux could be enhanced by more than an
order of magnitude in the energy decade between $10^{7 -8}$~GeV.
Therefore, the energy region $E_\nu\simeq 10^{7-7.5}$~GeV is
particularly amenable to probe high energy neutrino interactions:
commonly proposed non-atmospheric neutrino fluxes are still sizeable,
while the anisotropic background from atmospheric neutrinos has become
insignificant. This insures a statistically significant sample of
neutrino collisions with center-of-mass energy $\sqrt{s}\simeq 6$~TeV.
It should be noted that this energy is well above the HERA domain
$(\sqrt{s}\simeq 500$~GeV), the highest accelerator energy at which
even indirect tests of the Standard Model cross sections are possible.

\subsection{Directional Signals}
\label{nDK}

MeV neutrino astronomy has been possible for about 40 years. Thus 
far two sources have been identified: the Sun~\cite{Fukuda:1998fd} and 
supernova 1987a (SN87a)~\cite{Hirata:1987hu}.  
During the next decade, the  large sensitivity of IceCube to 
extraterrestrial neutrino fluxes gives a hope to open a new era in neutrino 
astronomy through detection of high energy neutrinos.

One of the key features of high energy neutrinos is that they are unique
carriers of unambiguous information about hadronic process in cosmic
accelerators~\cite{Anchordoqui:2004eb}. Since generally these processes also
result in TeV gamma-rays of comparable fluxes~\cite{Alvarez-Muniz:2002tn}, the
best candidates to be discovered by neutrino telescopes are sources of
gamma-rays with fluxes above 1~TeV around $(3-10) \times 10^{-12}~{\rm ph} \,
{\rm cm}^{-2}\, {\rm s}^{-1}.$ Presently several TeV gamma-ray sources, in
particular persistant galactic sources like Crab Nebula~\cite{Aharonian:2000pz}, shell type SNRs
RXJ1713~\cite{Aharonian:2004vr} and Vela Jr~\cite{Aharonian:2005sz}, as well as
strongly variable Active Galactic Nuclei Mrk 421, Mkn 501, and 1ES
1959~\cite{Aharonian:2004kd} do show fluxes of TeV gamma-rays above
$10^{-11}~{\rm ph} \, {\rm cm}^{-2}\, {\rm s}^{-1}.$ Therefore, if significant
fractions of these gamma-ray fluxes are of hadronic origin IceCube should be
able to detect the neutrino counterparts of TeV gamma-rays. Moreover, in the
case of sources with heavily absorbed TeV gamma-ray emission, the chances of
detection of countrerpart neutrino emission could be quite high even for
relatively weak TeV sources with a flux well below the level 0.1
Crab~\cite{Aharonian:2005cx}.

Antineutrinos are also produced through the $\beta$-decay process, but since 
cosmic neutrino production is dominated by $\pi^\pm$ decay, 
the search for the $\overline \nu_e$ channel is not straightforward. A 
potential way is to identify neutron-emitting-sources. However,
the decay mean free path of a neutron
is $c\,\Gamma_n\,\overline\tau_n=9.15\,(E_n/10^9~{\rm GeV})$~kpc (the
lifetime being boosted from its rest-frame value,
$\overline\tau_n=886$~seconds, to its lab value by
$\Gamma_n=E_n/m_n$), and so only neutrons with
energy $\agt 10^{9}$~GeV have a boosted $c\tau_n$ sufficiently large
to serve as Galactic messengers. 

Air shower arrays have observed a ``directional'' flux of cosmic rays
from the galactic plane~\cite{Bird:1998nu}, unlikely to be protons
since their directions are scrambled in the Galactic magnetic field. 
The flux appears
only in a narrow energy range from $10^{8.9}$ to $10^{9.5}$~GeV, the
energy where neutrons reach typical galactic kiloparsec distances
within their lifetime of minutes. Both the directionality and the
characteristic energy make a compelling case for electrically neutral
neutron primaries.  The galactic plane excess, which is roughly 4\% of
the diffuse flux, is mostly concentrated in the direction of the
Cygnus region~\cite{Teshimaicrc}.

Independent evidence may be emerging for a cosmic accelerator in the
Cygnus spiral arm. The HEGRA experiment has detected an extended TeV
$\gamma$-ray source, J2032+4130, in the Cygnus region with no clear
counterpart and a spectrum not easily accommodated with synchrotron
radiation by electrons~\cite{Aharonian:2002ij}. The difficulty to
accommodate the spectrum by conventional electromagnetic mechanisms
has been exacerbated by the failure of CHANDRA and VLA to detect
X-rays or radiowaves signaling acceleration of any
electrons~\cite{Butt:2003xc}.  The model proposed is that of a proton
beam, accelerated by a nearby mini-quasar or possibly Cygnus X-3,
interacting with a molecular cloud to produce pions that are the
source of the gamma rays. Especially intriguing is the possible
association of this source with Cygnus OB2, a cluster of more than
2700 (identified) young, hot stars with a total mass of $\sim 10^4$
solar masses~\cite{Knodlseder:2000vq}. Proton acceleration to explain
the TeV photon signal requires only 0.1\% efficiency for the
conversion of the energy in the stellar wind into cosmic ray
acceleration. Also, the stars in Cygnus OB2 could be the origin of
time-correlated, clustered supernova remnants forming a source of
cosmic ray nuclei. By cooperative acceleration their energies may even
exceed the $\sim 10^6$~GeV cutoff of individual remnants and
accommodate cosmic rays up to the ankle, where the steeply falling
($\propto E^{-3.16\pm 0.08}$) cosmic ray spectrum flattens to $E^{-2.8
  \pm 0.3}$~\cite{Nagano:1991jz}.  An immediate consequence of this
nucleus-dominance picture is the creation of free neutrons via nuclei
photodisintegration on background photon fields.  These liberated
neutrons are presumably responsible for the observed directional
signals.

The observed spectrum from the Cygnus region 
can be described by single power law
reflecting the average shape of the diffuse cosmic ray spectrum
between $10^{6}$ and $10^{8.5}$~GeV, 
\begin{eqnarray}
\frac{dF_n}{dE_n} & = 
& \frac{dF_n}{dE_n}\Big|_{\rm source}\,
e^{-D/(c\,\Gamma_n\,\overline \tau_n)} \nonumber \\
 & = &
C\, E_n^{-3.1}\, e^{- D/(c\,\Gamma_n\,\overline \tau_n)}\,,
\end{eqnarray}
where $D \approx 1.7~{\rm kpc}$ is the distance to Cygnus OB2.
By integrating the spectrum between $E_1 = 10^{8.9}~{\rm GeV}$ and
$E_2 = 10^{9.5}~{\rm GeV}$~\cite{Teshimaicrc}, normalization to the
observed integrated flux~\cite{Teshimaicrc},
\begin{equation}
\int_{E_1}^{E_2} C\, E_n^{-3.1} \,\,\,e^{- D/(c\,\Gamma_n\,\overline \tau_n)}
\,dE_n\, \approx 9~{\rm km}^{-2} {\rm yr}^{-1}\,\,,
\label{integ_n}
\end{equation}
leads to $C=1.15\times 10^{20}~{\rm km}^{-2}~{\rm yr}^{-1}$. 

For every surviving neutron at $\sim 10^9$~GeV, there are many
neutrons at lower energies that do decay via $n\rightarrow p+e^- +
\overline \nu_e.$ The proton is bent by the Galactic magnetic field
and the electron quickly loses energy via synchrotron radiation, but
the $\overline\nu_e$ travels along the initial neutron direction,
producing a directed TeV energy beam which is potentially observable.

\begin{figure}
\begin{center}
\postscript{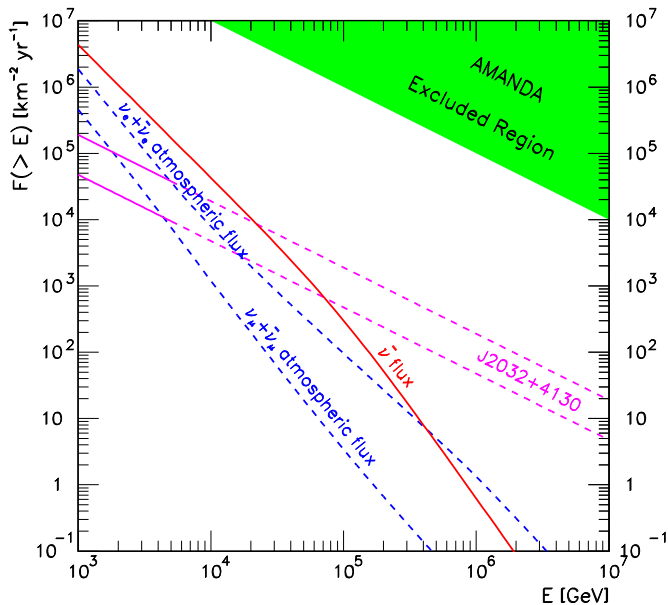}{0.55}
\caption{Integrated flux of $\overline \nu_\mu + \overline \nu_e +
\overline \nu_\tau$ (solid line) predicted to arrive at Earth from the
direction of the Cygnus region.  Also shown are the integrated
$\nu_\mu + \overline\nu_\mu$ and $\nu_e + \overline \nu_e$ atmospheric
fluxes for an angular bins of $1^\circ \times 1^\circ$ and $10^\circ
\times 10^\circ$, respectively.  The shaded band indicates the region
excluded by the AMANDA experiment~\cite{Ahrens:2003pv}.The fluxes of
neutrinos inferred from HEGRA measurements of the $\gamma$-ray flux
are also shown: the lower line is based on the assumption of $p\gamma$
interactions, whereas the upper line is based on $pp$ interactions
(the charged/neutral pion-production ratio depends on the
interaction). In each case the solid portion of the line indicates the
region where HEGRA data is available and the dashed part is an
extrapolation to unobserved energies~\cite{Anchordoqui:2005gj}.}
\label{cygOB2_nu}
\end{center}
\end{figure}

The basic formula that relates the neutron flux at the source to the
antineutrino flux observed at Earth is~\cite{Anchordoqui:2003vc}:
\begin{eqnarray}
\frac{d\Fnu}{d\Enu}(\Enu) & = &
\int  d\En\,\frac{ d\Fn}{d\En}(\En)\Big|_{\rm source}
\left(1- e^{-\frac{D\,m_n}{\En\,\tbar}}\right)\,
\int_0^Q d\Enubar\,\frac{ dP}{ d\Enubar}(\Enubar) \nonumber \\ 
 & \times & \int_{-1}^1 \frac{d\cth}{2}
\;\delta\left[\Enu-\En\,\Enubar\,(1+\cth)/m_n\right]
\,.
\label{nuflux}
\end{eqnarray}
The variables appearing in Eq.~(\ref{nuflux}) are the antineutrino and
neutron energies in the lab ($\Enu$, $\En$), the antineutrino angle
with respect to the direction of the neutron momentum in the neutron
rest-frame ($\thnu$), and the antineutrino energy in the neutron
rest-frame ($\Enubar$).  The last three variables are not observed by
a laboratory neutrino-detector, and so are integrated over.  The
observable $\Enu$ is held fixed.  The delta-function relates the
neutrino energy in the lab to the three integration variables. The
parameters appearing in Eq.~(\ref{nuflux}) are the neutron mass and
rest-frame lifetime ($m_n$ and $\tbar$). Finally, $ dP/
d\Enubar$ is the normalized probability that the decaying neutron
produces an antineutrino with energy $\Enubar$ in the neutron
rest-frame. Note that the maximum antineutrino energy in the neutron
rest frame is very nearly $Q \equiv m_n - m_p - m_e = 0.71$~MeV.

The integral neutrino flux, $\Fnu(>\Enu)\equiv\int_{\Enu} d\Enu\,
d\Fnu/d\Enu$, is particularly useful for
experiments having a neutrino detection-efficiency that is independent
of neutrino energy, or nearly so.  The integral flux,
normalized to the integrated neutron flux in Eq.~(\ref{integ_n}), is
shown in Fig.~\ref{cygOB2_nu}.  Note that the nuclear
photodisintegration threshold implies an infrared cutoff on the
primary neutron energy at the source, which in turn leads to a low
energy cutoff of ${\cal O}$(TeV) on the integral flux.

\vspace{1cm}

In summary, the AMANDA experiment using natural 1 mile-deep Antarctic ice as a
\v{C}erenkov detector is in steady operation collecting roughly 7 to 10
neutrinos per day.  Expansion into the IceCube kilometer scale neutrino
telescope will increase the event rate at least to 250 events per day. Because of this 
high-statistics data sample IceCube will be a powerful high-energy physics laboratory. 
We explore the science reach next.

\section{Neutrino Interactions beyond the Weak Scale}

The saga of the Standard Model~\cite{Weinberg:2004kv} is still
exhilarating because it leaves all questions of consequence
unanswered. The most evident of unanswered questions is why the weak
interactions are weak. Though electromagnetism and weak interactions
are unified, electromagnetism is apparent in day life while the weak
interactions are not. Already in 1934 Fermi~\cite{Fermi:1934hr}
provided an answer with a theory that prescribed a quantitative
relation between the fine structure constant and the weak coupling,
$G_F \sim \alpha/M_W^2.$ Although Fermi adjusted $M_W$ to accommodate
the strength and range of nuclear radioactive decays, one can readily
obtain a value of $M_W$ of 40 GeV from the observed decay rate of the
muon for which the proportionality factor is $\pi/\sqrt{2}.$ The
answer is off by a factor of 2 because the discovery of parity
violation and neutral currents was in the future and introduces an
additional factor $1 - M_W^2/M_Z^2,$
\begin{equation} G_F = \left[\frac{\pi \alpha}{\sqrt{2} M_W^2}
\right] \left[ \frac{1}{1 - M_W^2/M_Z^2} \right ] \, (1 + \Delta r)\,.
\end{equation}
Fermi could certainly not have anticipated that we now have a
renormalizable gauge theory that allows us to calculate the radiative
correction $\Delta r$ to his formula. Besides regular higher order
diagrams, loops associated with the top quark and the Higgs boson
contribute.

If one calculates the radiative corrections to the mass $M_H$
appearing in the Higgs potential, the same theory that withstood the
onslaught of precision experiments at LEP/SLC and the Tevatron yields
a result that grows quadratically
\begin{equation}
  \delta M_H^2 = \frac{3}{16 \pi v^2} \, 
(2 M_W^2 + M_Z^2 + M_H^2 - 4 M_t^2)\, \Lambda^2 \,\, ,
\end{equation}
where $M_H^2 = 2 \lambda v^2,$ $\lambda$ is the quadratic Higgs
coupling, $v = 246$~GeV, and $\Lambda$ is a cutoff. Upon minimization
of the potential, this translates into a dangerous contribution to the
Higgs vacuum expectation value which destabilizes the electroweak
scale~\cite{Casas:2004gh}. The Standard Model works amazingly well by fixing $\Lambda$ at the
electroweak scale $M_{\rm W}$. It is generally assumed that this
indicates the existence of new physics beyond the Standard Model. Following
Weinberg,
\begin{equation} {\cal L} (M_{\rm W}) = \frac{1}{2} \, M_H^2 H^\dagger H +
  \frac{1}{4} \, \Lambda (H^\dagger H)^2 + {\cal L}_{\rm SM}^{\rm gauge} +
  {\cal L}_{\rm SM}^{\rm Yukawa} + \frac{1}{\Lambda} \, {\cal L}^5 +
  \frac{1}{\Lambda^2} {\cal L}^6 + \dots \,,
\end{equation}
where the operators of higher dimension parametrize physics beyond
the Standard Model.  The optimistic interpretation of all this is
that, just like Fermi anticipated particle physics at 100 GeV in 1934,
the electroweak gauge theory requires new physics to tame the
divergences associated with the Higgs potential. By the most
conservative estimates this new physics is within our reach. Avoiding
fine tuning requires $\Lambda \alt 2 - 3$~TeV to be revealed by the
CERN Large Hadron Collider (LHC), possibly by the Tevatron. For
instance, for $M_H = 115 - 200$~GeV,
\begin{equation}
\left|\frac{\delta M_H^2}{M_H^2}\right| = \frac{\delta v^2}{v^2} \leq 10
\Rightarrow \Lambda = 2 -3~{\rm TeV} \,\,.
\end{equation}

Any physics which may turn on beyond the electroweak scale can easily
enhance weak interaction cross sections beyond Standard Model
predictions. In this context, neutrino telescopes provide an important
probe of new ideas in particle physics. One caveat for the realization
of such a program is that the event rates depend on the as yet unknown
extraterrestrial neutrino flux, generating an ambiguity in obtaining
cross sections from event rates. However, the event rates for
up-coming and down-going neutrinos have different responses to the
inelastic cross section~\cite{Kusenko:2001gj}.  Therefore, a full
study in the two-dimensional parameter space (cross section and flux)
can serve as a powerful testbed for probing anomalous neutrino
interactions~\cite{Anchordoqui:2001cg,Anchordoqui:2005pn}. Bounds (or
discovery potential) of neutrino fluxes emerge in a correlated fashion
with the cross section probes.

Physics at the high energy frontier is the physics of partons. We
master this physics with unforeseen precision because of a decade of
steadily improving HERA measurements of the nucleon
structure~\cite{Abramowicz:1998ii}. These now includes experiments
using targets of polarized protons and neutrons. HERA is our nucleon
microscope, tunable by the wavelength and the fluctuation time of the
virtual photon exchanged in the electron proton collision. The
wavelength of the virtual photons probing the nucleon is reduced with
increased momentum transfer $Q$. The proton has now been probed to
$10^{-3}$~fm, about one-thousandth of its size.  In the interaction,
the fluctuations of the virtual photons survive over distances $ct
\sim 1/x,$ where $x$ is the relative momentum of the parton. HERA now
studies the production of chains of gluons as long as 10~fm, an order
of magnitude larger than and probably insensitive to the proton
target. These are novel QCD structures, the understanding of which
has been challenging. The $x-Q^2$ region probed by cosmic neutrinos lies
in this challenging region, beyond the reach of the HERA experiments.
It is therefore neccesary to think of reliable methods for
extrapolating the structure functions to very low $x.$ For the purpose
of phenomenological studies, we have chosen to use the
parametrizations of the CC
\begin{equation}
\sigma_{\rm CC} = 5.53   
\left(\frac{E_\nu}{{\rm GeV}}\right)^{0.363}~{\rm pb}\,,
\label{sigmaCC}
\end{equation}
and NC 
\begin{equation}
\sigma_{\rm NC} = 2.31 
\left(\frac{E_\nu}{{\rm GeV}}\right)^{0.363}~{\rm pb}\,,
\end{equation}
neutrino-nucleon cross sections, as derived in~\cite{Gandhi:1998ri} using the
CTEQ4 parton distribution functions~\cite{Lai:1996mg}.

Deviations of the neutrino-nucleon cross sections from the simple
parton model can result from more sophisticated treatments of QCD or
from new physics beyond the Standard Model. Included in the former are
saturation effects which can substantially modify simple parton
results at small $x$~\cite{Gribov:1984tu}. These effects can
significantly alter the total cross section at high
energies~\cite{Kutak:2003bd}, softening the power law behavior
predicted by the simple parton model toward compliance with the
Froissart bound~\cite{Kwiecinski:1990tb}.  At the forefront of the new
physics category are scenarios based on TeV-scale quantum
gravity~\cite{Arkani-Hamed:1998rs}.  Enhancements of the neutrino
cross section in this framework may come from exchange of towers of
Kaluza-Klein gravitons~\cite{Nussinov:1998jt} and black hole
production~\cite{Feng:2001ib}. Similar enhancements may originate from
TeV-scale string excitations~\cite{Domokos:1998ry}, electroweak
instanton processes~\cite{Fodor:2003bn}, or in some supersymmetric
models through direct channel production of superpartner
resonances~\cite{Carena:1998gd}.

In what follows we discuss the potential of IceCube to reveal
deviations of the neutrino-nucleon cross section from Standard Model
predictions or to set limits on such deviations, without assuming
particular neutrino fluxes or cross sections. It will be convenient,
however, to present results relative to standard reference values.
For the reference cross section, we adopt $\sigmaSM$ the
neutrino-nucleon cross section of the simple parton model given in
Eq.~(\ref{sigmaCC}).\footnote{We neglect NC interactions, which at
these energies serve only to reduce the neutrino energy by
approximately 20\%.}  For the reference flux, we adopt the WB flux
$\phiWB$ given in Eq.~(\ref{WB}). Altogether, in the energy range
$10^{7 -7.5}$~GeV, after appling selection criteria neccesary for good
energy resolution, we estimate an effective integrated luminosity
(over the lifetime of the experiment) ${\cal L} \approx 30$~nb$^{-1}.$
At an average energy $\langle E \rangle = 10^{7.25}$~GeV, $\phi_{\rm
WB}^\nu \simeq 4 \times 10^{-23}$~GeV$^{-1}$ cm$^{-2}$ s$^{-1}$
sr$^{-1}$ per flavor and $\sigmaSM ( \langle E \rangle ) \simeq 2
\times 10^{-33}$~cm$^{2}.$ In order to separate the flux and cross
section contributions to the event rate, we require two independent
pieces of data, which hereafter we identify as up-coming and
down-going events.

For down-going events, the
probability of neutrino conversion is always small, barring
extraordinary enhancements to neutrino cross sections.  Letting
$\sigmanuN$ denote the total (NC $+$ CC) neutrino-nucleon cross
section, the down-going event rate is therefore
\begin{equation}
\Ndown = C_{\rm{down}} \ \frac{\phinu}{\phiWB} 
\ \frac{\sigmanuN}{\sigmaSM} \ ,
\label{down}
\end{equation}
where $\phinu$ indicates the average extraterrestrial $\nu + \overline
\nu$ flux per flavor in the energy bin of interest.  The constant
$C_{\text{down}}$ depends on exposure and varies according to neutrino
flavor from experiment to experiment. Here, $\sigmanuN$ indicates any
enhancement of the cross section which will increase the event rate
for down-going neutrinos, but because of absorption, also suppress the
up-coming events. The latter can be achieved through cuts on the
concommitant increase in shower energy fraction greater than or equal
to that characterizing the CC Standard Model process.

For up-going events, the dependence on cross section is completely
different.  At energies above $10^6~\gev$, neutrino interaction
lengths become smaller than the radius of the Earth, $R_\oplus \approx
6371$~km.  This implies that most upward-going neutrinos are shadowed
by the Earth, and only those that are traveling at large nadir angles
 along chords, $\ell = 2 R_\oplus \cos \theta$, with lengths
of order their interaction length, $l_\nu (E_\nu) = [\rho_{\rm crust}
N_A \sigma_{\nu N}(E_\nu)]^{-1}$, can produce a visible 
signal ($\rho_{\rm crust} \simeq 2.65\,\, {\rm g}\, {\rm cm}^{-3}$ 
is the density of the Earth's crust)~\cite{Halzen:1998be}.

The dependence of upward-going event rates on anomalous neutrino cross
sections depends on the source of the anomaly.  We consider two
prominent cases.  First, in many new physics cases, the Standard Model
CC cross section $\sigmaSM$ is not altered, but there are new neutrino
interactions that produce showers.  For $E_\nu > 10^7$~GeV, the 
neutrino interaction length satisfies $l_{l} \ll l_{\nu} < R_{\oplus},$ 
and so the up-going event rate 
is~\cite{Anchordoqui:2001cg}
\begin{equation}
\Nup = C_{\text{up}} \ \frac{\phinu}{\phiWB}
\ \frac{\sigmaSM^2}{\sigmanuN^2}
\quad \left(\frac{\sigmanuN}{\sigmaSM} > 1 \right) \ ,
\label{up}
\end{equation}
where $l_l$ is the typical lepton path length in Earth.  Extreme
enhancements to $\sigmanuN$ may reduce $l_{\nu}$ to $l_l$, leading to
a different parametric dependence in Eq.~(\ref{up}), but such cases
are neglected here. Next we consider the possibility of screening. In
order to probe deviations from the unscreened parton model
calculations of the cross section, it is necessary to note that the
screening correction affects CC and NC equally. In this case one
finds~\cite{Kusenko:2001gj,Anchordoqui:2001cg},
\begin{equation}
\Nup = C^{\rm{screen}}_{\rm{up}}
\ \frac{\phinu}{\phiWB} \ \frac{\sigmaSM}{\sigmanuN} \ ,
\label{upscreen}
\end{equation}
where $\sigmanuN$ and $\sigmaSM$ are CC cross sections with and
without screening, respectively.

\begin{figure}
\postscript{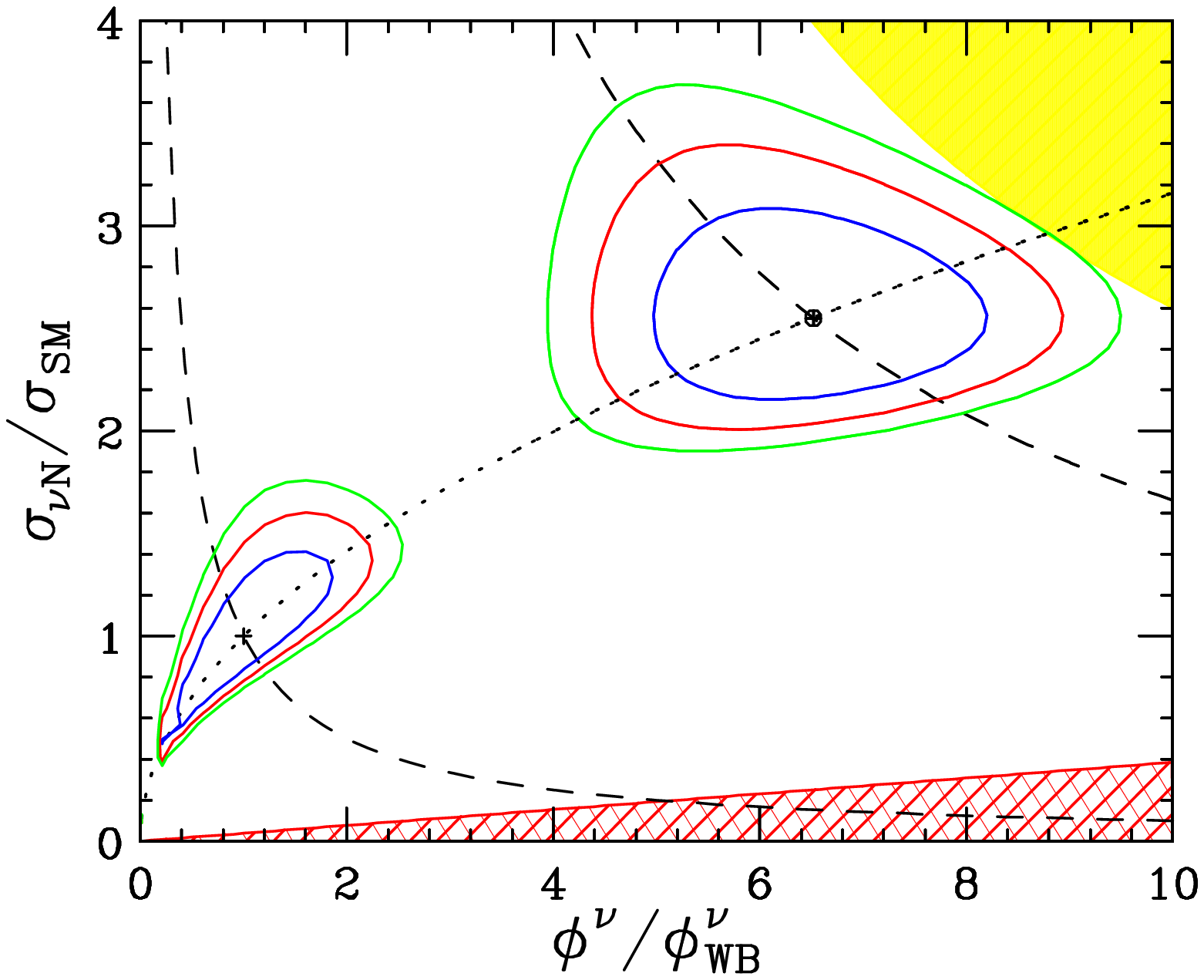}{0.6} 
\caption{Projected determinations of neutrino fluxes and cross
  sections at $\sqrt{s} \approx 6~\tev$ from future IceCube data.  Two
  cases with $(\Ndownobs, \Nupobs) = (3,25)$ and $(50,25)$ are
  considered.  In both cases, the best fit flux and cross sections are
  shown, along with the 90\%, 99\%, and 99.9\% CL exclusion contours.
  Contours of constant $\Nup = 25$ (dotted), $\Ndown = 3$ (left
  dashed), and $\Ndown = 50$ (right dashed) are also shown. Neutrino
  fluxes and cross sections excluded by AMANDA at
  90\%CL~\cite{Ackermann:2005sb} are also indicated: The shaded
  (cross-hatched) region is excluded by null results for down-going
  (up-going) events~\cite{Anchordoqui:2005pn}.}
\label{IceCubesigma}
\end{figure}

Constraints on the two-parameter space can be obtained using AMANDA data.
To see this, we note that the 90\% CL upper bound on the diffuse
neutrino flux 
\begin{equation}
E_\nu^2\,\phinumax = 3.3 \times
10^{-7}~\gev~\cm^{-2}~\s^{-1}~\sr^{-1}
\end{equation}
(per flavor) reported by the AMANDA Collaboration~\cite{Ackermann:2005sb} was
derived on the basis of CC neutrino interactions within Standard Model physics,
assuming an $E_\nu^{-2}$ dependence of the flux, valid for $10^6~\gev$ to
$10^{9.5}~\gev$ neutrinos. In full generality, the stated quantity $\phi_{\rm
max}^\nu$ constitutes two joint constraints on {\em two} unknowns: the diffuse
neutrino flux and any deviation of the simple parton model neutrino-nucleon
cross sections. Since the energy 
distribution of the AMANDA data peaks in the energy bin of interest, it is 
reasonable to use $\phinumax (\langle E \rangle)$ as the upper limit 
in the bin. Applied to down-going events, the constraint implies
\begin{equation}
\phinu  \ \frac{\sigmanuN}{\sigmaSM} < \phinumax \ . \label{Ad}
\end{equation}
Dividing Eq.~\eqref{Ad} by $\phiWB$ gives
\begin{equation}
\frac{\phinu}{\phiWB} \ \frac{\sigmanuN}{\sigmaSM} < 26
\label{downconstraint}
\end{equation}
at 90\% CL. A similar analysis for up-going events yields
\begin{equation}
\frac{\phinu}{\phiWB} \ \frac{\sigmaSM^2}{\sigmanuN^2} < 26 
\label{upconstraint}
\end{equation}
for the case of new physics contributions, and
\begin{equation}
\frac{\phinu}{\phiWB} \ \frac{\sigmaSM}{\sigmanuN} < 26
\label{upconstraintscreen}
\end{equation}
for the case of screening.
These constraints exclude the shaded and cross-hatched regions of
Fig.~\ref{IceCubesigma}.  The shaded region is excluded by down-going data,
and the cross-hatched region is excluded by the up-going data,
assuming screening.  The shaded and cross-hatched regions meet at
$\sigmanuN / \sigmaSM =1$.  As a result, irrespective of cross section
assumptions, one finds an upper bound on the neutrino flux in the energy
range $10^7~\gev$ to $10^{7.5}~\gev$ of $\phinu < 26 \phiWB$.

How will these results improve in the near future?  
To evaluate the prospects for IceCube, one must determine
$C_{\rm{down}}$ and $C_{\rm{up}}$.  We focus on the
case in which new neutrino interactions modify the NC cross sections,
but leave the CC cross sections invariant.  In NC processes most of
the energy is carried off by pions. At TeV energies, the interaction
mean free path of $\pi^\pm$ in ice is orders of magnitude smaller than
the pion decay length, and so nearly all energy is channeled into
electromagnetic modes through $\pi^0$ decay. To estimate the
efficiencies for down-going events in the NC channel, we therefore
adopt as our basis of comparison the electromagnetic showers induced
by $\nu_e$. With this in mind,
\begin{eqnarray}
C_{\rm{down}} & = & 2 \pi \, T\, \int \phiWB \,\, \sigmaSM \, \, 
n_T \,\,d E_\nu \nonumber \\
 & \approx &  2 \pi\, \phiWB (\langle E \rangle) \,\,
\sigmaSM (\langle E \rangle)\, \, n_T (\langle E \rangle)\,\, T 
   \Delta E \,,
\label{cdown}
\end{eqnarray}
where $n_T(\langle E \rangle) \simeq 6 \times 10^{38}$.  Inserting
these numbers into Eq.~\eqref{cdown}, along with a lifetime of the
experiment $T = 15~\yr$ and $\Delta E = 2.2 \times 10^7~\gev$, one
obtains $C_{\rm{down}} \simeq 3$.  The $\nu_e$ atmospheric background
is taken to be zero, (the potential ``prompt'' component can be
eliminated by taking advantage of the isotropic $\nu_\mu$-channel).
To determine the corresponding quantity for up-going events
$C_{\rm{up}}$, we first note that the absence of oscillation precludes
a $\nu_\tau$ atmospheric background.  For this reason the detection
prospects for up-going neutrinos (with $10^7~{\rm GeV} < E_\nu <
10^{7.5}~{\rm GeV}$) are brighter for $\nu_{\tau}$ than the other
flavors, and we focus on them below.  From Figs.~\ref{fig:aeff} and
\ref{ice3v}, after correcting for the Earth's absorption effects, one
obtains the $\nu_\tau$ effective aperture, $(A\Omega)_{\rm{eff}} \sim
\pi~\km^2~\sr$. Thus, the normalization constant for $\tau$'s
showering in IceCube is found to be
\begin{equation}
C_{\rm{up}} = (A
\Omega)_{\rm{eff}} \ T \int \phi^\tau(E) \,\,dE \approx 25,
\end{equation}
where~\cite{Dutta:2005yt} 
\begin{eqnarray}
\int
\phi^\tau(E) dE & \approx & \int 2 \times 10^{-3} \, 
\phi_{\rm WB}^{\nu} (E_\nu) \,\, dE_\nu
\nonumber \\
 & \approx & 5.5 \times 10^{-1}~\km^{-2}~\yr^{-1}~\sr^{-1}
\end{eqnarray}
is the $\tau$-lepton flux produced in $\nu_\tau$
interactions inside the Earth for $\sigmanuN = \sigmaSM$. 

Given these estimates of $C_{\rm{down}}$ and $C_{\rm{up}}$, one can
now determine projected sensitivities of IceCube to neutrino fluxes
and cross sections. The quantities ${\cal N}_{\rm up}$ and ${\cal
  N}_{\rm down}$ as defined in Eqs.~(\ref{down}) and (\ref{up}) can be
regarded as the theoretical values of these events, corresponding to
different points in the $\sigma_{\nu N}/\sigma_{\rm
  SM},\phi_\nu/\phi_{\rm WB}^\nu$ parameter space. For a given set of
observed rates $\Nupobs$ and $\Ndownobs$, two curves are obtained in
the two-dimensional parameter space by setting $\Nupobs = \Nup$ and
$\Ndownobs = \Ndown$.  These curves intersect at a point, yielding the
most probable values of flux and cross section for the given
observations.  Fluctuations about this point define contours of
constant $\chi^2$ in an approximation to a multi-Poisson likelihood
analysis. The contours are defined by
\begin{equation}
\chi^2  =  \sum_i 2 \,\left[{\cal N}_i  - {\cal N}_i^{\rm obs}\right] +
2\, {\cal N}_i^{\rm obs}\,
\ln \left[{\cal N}_i^{\rm obs}/{\cal N}_i \right] \,\,,
\label{baker}
\end{equation} 
where $i=$~up, down~\cite{Baker:1983tu}.

In Fig.~\ref{IceCubesigma}, we show results for two representative
cases that are consistent with the AMANDA bounds derived above.  In
the first case, we assume $\sigmanuN = \sigmaSM$ and $\phinu =
\phiWB$, leading to 3 down-going and 25 up-going events.  The 90\%,
99\%, and 99.9\% CL contours are those given in the lower left of the
figure.  (These contours will be slightly distorted for
$\sigmanuN/\sigmaSM < 1$, where Eq.~\eqref{up} receives corrections,
but we neglect this effect.)  We see that, even in the case that event
rates are in accord with standard assumptions, the neutrino-nucleon
cross section is bounded to be within 40\% of the Standard Model
prediction at 90\% CL. This is at a center-of-mass energy
$\sqrt{s}\simeq 6$ TeV, far beyond the reach of any future man-made
accelerator.\footnote{For small $x$ physics, Standard Model
predictions for neutrino-nucleon collisions at $\sqrt{s} \simeq 6$~TeV
probe $x$ down to $M_W^2/s \sim 2\times 10^{-4}$~\cite{Gandhi:1995za};
LHC will probe parton center-of-mass energies $\sqrt{\hat s}\agt
50$~GeV~(twice the minimum jet trigger threshold), corresponding to
$x\agt 50/14000\sim 3\times 10^{-3}.$ A future hadron collider would
require $\sqrt{s}\sim 200$ TeV to sample $x\approx 2\times 10^{-4}.$}
In the second case, we consider a scenario in which the number of
observed upcoming events remains at 25, but the number of down-going
events is 50. In the second case, clearly one has discovered new
physics at well beyond 5$\sigma$.

The analysis discussed in this section is very general as the
technique to extract the cross section is model independent and can be
readily extended to probe higher energy bins.  From Fig.~\ref{reachlum} one
can see that, if there are hidden sources in the heavens, we 
expect $C_{\rm{down}}$ and $C_{\rm{up}}$ to be about a factor of 15
higher, making it possible for IceCube to probe 40\% (70\%)
enhancements from Standard Model predictions at the 90\% (99.9\%)~CL
after only 1 year of observation.

\section{Flavor Metamorphosis}

\subsection{Neutrino Oscillations}
\label{FM}

The Standard Model is based on the gauge group
\begin{equation}
  G_{\rm SM}^{\rm gauge} = SU(3)_C \times SU(2)_L \times U(1)_Y \,,
\label{GSM}
\end{equation}
with three fermion generations. A single generation consists of five
different representations of the gauge group: $Q_L (3,2,1/6),$
$U_R(3,1,2/3),$ $D_R(3,1,-1/3),$ $L_L(1,2,-1/2),$ $E_R (1,1,-1);$
where the numbers in parenthesis represent the corresponding charges
under $G_{\rm SM}^{\rm gauge}.$ The left handed fermions belong to $SU_L(2)$
doublets while the right handed chiralities to $SU_L(2)$ singlets.
The model
contains a single Higgs boson doublet, $H(1,2,1/2),$ whose vacuum
expectation value breaks the gauge symmetry $G_{\rm SM}^{\rm gauge}$ into $SU(3)_C
\times U(1)_{\rm EM}.$ The Standard Model also comprises an accidental global
symmetry
\begin{equation}
G_{\rm SM}^{\rm global} = U(1)_B \times U(1)_e \times U(1)_\mu \times 
U(1)_\tau \,\,,
\end{equation}
where $U(1)_B$ is the baryon number symmetry, and $U(1)_{e,\mu,\tau}$
are three lepton flavor symmetries, with total lepton number given by
$L = L_e + L_\mu + L_\tau$. It is an accidental symmetry because we do
not impose it. It is a consequence of the gauge symmetries and the low
energy particle content. It is possible (but not necessary), however,
that effective interaction operators induced by the high energy
content of the underlying theory may violate sectors of the global
symmetry.

In the Standard Model charged lepton (and also quark) masses arise
from Yukawa interactions, which couple the right-handed fermion
singlets to the left-handed fermion doublets and the Higgs doublet
\begin{equation}
 {\cal L}_{\rm SM}^{\rm Yukawa} = \lambda_l \,\,\overline L_L \,\, H\,\, E_R + {\rm h.c.} 
\end{equation}
After spontaneous electroweak symmetry breaking these interactions
lead to charged fermion masses, $m_l = \lambda_l \,\, v/\sqrt{2},$ but
leave the neutrinos massless. No Yukawa interaction can be written
that would give a tree level mass to the neutrino because no
right-handed neutrino field exists in the Standard Model. One might 
think that neutrino masses could arise from loop corrections.  This,
however, cannot be the case, because any neutrino mass term that can
be constructed with Standard Model fields alone would violate the global 
lepton symmetry. In other words, the Standard Model predicts that
neutrinos are {\sl strictly} massless, and thus there is neither
mixing nor CP violation in the leptonic sector.

In recent years, stronger and stronger experimental evidence for
nonzero neutrino masses has been
accumulating~\cite{Gonzalez-Garcia:2004jd}. The weak interaction
coupling the $W$ boson to a charged lepton and a neutrino, can also
couple any charged lepton mass eigenstate $l_\alpha$ $(\alpha = e,
\mu, \tau)$ to any neutrino mass eigenstate $\nu_j$ ($j = 1, 2, 3$) .
The superposition of neutrino mass eigenstates produced in association
with the charged lepton of flavor $\alpha,$ $|\nu_\alpha\rangle =
\sum_j U_{\alpha j}^* |\nu_j\rangle,$ is the state we refer to as the
neutrino of flavor $\alpha$, where $U_{\alpha j}$'s are elements of the
neutrino mass-to-flavor mixing matrix, called the Maki-Nagakawa-Sakita
(MNS) matrix.  The density matrix of a flavor state, $\rho^\alpha $, can be
expressed in terms of mass eigenstates by
$\rho^\alpha=|\nu_\alpha\rangle\langle\nu_\alpha|=\sum_{i,j}U^*_{\alpha
  i}U_{\alpha j}|\nu_i\rangle\langle\nu_j| $. This is a pure quantum
system, therefore the density matrix satisfies ${\rm Tr}~\rho^2 \, =
\,{\rm Tr}~\rho = 1.$ The time evolution of the density matrix
\begin{equation}
\frac{\partial \rho}{\partial t} = - i\, [H,\, \rho] \,\,,
\label{liouville}
\end{equation}
is governed by the Hamiltonian $H$ of the system, thereby after traveling a
distance $L$ an initial state $\nu_\alpha$ becomes a superposition of all
flavors, with probability of transition to flavor $\beta$ given by
$P_{\nu_\alpha \to \nu_\beta} = {\rm Tr} [\rho_\alpha(t) \rho_\beta],$ or
equivalently~\cite{Eidelman:2004wy}
\begin{eqnarray}
 P_{\nu_\alpha \to \nu_\beta} 
& = & \delta_{\alpha \beta} - 4 \sum_{i>j} \Re {\rm e}\, 
(U_{\alpha i}^*\, U_{\beta i}\, 
U_{\alpha j} \, U_{\beta j}^*) \, \sin^2 \Delta_{ij} \nonumber \\
 & + & 2 \sum_{i>j} \Im {\rm m}\, (U_{\alpha i}^*\, U_{\beta i}\, 
U_{\alpha j} \, U_{\beta j}^*) \, \sin 2 \Delta_{ij} \,\,,
\label{pak}
\end{eqnarray}
where $\Delta_{ij} = 1.27\, \delta m_{ij}^2 L/E_\nu,$ when $L$ is measured in
km, $E_\nu$ in GeV and $\delta m_{ij}^2 = m_i^2 - m_j^2$ in eV$^2$.

The simplest and most direct interpretation of the atmospheric
data~\cite{Fukuda:1998tw} is that of muon neutrino oscillations. The
evidence of atmospheric $\nu_\mu$ disappearing is now at $> 15
\sigma$, most likely converting to $\nu_\tau$.  The angular
distribution of contained events shows that for $E_\nu \sim 1~{\rm
  GeV},$ the deficit comes mainly from $L_{\rm atm} \sim 10^2 -
10^4~{\rm km}.$ The corresponding oscillation phase must be maximal,
$\Delta_{\rm atm} \sim 1,$ which requires $\delta m_{\rm atm}^2 \sim
10^{-4} - 10^{-2}~{\rm eV}^2.$ Moreover, assuming that all upgoing
$\nu_\mu$'s which would yield multi-GeV events oscillate into a
different flavor while none of the downgoing ones do, the observed
up-down asymmetry leads to a mixing angle very close to maximal,
$\sin^2 2\theta_{\rm atm} > 0.85.$ These results have been confirmed
by the KEK-to-Kamioka (K2K) experiment which observes the
disappearance of accelerator $\nu_\mu$'s at a distance of 250~km and
find a distortion of their energy spectrum with a CL of
$2.5-4\sigma$~\cite{Ahn:2001cq}. The combined analysis leads to a best
fit-point and $1\sigma$ ranges, $\delta m^2_{\rm atm} =
2.2^{+0.6}_{-0.4} \times 10^{-3}~{\rm eV}^2$ and $\tan^2 \theta_{\rm
  atm} = 1^{+0.35}_{-0.26}$~\cite{Gonzalez-Garcia:2004jd}.  On the
other hand, reactor data~\cite{Apollonio:1999ae} requires $|U_{e3}|^2
\ll 1$. Thus, to simplify the discussion hereafter we use the fact
that $|U_{e3}|^2$ is nearly zero to ignore possible CP violation and
assume that the elements of the MNS matrix are real. The simple vacuum
relations between the neutrino states produced by the weak
interactions are then
\begin{eqnarray}
  |\nu_1 \rangle & = &  \frac{1}{\sqrt{2}}\sin 
\theta_\odot (|\nu_\mu\rangle - |\nu_\tau\rangle) - 
  \cos \theta_\odot |\nu_e\rangle    \nonumber\\
  |\nu_2 \rangle & = & \frac{1}{\sqrt{2}}\cos \theta_\odot 
(|\nu_\mu\rangle - |\nu_\tau\rangle) + \sin \theta_\odot |\nu_e\rangle \\
  |\nu_3 \rangle & = & \frac{1}{\sqrt{2}}(|\nu_\mu \rangle + 
|\nu_\tau \rangle) \nonumber
\label{3rd}
\end{eqnarray}
where $\theta_\odot$ is the solar mixing angle. Data collected by the
Sudbury Neutrino Observatory (SNO)~\cite{Ahmed:2003kj} in conjuction
with data from SuperKamiokande (SK)~\cite{Fukuda:1998fd} show that
solar $\nu_e's$ convert to $\nu_{\mu}$ or $\nu_\tau$ with CL of more
than 7$\sigma$. The KamLAND Collaboration~\cite{Araki:2004mb} has
measured the flux of $\overline \nu_e$ from distant reactors and find
that $\overline{\nu}_e$'s disappear over distances of about 180~km.
The combined analysis of Solar neutrino data and KamLAND data 
are consistent at the 
3$\sigma$ CL, with best-fit point and $1 \sigma$ ranges: $\delta
m^2_\odot = 8.2^{+0.3}_{-0.3} \times 10^{-5}~{\rm eV}^2$ and $\tan^2
\theta_\odot = 0.39^{+0.05}_{-0.04}$~\cite{Bahcall:2004ut}.

Altogether, neutrinos are massive and therefore the Standard Model
needs to be extended.  The most economic way to get massive neutrinos
would be to introduce the right-handed neutrino states (having no
gauge interactions, these sterile states would be essentially
undetectable and unproduceable) and obtain a Dirac mass term through a
 Yukawa coupling. However, in order to get reasonable neutrino
masses (below the eV range) the Yukawa coupling would have to be
unnaturally small $\lambda_\nu < 10^{-11}$ (note that for charged fermions 
the Yukawa couplings range
from $\lambda_t \simeq 1$ for the top quark down to $\lambda_e \simeq
10^{-5}$ for the electron). Other attempts to give masses to 
neutrinos require extension of the Standard Model in more radical ways.
Throughout this review we will not concern ourselves with the origin
of the neutrino masses (for a comprehensive discussion of this subject
see e.g.,~\cite{Gonzalez-Garcia:2002dz}), but simply assume that
neutrinos acquire mass by some mechanism and discuss the reach of
IceCube in probing physics beyond conventional neutrino mass-induced
oscillations. The methods discussed in this section are readily
applicable to speculations concerning other non-conventional physics
associated with neutrino flavor mixing.

The only  interaction a neutrino can feel, besides the weak one, is
the gravitational interaction. A gravitational field, could contribute
to neutrino oscillations if the different neutrino flavors are affected
differently  by gravity, i.e., if the Equivalence Principle is 
violated~\cite{Gasperini:1988zf}. The deviation
can be parametrized by assuming that the post-Newtonian parameters in
the expansion of a given metric have a species dependent value,
representing thus the effective coupling constant of the different
flavors to a given geometry.  In the weak field limit of a static
source, to first order in the Newtonian potential $\phi = GM/r \ll 1,$
one can consider the metric to be $g_{tt}= 1 - 2 \gamma \phi$ and
$g_{ij} = -\delta_{ij} (1 + 2 \tilde \gamma \phi).$ The parameters
$\gamma$ and $\tilde \gamma$ are theory dependent (e.g., $\gamma =
\tilde \gamma = 1$ in General Relativity), but for any given theory,
their value is universal (the same for all kinds of particles) only if
the Principle of Equivalence is satisfied. Otherwise their value could
be different for test particles with different internal quantum
numbers and or different energies.

Of particular interest here, it has been shown~\cite{Longo:1987gc}
that the value of $\tilde \gamma$ is the same to an accuracy of about
0.1\% for neutrinos and photons received from
the SN87a. Moreover,
if neutrinos are light, $\tilde \gamma$ is the same for neutrinos (and
antineutrinos~\cite{LoSecco:1988hb}) of different energies (ranging
from 7 to 40~MeV) up to an accuracy of one part in $10^6.$ It is
worthwhile to note that the above results have been obtained with the
hypothesis $\gamma =1$. This hypothesis holds with a data set
containing only one type of particle, as possible deviations of
$\gamma$ from 1 can be absorbed by redefining the mass of the source.
The hypothesis may not be justified however, if we compare different
particles or particles with different energies, since the value of
$\gamma$ can be particle dependent or energy dependent just like
$\tilde \gamma$.  The laboratory experiments on neutrino oscillations
testing the universality of the gravitational redshift provide
information on the possibility that the value of $\gamma$ be different
for different neutrino flavors. If there are deviation from
universality in the value of $\gamma$, in fact, the energy splitting
induced by gravity may contribute in general to the transition
probability between different flavors, even if gravity is not the
primary source of oscillations, i.e., even if the energy eigenstates
are unchanged by the weak gravitational field present in the
laboratory.

  At this stage it is worthwhile to point out that for a constant
  potential $\phi$, a violation of the equivalence principle is
  phenomenologically equivalent to the breakdown of Lorentz invariance
  resulting from different asymptotic values of the velocity of the
  neutrinos, $c_1\neq c_2$, with $\nu_1$ and $\nu_2$ being related to
  $\nu_\mu$ and $\nu_\tau$ by a rotation of an angle
  $\xi_{\rm vli}$~\cite{Coleman:1997xq}.

  Gravity induced neutrino oscillations can also arise if the
  spacetime manifold has a non-symmetric Christoffel
  connection~\cite{DeSabbata:1981ek}. This is because the flavor
  eigenstates produce in the charged weak interactions are not, in
  general, the eigenstates of the torsion Hamiltonian, and then the
  oscillations can be strongly damped or enhanced in the presence of
  strong gravitational fields.  The torsion eigenstates can be
  interpreted as different mixtures of right-handed and left-handed
  fields, yielding a deviation of the weak neutrino current from the
  standard $V-A$ form. For this vector-like interaction the
  gravity-induced oscillation wavelength is energy independent.
  Violation of CPT resulting from Lorentz-violating effects also 
  leads to an energy independent contribution to
  the oscillation wavelength~\cite{Colladay:1996iz} which is a
  function of the eigenvalues of the Lorentz violating CPT-odd
  operator and the rotation angle between the
  corresponding neutrino eigenstates $\nu_i$ and the flavor
  eigenstates $\nu_\alpha$.

  In the most general situation in which neutrinos are massive and
  gravity may contribute to the transition probability between
  different flavors, we may have $|\nu_\alpha\rangle \neq
  |\nu_i\rangle \neq |\nu_G \rangle,$ where $|\nu_G \rangle$ denotes
  hypothetical gravity eigenstates. The eigenstates of the total
  energy are obtained by diagonalizing the matrix which includes the
  contributions of the mass and gravitational terms. In the case of 2
  flavor mixing, the total matrix for neutrino ($+$) or antineutrino
  ($-$) oscillations can be written as~\cite{Gasperini:1989rt}
\begin{equation} \label{eq:hamil}
    {\rm H}_\pm \equiv
    \dfrac{\Dmq}{4 E_\nu}
    \mathbf{U}_\theta
    \begin{pmatrix}
	-1 & ~0 \\
	\hphantom{-}0 & ~1
    \end{pmatrix}
    \mathbf{U}_\theta^\dagger
    + \sum_n
    \sigma_n^\pm \dfrac{\delta_n\, E_\nu^n}{2}
    \mathbf{U}_{\xi_n,\pm\eta_n}
    \begin{pmatrix}
	-1 & ~0 \\
	\hphantom{-}0 & ~1
    \end{pmatrix}
    \mathbf{U}_{\xi_n,\pm\eta_n}^\dagger \;,
\end{equation}
where $\sigma_n^\pm$ accounts for a possible relative sign of the
gravity effects between neutrinos and antineutrinos\footnote{For the
  CPT violating case $\sigma^+ = -\sigma^-,$ otherwise $\sigma^+ =
  \sigma^-$.} and $\delta_n$ parametrizes the size of the new physics
terms. The matrices $\mathbf{U}_\theta$ and
$\mathbf{U}_{\xi_n,\pm\eta_n}$ are given by
\begin{equation} \label{eq:rotat}
    \mathbf{U}_\theta =
    \begin{pmatrix}
	\hphantom{-}\cos\theta & ~\sin\theta \\
	-\sin\theta & ~\cos\theta
    \end{pmatrix}\,,
    \qquad
    \mathbf{U}_{\xi_n,\pm\eta_n} =
    \begin{pmatrix}
	\hphantom{-}\cos\xi_n\hphantom{e^{-i\eta_n}} 
	& ~\sin\xi_n e^{\pm i\eta_n} 
	\\
	-\sin\xi_n e^{\mp i\eta_n} 
	& ~\cos\xi_n\hphantom{e^{-i\eta_n}}
    \end{pmatrix}\,;
\end{equation}
where $\eta_n$ denotes the possible non-vanishing relative phases.
Note that in contrast to the conventional oscillation length, new
physics predicts neutrino oscillations with wavelengths that are either 
constant or decrease with energy. Therefore, IceCube with an energy
reach in the $10^2~{\rm GeV} < E_\nu < 10^6~{\rm GeV}$ range for
atmospheric neutrinos, is the ideal experiment to search for new
physics. Moreover, these neutrinos have high enough energy for the
standard  oscillations (due to $\delta m_{ij}^2$) to be very  suppressed so
the $\nu$'s behave essentially as flavor eigenstates.  For simplicity in what
follows we concentrate on oscillations resulting from tensor-like
interactions that lead to an oscillation wavelength inversely
proportional to the neutrino energy. The results can be directly
applied to oscillations due to violations of the equivalence principle
with the identification, $\xi_1 = \xi_{\rm vep}$ and $\delta_1 = 2
|\phi|(\gamma_1- \gamma_2) \equiv 2 |\phi| \Delta \gamma,$ as well as
oscillations due to violations of local Lorentz invariance with the
identification, $\xi_1 = \xi_{\rm vli}$ and $\delta_1 = \Delta c/c.$

If the effects of new physics are constant along the neutrino
trajectory, the survival probability for total oscillation length
$L_{\rm tot}$ takes the form~\cite{Gasperini:1989rt}
\begin{equation} \label{eq:prob}
    P_{\nu_\mu \to \nu_\mu} = 1 - P_{\nu_\mu \to \nu_\tau} =
    1 - \sin^2 2\Theta \, \sin^2 \left(\Delta_{ij} \, \mathcal{R} \right) \,,
\end{equation}
where ${\cal R} = L_{\rm atm}/L_{\rm tot};$ 
if the contributions of mass and gravity are both non-vanishing then
\begin{equation}
\mathcal{R} = \sqrt{1 + R^2 + 2 R \left( \cos 2\theta_{\rm atm} \cos 2\xi_1
      + \sin 2\theta_{\rm atm} \sin 2\xi_1 \cos\eta_1 \right)}\, .
\end{equation}
Here, $\Theta$ is the rotation angle which diagonalizes the matrix in
Eq.~(\ref{eq:hamil}),
\begin{equation}
    \label{eq:Theta}
    \sin^2 2\Theta = \frac{1}{\mathcal{R}^2} \left(
    \sin^2 2\theta_{\rm atm} + R^2 \sin^2 2\xi_1
    + 2 R \sin 2\theta_{\rm atm} \sin 2\xi_1 \cos\eta_1 \right) \,,
\end{equation}
and
\begin{equation}
    R \equiv \frac{L_{\rm atm}}{L_{\rm G}} =  
\frac{\delta_1 E_\nu}{2} \, \frac{4E_\nu}{\Dmq} \,,
\end{equation}
where $L_G$ defines the gravity oscillation length.

\begin{figure}
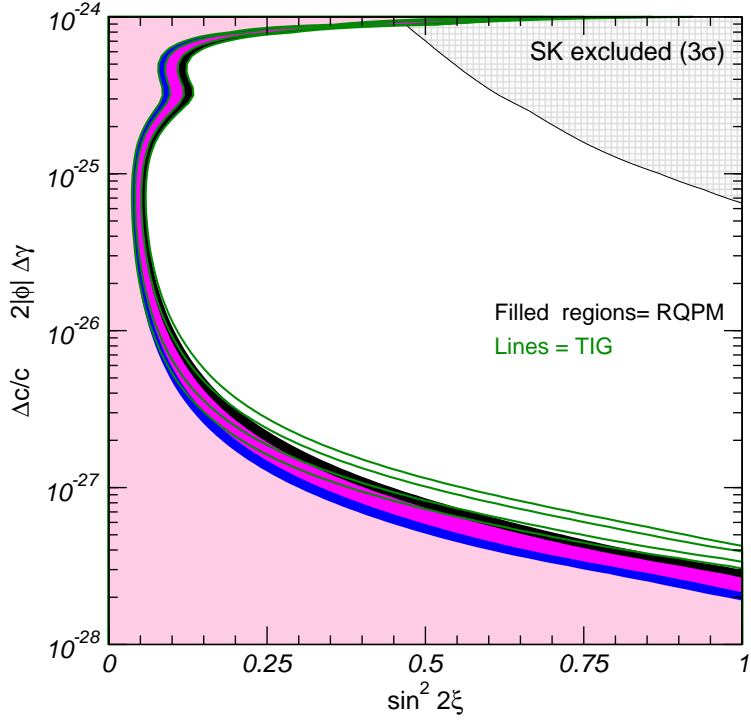

\postscript{fig.chisq.eps}{0.6} 
\caption{\label{fig:chisq} Sensitivity limits in the ($\Delta c/c, \xi_{\rm
vli}$) and ($2 |\phi| \Delta \gamma, \xi_{\rm vep}$) planes at 90\%, 95\%, 99\%
and $3\sigma$~CL. To account for the uncertainty associated with the poorly
known prompt neutrino fluxes results for two extreme models are shown:
TIG~\cite{Zas:1992ci} (full lines) and RQPM~\cite{Bugaev:1998bi} (filled
regions).  The difference is about 50\% in the strongest bound on $\delta_1$.
The hatched area in the upper right corner is the present $3\sigma$ bound from
the analysis of SK and K2K data~\cite{Gonzalez-Garcia:2005xw}.}
\end{figure}

The experimental probes of neutrino oscillations, measuring $P$,
provide simultaneously information on $\Theta$ and $L_{\rm tot}.$ In
the absence of positive results, however, an upper bound on $P$ does
not fix any value for $\Theta$ and $L_{\rm tot}$ separately: it
determines only an allow region in the $(\Theta, L_{\rm tot})$ plane,
or equivalently in the $(\xi_1, \delta_1)$ plane.  At present, the
strongest limits on neutrino oscillations due to new physics come from
the non-observation of departures from the standard oscillations seen
in atmospheric neutrinos at SK, and the confirmation of $\nu_\mu$
oscillations with the same oscillation parameters (i.e., $\delta
m_{\rm atm}^2$) from K2K. The $3\sigma$ bounds from the updated
combined analysis~\cite{Gonzalez-Garcia:2004wg} of SK and K2K data are
shown in Fig.~\ref{fig:chisq}.

For most of the neutrino energies considered here, the standard
mass-induced oscillations are suppressed and the gravity effect is
directly observed. As a consequence, the results will be independent
of the phase $\eta_1$ and one can set the new physics parameters to be
in the range $\delta_1 \geq 0$ and $0 \leq \xi_1 \leq \pi/4.$

The Hamiltonian of Eq.~(\ref{eq:hamil}) describes the coherent
evolution of the $\nu_\mu$--$\nu_\tau$ ensemble for any neutrino
energy.  High-energy neutrinos propagating in the Earth can also
interact inelastically with the Earth matter and as a consequence the
neutrino flux is attenuated.  For atmospheric neutrinos oscillation,
attenuation, and regeneration effects occur simultaneously when the
neutrino beam travels across the Earth's matter. For the
phenomenological analysis of conventional neutrino oscillations this
fact can be ignored because the neutrino energies covered by current
experiments are low enough for attenuation and regeneration effects to
be negligible. Especially for non-standard scenario oscillations,
future experiments probe high-energy neutrinos for which the
attenuation and regeneration effects have to be accounted for
simultaneously. Attenuation effects due to CC and NC interactions can
be introduced by relaxing the condition ${\rm Tr}(\rho)=1$, so that
Eq~(\ref{liouville}) can be re-written as~\cite{Gonzalez-Garcia:2005xw}
\begin{equation}
\frac{\partial \rho}{\partial t}=-i[H, \rho]
-\sum_\alpha \frac{1}{2 l_\nu}
\left\{\Pi_\alpha,\rho \right\} \, ,
\end{equation}
where $l_{\nu}$ is the neutrino mean free path and $\Pi_\alpha =
\nu_\alpha \otimes \nu_\alpha^\dagger$ is the $\nu_\alpha$ state
projector.

\begin{table}
\begin{tabular}{|c|cc|cc|}
\hline
\hline
& \multicolumn{2}{c|}{RPQM} 
& \multicolumn{2}{c|}{TIG}\\
\hline
 ~$\log_{10}(E_\mu^{\rm fin}/{\rm GeV})~ $ 
&$-1\leq\cos\theta\leq -0.6$ 
&$-0.6\leq\cos\theta\leq -0.2$ 
&$-1\leq\cos\theta\leq -0.6$ 
&$-0.6\leq\cos\theta\leq -0.2$ \\
\hline
 2.00 -- 2.20 & 52474& 61806 &  51427& 60920\\
 2.20 -- 2.40 & 46234& 55598 &  44987& 54539\\
 2.40 -- 2.60 & 35965& 44586 &  34634& 43422\\
 2.60 -- 2.80 & 26001& 33588 &  24647& 32415\\
 2.80 -- 3.00 & 17358& 23400 &  16107& 22294\\
 3.00 -- 3.20 & 10710& 15126 &  9630 &14141\\
 3.20 -- 3.40 &  6172&  9054 &  5320 & 8250 \\
 3.40 -- 3.60 &  3330&  5099 &  2701 & 4494\\
 3.60 -- 3.80 &  1721&  2722 &  1289 & 2291\\
 3.80 -- 4.00 &   856&  1388 &  578  &1098 \\
 4.00 -- 4.20 &   410&   685 &  242  & 498\\
 4.20 -- 4.40 &   191&   330 &   96  & 215 \\
 4.40 -- 4.60 &    86&   156 &  36   & 89\\
 4.60 -- 4.80 &    38&    74 &  13   & 36\\
 4.80 -- 5.00 &    16&    34 &  5    &14\\
 5.00 -- 9.00 &    10&    28 &  2    & 8\\\hline \hline
\end{tabular}
\caption{Number of expected atmospheric $\nu_\mu$-induced muon events 
in 10 years of IceCube operaton in the 16 energy bins and
angular bins discuss in the text, assuming no new physics effect is observed.}
\label{tab:nevents}
\end{table}

Neutrino oscillations introduced by new physics effects result in an
energy dependent distortion of the zenith angle distribution of
atmospheric muon events. This effect can be quantify by evaluating the
expected angular and $E_\mu^{\rm fin}$ distributions in the IceCube
detector using Eq.~(\ref{eq:numuevents}) in conjunction with $\nu_\mu$
(and $\bar\nu_\mu$) fluxes obtained after evolution in the Earth for
different sets of new physics oscillation parameters.  Together with
$\nu_\mu$-induced muon events, oscillations also generate $\mu$ events
from the CC interactions of the $\nu_\tau$ flux which reaches the
detector producing a $\tau$ that subsequently decays as
$\tau\rightarrow \mu \bar \nu_\mu \nu_\tau$ and produces a $\mu$ in
the detector. Generalizing the formulae given in Sec.~\ref{ODP} it is
straightforward to see that the number of $\nu_\tau$-induced muon
events is
\begin{eqnarray}
{\cal N}^{\nu_\tau}_{\nu_\mu}
&=& T \int^{1}_{-1} d\cos\theta\,  
\int^\infty_{l_\mu^{\rm min}} dl_\mu\,
\int_{m_\mu}^\infty dE_\mu^{\rm fin}\,
\int_{E_\mu^{\rm fin}}^\infty dE_\mu^0\, 
\int_{E_\mu^0}^\infty dE_\tau 
\int_{E_\tau}^\infty dE_\nu \\ \nonumber
&&\frac{dF^{\nu_\tau}}{dE_\nu \,d\cos\theta}(E_\nu,\cos\theta)
\frac{d\sigma_{\rm CC}}{dE_\tau}(E_\nu,E_\tau)\, n_T\, 
\frac{dN_{\rm dec}}{dE_\mu^0}(E_\tau,E_\mu^0)
F(E^0_\mu,E_\mu^{\rm fin},l_\mu)\, A^0_{\rm eff}\,  \; ,
\label{eq:nutauevents}
\end{eqnarray}
where $d N_{\rm dec} /d E_\mu^0$ is given in Ref.~\cite{Gaisser:1990vg}. 
To quantify the energy-dependent angular distortion one defines 
the vertical-to-horizontal double ratio 
\begin{equation}
R_{\rm hor/ver}(E_\mu^{{\rm fin},i})\equiv
\frac{P_{\rm hor}}{P_{\rm ver}}(E_\mu^{{\rm fin},i})
=\frac
{\frac
{\displaystyle 
{\cal N}^{\delta_1 \neq 0}_\mu(E_\mu^{{\rm fin},i}, -0.6<\cos\theta<-0.2)}
{\displaystyle 
{\cal N}^{\rm \delta_1 =0}_\mu(E_\mu^{{\rm fin},i}, -0.6<\cos\theta<-0.2)}}
{\frac
{\displaystyle 
{\cal N}^{\rm \delta_1 \neq 0}_\mu(E_\mu^{{\rm fin},i}, -1<\cos\theta<-0.6)}
{\displaystyle 
{\cal N}^{\delta_1 = 0}_\mu(E_\mu^{{\rm fin},i}, -1<\cos\theta<-0.6)}} \; ,
\label{eq:dblratio}
\end{equation} 
where $E_\mu^{{\rm fin},i}$ denotes integration in an energy bin 
of width $0.2\,\log_{10}(E_\mu^{{\rm fin},i})$. This leads to 16 $E_\mu^{\rm fin}$ bins 
in the data sample: 15 bins between $10^2$ and $10^5$~GeV and one containing all events above $10^5$ GeV.
 Note that the double 
ratio eliminates uncertainties associated 
with the overall  normalization of the atmospheric fluxes at high energies. 
Additionally, in the definition of the double ratio we have conservatively 
included only events well below the horizon $\cos\theta<-0.2$ to 
avoid the possible contamination from missreconstructed atmospheric 
muons which can still survive after level 2 cuts in the angular 
bins closer to the  horizon.  
In order to estimate the expected sensitivity we assume that no new physics effect 
is observed and define a simple $\chi^2 (\delta_1,\xi_1)$ function as
\begin{equation}
\chi^2 (\delta_1,\xi_1) =\sum_{i=1}^{16} 
\frac{(R_{\rm hor/ver}(E_\mu^{{\rm fin},i})-1)^2}
{\sigma_{{\rm stat},i}^2} \,
\end{equation}
where $\sigma_{{\rm stat},i}$ is computed from the expected number of events
in the absence of new physics effects. The expected event rates are given in  Table~\ref{tab:nevents}. 

Figure~\ref{fig:chisq} shows
the sensitivity limits
in the $[\delta_1, \xi_1]$-plane at 90\%, 95\%, 99\% and $3\sigma$~CL
obtained from the condition 
$\chi^2(\delta_1, \xi_1)<\chi^2_{\rm max}({\rm CL,2dof})$.
The figure illustrates the improvement on the present bounds
by more than two orders of magnitude even within the context of this very 
conservative analysis. The loss of sensitivity at large 
$\delta_1$ is a consequence of the use of a double ratio as
an observable. Such an observable is insensitive to new physics effects 
if $\delta_1$ is large enough for the oscillations
to be always averaged leading only to an overall suppression.

\subsection{Quantum Decoherence}

What is the meaning of quantum gravity? It means that spacetime itself
is subject to quantum laws, necessitating inherent fluctuations in the
fabric (metric and topology) of space and time. These microscopic
boiling bubbles force on spacetime a foam-like
structure~\cite{Wheeler:1957mu}. A heuristic example pictures
spacetime to be filled with tiny virtual black holes that pop in and
out of existence on a timescale allowed by Heisenberg's uncertainty
principle~\cite{Hawking:1995ag}. These black holes conserve energy,
angular momentum, and electric and color charge (unbroken gauged
quantum numbers), but they {\it are believed} not to conserve global
quantum numbers. The transition between initial and final density
matrices associated with black hole formation and evaporation is not
factorizable into products of $S$-matrix elements and their hermitian
conjugates. The evolution of such a quantum system is characterized by
a superscattering operator $\mathbb{S}$ that maps initial mixed states
to final mixed states, $\rho_{\rm out} = \mathbb{S} \rho_{\rm in},$
with $\mathbb{S} \neq S^\dagger S$~\cite{Hawking:1976ra}. In other
words, there {\em may be} a loss of quantum information across the
black hole event horizons, providing an environment that can induce
decoherence of apparently isolated matter
systems~\cite{Hawking:2005kf}.\footnote{Recent developments in string
  theory show that there is in fact no information loss in the
  fundamental underlying theory, nevertheless such information loss
  might appear in the effective low energy theory. This is clearly an
  important issue, worthy of investigation.}

Of particular interest is the question whether fast baryon decay can
proceed via virtual black hole states in the spacetime
foam~\cite{Adams:2000za}. The process is envisioned as the
simultaneous absorption of two quarks into the black hole, followed
without memory of the initial state by the thermal emission of an
antiquark and a lepton, entailing a change in the global baryon and
lepton quantum numbers $qq \to \overline q l$. The probability that
two quarks in a proton of size $\Lambda_{\rm QCD}^{-1}$ pass within a
fundamental Planck length, within the Heisenberg lifetime uncertainty
of the black hole is $\propto (\Lambda_{\rm QCD}/M_{\rm QG})^4,$ where
$M_{\rm QG}$ is the gravitational UV cutoff. Thus, the present limit
on the proton lifetime, $\tau_p \sim
10^{33}$~yr~\cite{Shiozawa:1998si}, implies $M_{\rm QG} > 10^{16}~{\rm
GeV}$~\cite{Adams:2000za}.\footnote{Note that interactions through the
higher dimensional QQQL operator can be prevented if one separates the
quark and lepton fields far enough in an extra dimension, so that their
wave function overlap is exponentially
suppressed~\cite{Arkani-Hamed:1999dc}.}

Measurements of flavor transformations in a neutrino beam can
provide a clean and sensitive probe of interactions with the
spacetime foam. Without interference from the gravitational
sector, oscillations in the neutrino sector provide a pure quantum
phenomenon in which the density matrix has the properties of a
projection operator.  Because black
holes do conserve energy, angular momentum (helicity), color and
electric charge, any neutrino interacting with the virtual black
holes needs to re-emerge as a neutrino. As an example, if
spacetime foam black holes do not conserve $U(1)_e \times
U(1)_\mu \times U(1)_\tau,$ neutrino flavor is randomized by
interactions with the virtual black holes. The result of many
interactions is to equally populate all three possible flavors.

The Hamiltonian evolution in Eq.~(\ref{liouville}) is a characteristic
of physical systems isolated from their surroundings. The time
evolution of such a quantum system is given by the continuous group of
unitarity transformations, $U_t = {\rm e}^{-i H\, t},$ where $t$ is
the time. The existence of the inverse of the infinitesimal generator,
$H,$ which is a consequence of the algebraic structure of the group,
guarantees {\em reversibility} of the processes. For open
quantum-mechanical systems, the introduction of dissipative effects
lead to modifications of Eq.~(\ref{liouville}) that account for the
{\em irreversible} nature of the evolution. The transformations
responsible for the time evolution of these systems are defined by the
operators of the Lindblad quantum dynamical
semi-groups~\cite{Lindblad:1975ef}. Since this does not admit an
inverse, such a family of transformations has the property of being
only forward in time.

The Lindblad approach to decoherence does not require any detailed
knowledge of the environment.  The corresponding time evolution
equation for the density matrix takes the form
\begin{equation}
\frac{\partial \rho}{\partial t} = - i [H_{\rm eff},\, \rho] + {\cal D}
[\rho] \,\,,
\label{liodeco}
\end{equation}
where the decoherence term is given by
\begin{equation}
{\cal D} [\rho] = - \frac{1}{2} \sum_j \left([b_j,\, \rho\, b_j^\dagger] +
[b_j\, \rho,\, b_j^\dagger]\right)\,\,.
\end{equation}
Here, $H_{\rm eff} = H + H_{\rm d}$ is the effective Hamiltonian of
the system, $H$ is its free Hamiltonian, $H_{\rm d}$ accounts for
possible additional dissipative contributions that can be put in the
Hamiltonian form, and $\{b_j\}$ is a sequence of bounded operators
acting on the
Hilbert space of the open quantum system, ${\cal H}$, and satisfying $\sum_j
b^\dagger_j b_j \in {\cal B} ({\cal H}),$ where ${\cal B} ({\cal H})$
indicates the space of bounded operators acting on ${\cal H}.$

The intrinsic coupling of a microscopic system to the space-time foam
can then be interpreted as the existence of an arrow of time which in
turn makes possible the connection with thermodynamics via an
entropy. The monotonic increase of the von Neumann entropy, $S(\rho) =
- {\rm Tr}\,\, (\rho\, \ln \rho)$, implies the hermiticity of the
Lindblad operators, $b_j = b_j^\dagger$~\cite{Benatti:1987dz}. In
addition, the conservation of the average value of the energy can be
enforced by taking $[H, b_j] = 0$~\cite{Banks:1983by}.

In a 2-level system with mixing angle $\theta$ (relevant for
atmospheric neutrinos), the $\nu_\mu$ survival probability after
propagation of a distance $L$ is~\cite{Lisi:2000zt}
\begin{equation}
P_{\nu_\mu \to \nu_\mu} = 1 - P_{\nu_\mu \to \nu_\tau} = 
1- \sin^2 {2 \theta}\,\, \sin^2 [e^{-\gamma L} \Delta_{ij}] \,, 
\label{pdeco}
\end{equation}
where $\gamma$ has dimension of energy, and its inverse defines the
typical (coherence) length after which the system gets
mixed.\footnote{The parametrization that leads to Eq.~(\ref{pdeco}) is
  appropriate for including dissipation as a perturbation on the
  standard oscillations~\cite{Benatti:2000ph}. In the most general
  solution of Eq.~(\ref{liodeco})~\cite{Hooper:2005jp}, after
  propagation over extreme long-baselines (i.e., $L\to \infty$) the
  system evolves to an equal flavor ratio regardless of the initial
  flavor content and mixing angle.}  Thus, for $\gamma L \sim {\cal
  O}(1)$ one expects significant deviations from standard mass-induced
oscillations formula $P_{\nu_\mu \to \nu_\mu} = 1 - \sin^2 {2
  \theta}\,\, \sin^2 \Delta_{ij}.$ For dissipative scenarios in which
decoherence effects vanish in the weak gravitational limit, $M_{\rm
  QG}\rightarrow \infty,$ $\gamma$ can be parametrized as
\begin{equation}
\gamma  = \tilde \kappa \,\left(\frac{E_\nu}{{\rm GeV}} \right)^n \,\left(\frac{M_{\rm
QG}}{{\rm GeV}}\right)^{-n+1}~{\rm GeV} \,\,,
\label{osde}
\end{equation}
where $\tilde\kappa$ is a dimensionless parameter, which by
naturalness is expected to be ${\cal O}(1),$ and $n \geq 2.$

The energy behavior of $\gamma$ depends on the dimensionality
of the operators $b_j.$ But care must be taken, since ${\cal D}$ is
bilinear in the $b_j,$ and due to the hermiticity requirement, each
$b_j$ is itself at least bilinear in the neutrino fields
$\psi$. Examples are
\begin{equation}
b_j \propto \int d^3x\,\, \psi^\dagger\,\, (i\partial_t)^j \psi\ \ ,
\label{bj}
\end{equation}
where $j=0,\,1,\,2\dots .$ A Fourier expansion of the fields
$\psi,\,\psi^\dagger$, inserted into Eq.~(\ref{bj}), gives the energy
behavior $b_j\propto E_{\nu}^j,$ and hence
$\gamma\propto E_{\nu}^{2j}.$ This restriction of
the energy behavior to non-negative even powers of $E_{\nu}$
may possibly be relaxed when the dissipative term is directly
calculated in the most general space-time foam background.

An interesting example is the case where the dissipative term is
dominated by the dimension-4 operator $b_1,$ $\int d^3x\,
\psi^\dagger\, i\partial_t \psi\ ,$ yielding the energy dependence
$\gamma \propto E_{\nu}^{2}/M_{\rm QG}.$ This is
characteristic of non-critical string theories where the space-time
defects of the quantum gravitational ``environment'' are taken as
recoiling $D$-branes, which generate a cellular structure in the
space-time manifold~\cite{Ellis:1996bz}.

A best fit to data collected by the SK atmospheric
neutrino experiment~\cite{Fukuda:1998fd}, allowing for both
oscillation and decoherence yields, for $n=2,$
\begin{equation}
\tilde \kappa \,\, \left(\frac{M_{\rm QG}}{{\rm GeV}}\right)^{-1}
 < 0.9 \times 10^{-27} \,\,, \label{tkappa}
\end{equation}
at the 90\% CL~\cite{Lisi:2000zt}. A similar lower limit on $M_{\rm
  QG}$ has been obtained at Fermilab.  The CCFR detector is sensitive
to $\nu_\mu\leftrightarrow\nu_e$~\cite{Romosan:1996nh} and
$\nu_e\leftrightarrow\nu_\tau$ \cite{Naples:1998va} flavor
transitions.  Neutrino energies range from 30 to 600 GeV with a mean
of 140 GeV, and their flight lengths vary from 0.9 to 1.4~km. A best
fit to the data allowing for both oscillation and decoherence yields
\begin{equation}
\tilde \kappa \,\, \left(\frac{M_{\rm QG}}{{\rm GeV}}\right)^{-1} 
< 2.0 \times 10^{-24} \,\,,
\end{equation}
at the 99\% CL~\cite{Gago:2000qc}. It is clear that for $n =2$ existing
experiments are probing $M_{\rm QG}$ all the way up to the Planck scale
$M_{\rm Pl} \sim 10^{19}$~GeV. For larger $n,$ Eq.(\ref{tkappa}) generalizes 
to
\begin{equation}
\tilde \kappa \,\, \left(\frac{M_{\rm QG}}{{\rm
GeV}}\right)^{-n+1}
 < 0.9 \times 10^{-27} \,\, . \label{tkappan}
\end{equation}
This can be seen from Eq.~(\ref{osde}): the analysis of data places bounds on
$\gamma,$ and the neutrino energies are well above 1~GeV. Thus, for $n =3,$ 
the lower bound on $M_{\rm QG} > 10^{13}~{\rm GeV}$ is comparable to the 
one obtained from limits on proton 
decay~\cite{Adams:2000za}.\footnote{Note that for $\tilde
\kappa \sim 1,$  the lower limit on the UV cutoff is well
beyond the electroweak scale. Thus, if TeV-scale gravity is realized in
nature, interactions with virtual black holes would be non-dissipative. Non-dissipative
interactions are expected when gravity is embedded in string
theory, so that an $S$-matrix description is possible. The
existence of an $S$-matrix makes it no longer automatic that, {\em
e.g.,} the $B$-violating  $\overline q l$ and $B$-conserving $qq$
outgoing channels have the same probability, as they would in
thermal evaporation. Thus, the problem of avoiding rapid baryon
decay  is shifted to the
examination of symmetries~\cite{Krauss:1988zc} in the underlying
string theory which would suppress the appropriate
non-renormalizable operators at low energies.}
Although the cubic energy dependence $\gamma\propto
E_{\nu}^3$ is not obtainable from the simple operator
analysis presented above, it may be heuristically supported by a
general argument that each of the $b_j$ must be suppressed by at least
one power of $M_{\rm Pl},$ giving a leading behavior
\begin{equation}
\gamma = \tilde\kappa\ E_{\nu}^{3}/ M_{\rm Pl}^2\,\, .
\end{equation}

IceCube will collect a data set of order one million atmospheric
neutrinos over 10 years. Not surprisingly, because of the increased
energy and statistics over present experiments, the telescope will be
sensitive to effects of quantum decoherence at a level well below
current limits. Additionally, since the loss of quantum coherence is
weighted by the distance traveled by the neutrinos, IceCube data
analysis of high energy extraterrestrial neutrinos can be used to probe
decoherence effects arising only after long propagation lengths.  Such
an analysis can be carried out by measuring the ratios of neutrino
flavors present in the cosmic spectrum.  

Let the ratios of neutrino flavors at production in the cosmic
sources be written as $w_e : w_\mu : w_\tau$ with $\sum_\alpha
w_\alpha = 1,$ so that the relative fluxes of each mass eigenstate are
given by $w_j = \sum_\alpha \omega_\alpha \,\,U_{\alpha j}^2$.
For $\Delta_{ij} \gg 1$, the
phases in Eq.~(\ref{pak}) will be erased by uncertainties in $L$ and
$E_\nu$.  Consequently, averaging over $\sin^2 \Delta_{ij}$ one finds the
transition probability between flavor states $\alpha$ and
$\beta$~\cite{Learned:1994wg}
\begin{equation}
P(\nu_\alpha \to \nu_\beta) = \sum_{i} U_{\alpha i}^2 \,\,U_{\beta i}^2 \,\,. 
\end{equation}
Thus, we conclude that the probability of measuring
on Earth a flavor $\alpha$ is given by
\begin{equation}
P_{\nu_\alpha \,\,{\rm detected}} = \sum_j U_{\alpha j}^2 \,\, \sum_\beta w_\beta \,\,U_{\beta j}^2 \,\,.
\end{equation}
Straightforward calculation shows that any initial flavor ratio that
contains $w_e = 1/3$ will arrive at Earth with equipartition on the
three flavors. Since neutrinos from astrophysical sources are expected
to arise dominantly from pions (and kaons) and their muon daughters,
their initial flavor ratios of $1/3:2/3:0$ should arrive at Earth
democratically distributed. Consequently there is a fairly robust
prediction of $1/3:1/3:1/3$ flavor ratios for measurement of astrophysical
neutrinos. 

The prediction for the flavor population at Earth due to standard
flavor-mixing (i.e. with no spacetime dynamics) of a  pure
$\overline\nu_e$ beam is $\sum_j |U_{ej}|^2 |U_{\alpha j}|^2 \sim
\frac{1}{3} |U_{\alpha 2}|^2 + \frac{2}{3} |U_{\alpha 1}|^2$ for
flavor $\alpha,$ which leads to the flavor ratios $\sim 3/5:1/5:1/5$. This
is very different from the democratic $1/3:1/3:1/3$. Since the effects of
quantum decoherence would alter the flavor mixture to the ratio $1/3:1/3:1/3$
(regardless of the initial flavor content) and the loss of quantum
coherence is weighted by the distance travelled by the (anti)
neutrinos, by measuring the $\overline \nu$-Cygnus beam IceCube will
be sensitive to the effects of quantum decoherence at a level well
below current limits~\cite{Hooper:2004xr}.

The Lindblad
operators of an $N$-level quantum mechanical system can be expanded in
a basis of matrices satisfying standard commutation relations of Lie
groups. For a 3-level system, the basis comprises the eight Gell-Mann
SU(3) matrices plus the $3 \times 3$ identity
matrix. As mentioned above, the 
theoretical approach provided by Lindblad
quantum dynamical semi-groups is a very general in the sense that no
explicit hypothesis has been made about the actual interactions
causing the loss of coherence. In what follows  we
adopt an expansion in a 3 flavor basis with a diagonal form for the $9
\times 9$ decoherence matrix, ${\cal D}$. Note that neutrinos
oscillate among flavors separately between particle and antiparticle
sectors and so the respective decoherence parameters for antineutrinos
can be different from the corresponding ones in the neutrino
sector. Upon averaging over the rapid oscillation for propagation
between Cygnus OB2 and the Earth, only the diagonal Gell-Mann matrices 
survive, and so  the transition probability for
antineutrinos takes the form~\cite{Gago:2002na}
\begin{eqnarray}
P_{\overline\nu_\alpha \to \overline\nu_\beta} & = & \frac{1}{3} +
\left[\frac{1}{2}\,\, {\rm e}^{-\overline\gamma_3 \,d}\,\, (U_{\alpha 1}^2 -
U_{\alpha2}^2) ( U_{\beta 1}^2 - U_{\beta2}^2) \right. \nonumber \\
 &
+ & \left. \frac{1}{6}
\,\,{\rm e}^{-\overline \gamma_8 \,d}\,\, (U_{\alpha1}^2 + U_{\alpha 2}^2 -
2 U_{\alpha 3}^2) (U_{\beta 1} + U_{\beta 2}^2 - 2 U_{\beta 3}^2)
\right] \, \, ,
\label{Prob}
\end{eqnarray}
where $\overline \gamma_3$ and $\overline \gamma_8$ are eigenvalues of
the decoherence matrix for antineutrinos. Note that in Eq.~(\ref{Prob})
we set the CP violating phase to zero, so that all mixing matrix
elements become real. Furthermore, under the assumptions that CPT is
conserved and that decoherence effects are negligible at present
experiments, the values of the mixing angle combinations appearing in
Eq.~(\ref{Prob}) can be well determined by the usual oscillation analysis
of solar, atmospheric, long-baseline  and reactor
data~\cite{Gonzalez-Garcia:2004jd}. In what follows, we will assume
that CPT is conserved both by neutrino masses and mixing as well as
in decoherence effects. Note however that since the decoherence
effects in the present study primarily affect antineutrinos, the
result will hold for the antineutrino decoherence parameters
exclusively if CPT is violated only through quantum-gravity effects.

Now, we require further $\overline \gamma_3 = \overline \gamma_8\equiv
\overline \gamma$ ( $=\gamma_3 = \gamma_8$ under CPT conservation)
so that Eq.~(\ref{Prob}) can be rewritten for the case of interest as:
\begin{eqnarray}
&& P_{\overline \nu_e \to \overline \nu_\mu}=
P_{\overline \nu_\mu \to \overline \nu_e}=
P_{\nu_e \to \nu_\mu}=
P_{\nu_\mu \to \nu_e}
= \frac{1}{3} + f_{\nu_e \to \nu_\mu}
e^{-\overline\gamma\,L} \,\,,
\nonumber\\
&& P_{\overline \nu_e \to \overline \nu_\tau} =
P_{\overline \nu_\tau \to \overline \nu_e} =
P_{\nu_e \to \nu_\tau} =
P_{\nu_\tau \to  \nu_e}
= \frac{1}{3} + f_{\nu_e \to \nu_\tau} 
e^{-\overline\gamma\,L} \,\,,
\nonumber
\\
&& P_{\overline \nu_\mu \to \overline \nu_\tau}=
P_{\overline \nu_\tau \to \overline \nu_\mu}=
P_{\nu_\mu \to \nu_\tau}=
P_{\nu_\tau \to  \nu_\mu}
= \frac{1}{3} + f_{\nu_\mu \to \nu_\tau}
e^{-\overline\gamma\,L} \,\,, \label{eq:probdeco}\\
&&
P_{\overline \nu_e \to \overline \nu_e} =
P_{\nu_e \to \nu_e} =
\frac{1}{3}-(f_{\nu_e \to \nu_\mu}+f_{\nu_e \to \nu_\tau})\,
e^{-\overline\gamma\,L} \,\,,\nonumber\\
&&
P_{\overline \nu_\mu \to \overline \nu_\mu} =
P_{\nu_\mu \to \nu_\mu} =
\frac{1}{3}-(f_{\nu_e \to \nu_\mu}+f_{\nu_\mu \to \nu_\tau})\,
e^{-\overline\gamma\,L} \,\,,\nonumber\\
&&
P_{\overline \nu_\tau \to \overline \nu_\tau} =
P_{\nu_\tau \to \nu_\tau} =
\frac{1}{3}-(f_{\nu_e \to \nu_\tau}+f_{\nu_\mu \to \nu_\tau})\,
e^{-\overline\gamma\,L}\,\,. \nonumber
\end{eqnarray}
We make this simplification only to emphasize the primary signature of
quantum decoherence, namely that after travelling a sufficiently long
distance the flavor mixture is altered to the ratio $1/3:1/3:1/3,$
regardless of the initial flavor content. Consequently, if a flux of
antineutrinos were to be observed from the Cygnus spiral arm with a
flavor ratio $\neq 1/3:1/3:1/3,$ strong constraints can be placed on the
decoherence parameter $\overline \gamma$.

Using the results of the up-to-date 3-$\nu$ oscillation analysis of
solar, atmospheric, long-baseline and reactor 
data~\cite{Gonzalez-Garcia:2004jd}
we obtain the following values and 95\% confidence ranges~\cite{Anchordoqui:2005gj}
\begin{eqnarray}
f_{\nu_e \to  \nu_\mu}&=&-0.106^{+0.060}_{-0.082} \,\,,\nonumber\\
f_{\nu_e \to  \nu_\tau}&=&-0.128^{+0.089}_{-0.055} \,\,, \label{fit} \\
f_{\nu_\mu \to  \nu_\tau}&=&\phantom{-}0.057^{+0.011}_{-0.035} \,\, . \nonumber
\end{eqnarray}
The numbers given in Eq.~(\ref{fit}) are obtained using the same
techniques as described in Ref.~\cite{Gonzalez-Garcia:2004jd} but
including the final SNO salt phase data~\cite{Miknaitis:2005rw}.

Equipped with Eqs.~(\ref{osde}), (\ref{eq:probdeco}), and (\ref{fit}),
we now proceed to determine the sensitivity of IceCube to decoherence
effects. We estimate the expected number of $\nu_\mu$ induced tracks
from the Cygnus OB2 source antineutrino flux using the semi-analytical
calculation discussed in Sec.~\ref{MT}. Namely, we replace the
antineutrino flux shown in Fig.~\ref{cygOB2_nu} into
Eq.~(\ref{eq:numuevents}) and obtain, for standard mass-induced
oscillations, a total of $212\times
P_{\overline\nu_e\to\overline\nu_\mu}=48$ $\overline\nu_\mu$-induced
tracks in 15 years of observation (cluster within $1^\circ$ of the
source direction).  For showers, the angular resolution is
significantly worse than for muon tracks. Normally, a reduction of the
muon bremsstrahlung background is effected by placing a cut of $4
\times 10^4$~GeV on the minimum reconstructed
energy~\cite{Ackermann:2004zw}. For Cygnus OB2, this strong energy cut
is not needed since this background is filtered by the Earth. Thus we
account for all events with shower energy $E_{\rm sh}\geq E_{\rm
  sh}^{\rm min}=10^3$~GeV, trigger level. The directionality
requirement, however, implies that the effective volume for detection
of showers is reduced to the instrumented volume of the detector, 1
${\rm km}^3$, because of the small size of the showers (less than 200
m in radius) in this energy range. Replacing the antineutrino flux
shown in Fig.~\ref{cygOB2_nu} into Eq.~(\ref{FTjpg}), within the
framework of standard oscillations, one expects 25 showers from the
Cygnus OB2 source in 15 years of operation.  We now turn to the
estimate of the background. There are two different contributions ---
atmospheric neutrinos and additional fluxes of extraterrestrial
neutrinos. We obtain the number of expected track and shower events
from atmospheric neutrinos as in Eqs~(\ref{eq:numuevents}),
(\ref{eq:shsourcecc}), and (\ref{eq:shsourcenc}) with
$dF^{\nu_\alpha}_{\rm atm}/dE_\nu$ as given in Fig.~\ref{cygOB2_nu},
being the $\nu_e$ and $\nu_\mu$ atmospheric neutrino fluxes integrated
over a solid angle of of $1^\circ\times 1^\circ$ (for tracks) and
$10^\circ\times 10^\circ$ (for showers) width around the direction of
the Cygnus OB2 source ($\theta=131.2^\circ$). We get an expected
background of 14 atmospheric tracks and 47 atmospheric showers in 15
years. Of the 47 showers, 16 correspond to $\nu_e$ CC interactions
while 31 correspond to $\nu_\mu$ NC interactions. The large yield of
NC events is due to the fact that at these energies, the atmospheric
flux contains a very unequal mix of neutrino flavors (with ratios
$\approx 1/20:19/20:0$). Interestingly, by increasing the minimum
shower energy cut to $5 \times 10^3$~GeV, $\nu_e$ CC and $\nu_\mu$ NC
contribute in equal amounts to the number of atmospheric showers. This
is in agreement with simulations by the AMANDA
Collaboration~\cite{Ahrens:2002wz}.  We turn now to the discussion of
background events from other extraterrestrial sources.  As discussed
in Sec.~\ref{nDK}, the TeV $\gamma$-ray flux reported by the HEGRA
Collaboration~\cite{Aharonian:2002ij} in the vicinity of Cygnus OB2 is
likely to be due to hadronic processes. For the purposes of setting an
upper bound on the neutrino flux we ignore all other sources near
J2032+4130 because their steady emission in $\gamma$-rays is estimated
to be smaller by more over a factor of 5 than the source of
interest.\footnote{This includes the famous microquasar Cygnus X-3,
  for which HEGRA Collaboration reported~\cite{Aharonian:2002ij} an 
  upper limit for steady emission of $F_\gamma
  (E_\gamma > 0.7~{\rm TeV}) = 1.7 \times 10^{-13}~{\rm cm}^{-2} \,
  {\rm s}^{-1}$.}  Using the expected flux of neutrinos from
J2032+4130 given in Fig.~\ref{cygOB2_nu} we obtain the corresponding
background from neutrinos with flavor ratios $1/3:1/3:1/3$. As can be
seen in Fig.~\ref{cygOB2_nu}, the background is dominated by
atmospheric neutrinos. Thus after 15~years of data collection we
expect 18 tracks and 1 shower from J2032+4130 for standard
oscillations. 

We will now discuss how to isolate the possible signal due to
decoherence in the antineutrinos from Cygnus OB2 from the atmospheric
background and possible fluctuations in the event rate due to unknown
diffuse fluxes of extraterrestrial neutrinos.  In general, we can
predict that the expected number of track and shower events in the
direction of the Cygnus OB2 source to be
\begin{eqnarray}
{\cal N}_{\rm tr} &=& {\cal N}^{\rm S}_{\rm tr} + {\cal N}^{\rm atm}_{\rm tr} +
  {\cal N}^{\rm HEGRA}_{\rm tr} \,\,,
\label{pepetr}\\
{\cal N}_{\rm sh} &=& {\cal N}^{\rm S}_{\rm sh} + {\cal N}^{\rm atm}_{\rm sh} +
  {\cal N}^{\rm HEGRA}_{\rm sh}.
\label{pepesh}
\end{eqnarray}
The first term corresponds to antineutrinos from neutron
$\beta$-decay. In the presence of decoherence effects these event
rates can be computed from
Eqs.~(\ref{eq:numuevents}),~(\ref{eq:shsourcecc})~and~(\ref{eq:shsourcenc})
with flavour transition probabilities given in Eq.~(\ref{eq:probdeco})
with $L=1.7$ kpc.  The second term refers to atmospheric
(anti)neutrinos (${\cal N}^{\rm atm}_{\rm tr}=14$, ${\cal N}^{\rm
atm}_{\rm sh}=47$ for 15 years of exposure). The third term takes into
account additional contributions from a diffuse flux of
(anti)neutrinos produced via charged pion decay. In principle,
decoherence effects may also affect the expected number of events from
this diffuse flux. However given that the flavour ratios both from
oscillation and decoherence are very close to $1/3:1/3:1/3$ for the case of
neutrinos produced via charged pion decay, there is no
difference in the sensitivity region if decoherence effects are
included or not in the evaluation of $ {\cal N}^{\rm HEGRA}_{\rm tr}$ and
$ {\cal N}^{\rm HEGRA}_{\rm sh} $.  They are normalized to the maximum
expected flux from J2032+4130 by a factor $x= {\cal N}^{\rm HEGRA}_{\rm
tr}/18= {\cal N}^{\rm HEGRA}_{\rm sh}/1$.

Altogether, the quantities ${\cal N}_{\rm tr}$ and ${\cal N}_{\rm
  sh}$, as defined in Eqs.~(\ref{pepetr}) and (\ref{pepesh}), can be
regarded as the theoretical expectations of these event rates,
corresponding to different points in the $x-\kappa_n$ parameter space.
For a given set of observed rates, ${\cal N}^{\rm obs}_{\rm tr}$ and
${\cal N}^{\rm obs}_{\rm sh}$, two curves are obtained in the
two-dimensional parameter space by setting ${\cal N}^{\rm obs}_{\rm
  tr}={\cal N}_{\rm tr}$ and ${\cal N}^{\rm obs}_{\rm sh}={\cal
  N}_{\rm sh}.$ These curves intersect at a point, yielding the most
probable values for the flux and decoherence scale for the given
observations.  Fluctuations about this point define contours of
constant $\chi^2$ in an approximation to a multi-Poisson likelihood
analysis. These contours can be obtained using Eq.~(\ref{baker}) with
the identification $i = {\rm sh},\, {\rm tr}.$
Marginalizing with respect to $x$, for $n=2$ we obtain the following 1
degree-of-freedom bound
\begin{equation}
\overline \gamma \leq 2.0\times 10^{-44} \;\; (5.5\times 10^{-42}) \;{\rm GeV} \, ,
\end{equation}
at the 90\% (99\%) CL~\cite{Anchordoqui:2005gj}. This  corresponds
to an improvement of about 17 orders of magnitude over the best
existing bounds from the SK and K2K data~\cite{Lisi:2000zt}. Moreover,
for $n=3,$ the 1-degree freedom bound is
\begin{equation}
\overline \gamma
\leq 3.0\times 10^{-47} \;\; (2.9\times 10^{-45}) \;{\rm GeV}
\end{equation}
at 90\% (99\%)~CL~\cite{Anchordoqui:2005gj}. For $M_{\rm QG} = M_{\rm Pl}$ this corresponds to
$\tilde \kappa \alt 3 \times 10^{-3}$ at the 99\% CL, well below the
natural expectation, $\tilde \kappa \sim 1$.

Finally we note that, for $\overline\gamma =
\kappa \,\, (E_\nu/{\rm GeV})^{-1},$ a non-vanishing $\overline \gamma$ in
Eq.~(\ref{eq:probdeco}) can be related to a finite $\overline \nu_e$
lifetime in the lab system~\cite{Barger:1998xk}
\begin{equation}
e^{-\overline \gamma \,L} \equiv e^{-L/\tau_{\rm lab}} =
 e^{-L\,m_{\overline\nu_e}/E_{\overline \nu}\, \overline
\tau_{\overline\nu_e}}\,\,,
\end{equation}
where $\overline \tau_{\overline\nu_e}$ is the antineutrino rest frame
lifetime and $m_{\overline\nu_e}$ its mass. By duplicating our discussion 
for $\overline \gamma \propto E^{-1}$ we obtain a 90\% (99\%) CL sensitivity~\cite{Anchordoqui:2005gj}
\begin{equation}
  \overline \gamma \leq 1.0 \times 10^{-34} \;\; (2.3\times 10^{-31}) \;{\rm GeV}\ ,
\end{equation}
which can  be translated into
\begin{equation}
\frac{\overline \tau_{\overline\nu_e}} {m_{\overline\nu_e}}> 10^{34}~{\rm GeV}^{-2} \equiv 6.5\,\,
{\rm s}~{\rm eV}^{-1}\, \,\, .
\label{mtau}
\end{equation}
This corresponds to an improvement of about 4 orders of magnitude over
the best existing bounds from solar neutrino
data~\cite{Bahcall:1986gq}, and of course gives results comparable to
the reach derived for neutrinos decaying over cosmic
distances~\cite{Beacom:2002vi}.  It should be noted that although the
similar algebraic structure of the decoherence term in
Eq.~(\ref{eq:probdeco}) and a decaying component in the neutrino beam
provide a bound on the neutrino lifetime, these are conceptually two
different processes. The decoherence case can be viewed as a
successive absorption and re-emission of a neutrino from the quantum
foam with change in flavor but no change in the average energy because
of the condition $[H,b_j]=0$. This contrasts with the decay process
which involves the emission of an additional particle.

\vspace{1cm}

In summary, IceCube will provide a major improvement in the sensitivity to
possible effects of quantum gravity. Because of the increase energy and
statistics over present experiments measuring the atmospheric neutrino flux,
existing bounds on the violation of the equivalence principle and of Lorentz
invariance can be improved by two orders of magnitude. On the other hand,
antineutrinos produced in the decays of neutrons from Cygnus OB2 provide an
excellent source in which to search for decoherence effects. Although the
precise conclusions depend on the model considered, IceCube will improve the
sensitivity to decoherence effects of ${\cal O} (E^2/M_{\rm Pl})$ by 17 orders
of magnitude over present limits and, moreover, it can probe decoherence effects
of ${\cal O} (E^3/M_{\rm Pl}^2)$ which are well beyond the reach of other
experiments.

\section{Dark Matter}

Over the past few years, a flood of high-quality data from the
Supernova Cosmology Project~\cite{Riess:1998cb}, the Supernova Search
Team~\cite{Perlmutter:1998np}, the Wilkinson Microwave Anisotropy
Probe (WMAP)~\cite{Spergel:2003cb}, the Two Degree Field Galaxy
Redshift Survey (2dFGRS)~\cite{Colless:2003wz}, and the Sloan Digital
Sky Survey (SDSS)~\cite{Tegmark:2003uf} pin down cosmological
parameters to-percent level precision, establishing a new paradigm of
cosmology.  A surprisingly good fit to the data is provided by a
simple geometrically flat Friedman-Robertson-Walker Universe, in which
30\% of the energy density is in the form of non-relativistic matter
$(\Omega_{\rm m} = 0.30 \pm 0.04$) and 70\% in the form of a new,
unknown dark energy component with strongly negative pressure
$(\Omega_\Lambda = 0.70 \pm 0.04)$~\cite{Tegmark:2003ud}. The matter
budget has only 3 free parameters: the Hubble parameter $h = 0.70^{+
  0.04}_{-0.03},$ the matter density $ \Omega_{\rm m} h^2 = 0.138 \pm
0.012,$ and the density in baryons, $ \Omega_{\rm b} h^2 =
0.0230^{+0.0013}_{-0.0012}.$\footnote{The latter is consistent with
  the estimate from Big Bang nucleosynthesis, based on measurements of
  deuterium in high redshift absorption systems, $\Omega_{\rm b} h^2 =
  0.020 \pm 0.002$~\cite{Burles:2000zk}.} This implies that the
structure of the Universe is dictated by the physics of
as-yet-undiscovered cold dark matter ($\Omega_{\rm CDM} h^2 = 0.115
\pm 0.012$) and the galaxies we see today are the remnants of
relatively small overdensities in the nearly uniform distribution of
matter in the very early Universe. Overdensity means overpressure that
drives an acoustic wave into the other components making up the
Universe: the hot gas of nuclei and photons and the neutrinos. These
acoustic waves are seen today in the temperature fluctuations of the
microwave background as well as in the distribution of galaxies on the
sky.

\begin{figure}
\postscript{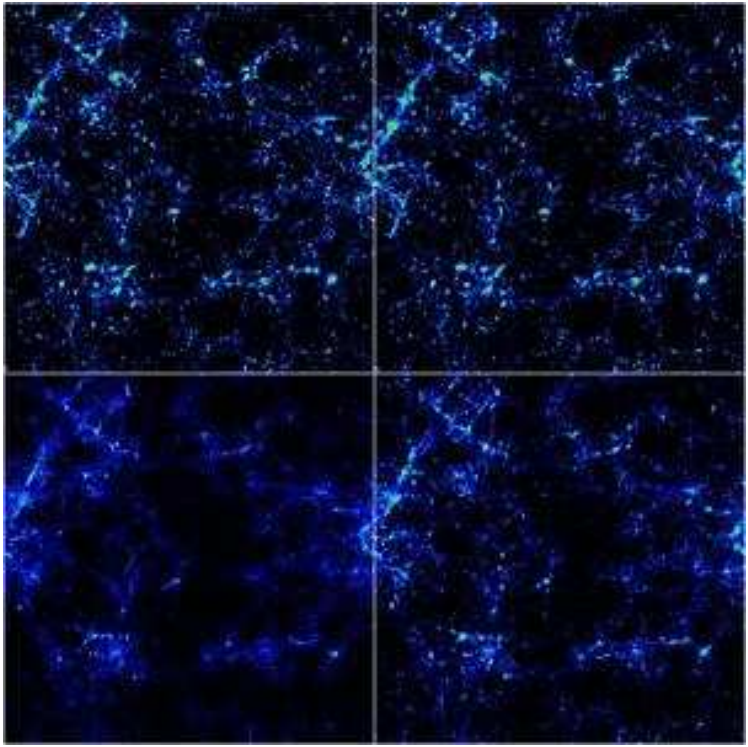}{0.6} 
\caption{\label{numass} Simulations of structure formation with
  varying amounts of matter in the neutrino component, i.e., varying
  neutrino mass: (top left) massless neutrinos, (top right)
  $\sum_\alpha m_{\nu_\alpha} = 1$~eV, (bottom left) $\sum_\alpha
  m_{\nu_\alpha} = 7$~eV, (bottom right) $\sum_\alpha m_{\nu_\alpha}
  = 4$~eV.}
\end{figure}

The only dark matter particle which is known to exist from experiment
is the neutrino. Neutrinos decouple from thermal equilibrium while
still relativistic, and consequently constitute the hot dark matter.
With a contribution to the Universe's matter balance similar to that
of light, neutrinos play a secondary role. The role is however
identifiable: because of their large mean free path, neutrinos prevent
the smaller structures in the cold dark matter from fully developing
and this is visible in the observed distribution of galaxies, see
Fig.~\ref{numass}. Simulations of structure formation with varying
amounts of matter in the neutrino component can match to a variety of
observations of today's sky, including measurement of galaxy-galaxy
correlations and temperature fluctuations on the surface of last
scattering. The results suggest a neutrino mass, summed over the three
neutrino flavors, $\sum_\alpha m_{\nu_\alpha} \alt
1$~eV~\cite{Hannestad:2005gj}, a range compatible with the one deduced
from oscillations.

The simplest model for cold dark matter consists of WIMP's - weakly
interacting massive particles~\cite{Bergstrom:2000pn}. Generic WIMP's
were once in thermal equilibrium, but decoupled while strongly
non-relativistic. Many approaches have been developed to attempt to
detect dark matter. Such endeavors include direct detection experiments
which hope to observe the scattering of dark matter particles with the target
material of the detector~\cite{Goodman:1984dc} and indirect detection
experiments which are designed to search for the products of WIMP
annihilation  into gamma-rays, anti-matter and
neutrinos~\cite{Silk:1985ax}.  Neutrino telescopes indirectly search
for the presence of dark matter by taking advantage of the Sun's
ability to capture large numbers of WIMP's over time. Over billions of
years, a sufficiently large number of WIMP's can accumulate in the
Sun's core to allow for their efficient annihilation. Such
annihilations produce a wide range of particles, most of which are
quickly absorbed into the solar medium. Neutrinos, on the other hand,
may escape the Sun and be detected in experiments on the Earth. The
prospects for such experiments detecting dark matter critically depend
on the capture rate of WIMP's in the Sun, which in turn depends on the
elastic scattering cross section of these particles.

The rate at which WIMP's are captured in the Sun depends on the nature
of the interaction the WIMP undergoes with nucleons in the Sun. These
elastic scattering processes are often broken into two
classifications: {\it spin dependent} interactions in which cross
sections increase with the spin of the target nuclei, and 
{\it spin independent} interactions in which cross sections increase with the
total number of nucleons in the target.  For spin-dependent
interactions, the WIMP capture rate in the Sun is given by
\cite{Jungman:1995df}
\begin{equation} 
C_{\mathrm{SD}}^{\odot} \simeq 3.35 \times 10^{20} \, \mathrm{s}^{-1}
\, f(\rho_{\mathrm{local}},\overline v_{\rm local}, m_{\rm WIPM})\,\, 
\left( \frac{\sigma_{\mathrm{H, SD}}} {10^{-6}\, \mathrm{pb}} \right) \,\,,
\label{c-eq}
\end{equation} 
where 
\begin{equation}
 f(\rho_{\mathrm{local}},\overline v_{\rm local}, m_{\rm WIPM}) = 
\left( \frac{\rho_{\mathrm{local}}}{0.3\, \mathrm{GeV}/\mathrm{cm}^3} \right) 
\left( \frac{\bar{v}_{\mathrm{local}}}{270\, \mathrm{km/s}} \right)^{-3}
\left( \frac{m_{\rm{WIMP}}}{100 \, \mathrm{GeV}} \right)^{-2} \,\,,
\end{equation}
$\rho_{\mathrm{local}}$ is the local dark matter density,
$\sigma_{\mathrm{H,SD}}$ is the spin-dependent WIMP-on-proton
(hydrogen) elastic scattering cross section,
$\bar{v}_{\mathrm{local}}$ is the local rms velocity of halo dark
matter particles, and $m_{\rm{WIMP}}$ is the dark matter particle's mass.
The analogous formula for the capture rate from spin-independent
(scalar) scattering is \cite{Jungman:1995df}
\begin{equation}
C_{\mathrm{SI}}^{\odot} \simeq 3.35 \times 10^{20} \, \mathrm{s}^{-1} \,\,
 f(\rho_{\mathrm{local}},\overline v_{\rm local}, m_{\rm WIPM})\,\,
\left( \frac{ \, \sigma_{\mathrm{H, SI}}
+ 0.07 \, \sigma_{\mathrm{He, SI}}}{10^{-6} \, \mathrm{pb}} \right) 
 \, ,
\label{c-eq2}
\end{equation}
where $\sigma_{\mathrm{H, SI}}$ is the spin-independent 
WIMP-on-proton elastic scattering cross section and
$\sigma_{\mathrm{He, SI}}$ is the spin-independent WIMP-on-helium,
elastic scattering cross section.  Typically, $\sigma_{\mathrm{He,
    SI}} \simeq 16.0 \, \sigma_{\mathrm{H, SI}}$.  The factor of
$0.07$ reflects the solar abundances of helium relative to hydrogen
as well as dynamical factors and form factor suppression.
Note that these capture rates are suppressed by two factors of the
WIMP mass. One of these is simply the result of the depleted number
density of WIMP's ($n \propto m_{\rm WIMP}^{-1}$) 
while the second factor is the
result of kinematic suppression for the capture of a WIMP much heavier
than the target nuclei, in this case hydrogen or helium.

If the capture rate and annihilation cross section is sufficiently
high, equilibrium will be reached between these two processes.  
For $N$ WIMP's in the Sun, the rate of change of this
quantity is given by
\begin{equation}
\frac{dN}{dt} = C^{\odot} - A^{\odot} N^2  ,
\end{equation}
where $C^{\odot} = C_{\mathrm{SD}}^{\odot} + C_{\mathrm{SI}}^{\odot}$ and 
$A^{\odot} = \langle \sigma v \rangle/V_{\mathrm{eff}}.$ Here, 
$V_{\rm eff} = 5.7 \times 10^{27} \, \mathrm{cm}^3 \, 
(m_{\rm{WIMP}}/100~\mathrm{GeV})^{-3/2},$ is
the effective volume of the core of the Sun, which is
estimated by matching the core temperature with
the gravitational potential energy of a single WIMP at the core
radius~\cite{Griest:1986yu}.  The present
WIMP annihilation rate is given by
\begin{equation} 
\Gamma = \frac{1}{2} A^{\odot} N^2 = \frac{1}{2} \, C^{\odot} \, 
\tanh^2 \left( \sqrt{C^{\odot} A^{\odot}} \, t_{\odot} \right) \;, 
\end{equation}
where $t_{\odot} \simeq 4.5$ billion years is the age of the solar system.
The annihilation rate is maximized when it reaches equilibrium with
the capture rate.  This occurs when 
\begin{equation}
\sqrt{C^{\odot} A^{\odot}} t_{\odot} \gg 1 \; .
\end{equation}
For many particle dark matter candidates, this condition can be met.
If this is the case, the final annihilation rate (and corresponding
event rate or neutrino flux) has no further dependence on the dark
matter particle's annihilation cross section.

The sensitivity of direct detection experiments has been improving at
a steady rate. The Cold Dark Matter Search (CDMS) experiment,
operating in the Soudan mine in northern Minnesota, currently has
produced the strongest limits on spin-independent scattering cross
sections of WIMP's with nucleons~\cite{Akerib:2005kh}, as well as on
spin-dependent scattering cross sections of WIMP's with
neutrons~\cite{Akerib:2005za}. CDMS data exclude spin-independent
cross sections larger than approximately $2 \times 10^{-7}$ pb for a
50-100 GeV WIMP or $7 \times 10^{-7}$~pb ($m_{\rm{WIMP}}$/500 GeV) for
a heavier WIMP~\cite{Akerib:2005kh}. The Zeplin-I~\cite{Alner:2005pa} and 
Edelweiss~\cite{Sanglard:2005we} experiments currently have spin-independent bounds
that are roughly a factor of 5 weaker over this mass range. The NAIAD
experiment~\cite{Alner:2005kt} has placed the strongest constraints on
spin-dependent WIMP-proton scattering. The data limit the
spin-dependent cross section with protons to be less than
approximately 0.3~pb for a WIMP in the mass range of 50-100 GeV and
less than 0.8 pb ($m_{\rm{WIMP}}$/500 GeV) for a heavier WIMP.  The
PICASSO~\cite{Barnabe-Heider:2005pg} and CDMS~\cite{Akerib:2005za} experiments have placed
limits on the spin-dependent WIMP-proton cross section roughly one
order of magnitude weaker than the NAIAD bound.  All of these results
(of course) impact the prospects for detecting neutrinos produced by
the annihilation of WIMP's in the Sun~\cite{Kamionkowski:1994dp}.
Currently, the SK experiment has placed the strongest bounds on
high-energy neutrinos from the direction of the
Sun~\cite{Desai:2004pq}. SK has two primary advantages over other
experiments. Firstly, they have analyzed data over a longer period
than most of their competitors, a total of nearly 1700 live days.
Secondly, SK was designed to be sensitive to low energy ($\sim$~GeV)
neutrinos, which gives them an advantage in searching for lighter
WIMP's. SK's limit on neutrino-induced muons above 1 GeV from WIMP
annihilations in the Sun is approximately 1000 to 2000 per square
kilometer per year for WIMP's heavier than 100~GeV, and approximately
2000 to 5000 per square kilometer per year for WIMP's in the 20 to 100
GeV range. The precise value of these limits depends on the WIMP
annihilation modes considered. The AMANDA~\cite{Ackermann:2005fr} and
MACRO~\cite{Ambrosio:1998qj} collaborations have reported limits on
the flux of neutrino-induced muons from the Sun that are only slightly
weaker than SK. The limit placed by the AMANDA experiment resulted
from only 144 live days of data. Having operated the detector for five
years, AMANDA is expected to produce significantly improved bounds in
the future.

The supersymmetric extension of the Standard Model is the leading
candidate to avoid 't~Hooft naturalness problem with the Higgs
mass~\cite{Dimopoulos:1981zb}.  Supersymmetry (SUSY) posits a
``complete democracy'' between integral and half-integral spins,
implying the existence of many as-yet-undiscovered superpartners.
Thus, if SUSY can serve as a theory of low energy interactions, it
must be a broken symmetry. The most common assumption is that the
minimal low energy effective supersymmetric theory (MSSM) has a
breaking scale of order $\Lambda_{\rm SUSY} \sim 1$~TeV. MSSM has a
concrete advantage in embedding the Standard Model in a GUT: the
supersymmetric beta functions for extrapolating the measured strengths
of the strong, electromagnetic, and weak couplings lead to convergence
at a unified energy value of the order $M_{\rm GUT }\sim
10^{16}$~GeV~\cite{Dimopoulos:1981yj}. Moreover, by imposing
$R$-parity conservation one obtains as a byproduct the stability of
the lightest SUSY particle $\tilde \chi$, making it a possible
candidate for cold dark matter~\cite{Goldberg:1983nd}.

The relic abundance of SUSY WIMP's  can be found by integrating the Boltzman 
equation~\cite{Scherrer:1985zt}, 
\begin{equation}
\frac{dn}{dt} + 3 H\, n = - \langle \sigma v \rangle (n^2 - n_{\rm{eq}}^2),
\label{expansion}
\end{equation}
where $n$ is the present number density of SUSY WIMP's, $n_{\rm{eq}} =
g\, (m_{\tilde \chi}\, T/2 \pi)^{3/2}\,e^{-m_{\tilde \chi}/T}$ is the
equilibrium number density, $H$ is the expansion rate of the Universe
at temperature $T$, $g$ is the number of internal degrees of freedom
of the WIMP, $\langle \sigma v\rangle$ is the thermally averaged
annihilation cross section, and $m_{\tilde \chi}$ is the neutralino
mass. Note that in the very early Universe, when $n \simeq
n_{\rm{eq}}$, the right hand side of Eq.~(\ref{expansion}) is small
and the evolution of the density is dominated by Hubble expansion.  As
the temperature falls below $m_{\tilde \chi}$, however, the
equilibrium number density becomes suppressed and the annihilation
rate increases, rapidly reducing the number density.  Finally, when
the number density falls enough, the rate of depletion due to
expansion becomes greater than the annihilation rate, $H \agt n
\langle \sigma v \rangle$, and the neutralinos {\it freeze out} of
thermal equilibrium. The freeze-out temperature $T_{\rm f}$ depends
logarithmically upon $\langle \sigma v \rangle,$ but for models with
TeV scale SUSY breaking, one finds that $T_{\rm f}/m_{\tilde \chi}
\sim 0.05$. SUSY WIMPs are, by far, the favored candidate for cold
dark matter, because for masses of order 100~GeV to 10~TeV a present
density of $\Omega_{\tilde \chi}\ h^2 \sim 0.1,$ comes out fairly
naturally~\cite{Griest:1989zh}.\footnote{An alternative dark matter
  candidate, which has received quite some attention recently, arises
  in models where all of the particles of the Standard Models can
  propagate through the bulk of an extra dimensional space that is
  compactified on a scale around $\sim {\rm TeV}^{-1}.$ Within this
  set up each Standard Model particle is accompanied by a tower of
  Kaluza Klein states. The lightest KK particle can be naturally
  stable in a way anologous to how the lightest superpartner is
  stabilized in $R$-parity conserving models of
  SUSY~\cite{Servant:2002aq}.} In what follows we mantain our
discussion as general as possible but keep in mind the specific needs
of the MSSM.

As they annihilate, WIMPs can generate neutrinos through a wide range
of channels. Annihilations to heavy quarks, tau leptons, gauge bosons
and Higgs bosons can all generate neutrinos in the subsequent 
decay~\cite{Jungman:1994jr}.
The total flux of neutrinos emitted by the Sun due to WIMP 
annihilition is then
\begin{equation} 
\left. \frac{dF^{\nu_\mu}_\odot}{dE_\nu}\right|_{\rm source} = C_\odot \, F_{\rm Eq} \sum_j 
\left(\frac{dN_{\nu_\mu}}{dE_\nu}\right)_{\! j}\,\,\,  
e^{-E_\nu/150~{\rm GeV}}\, \, ,
\end{equation}
where $F_{\rm{Eq}}$ is the non-equilibrium suppression factor
($\approx 1$ for capture-annihilation equilibrium),
$(dN_{\nu_\mu}/dE_\nu)_j$ is the $\nu_\mu$ flux produced by the $j$
decay channel per WIMP annihilation~\cite{Jungman:1994jr}.  Note that
neutrinos produced near the center of the Sun interact with the solar
medium~\cite{Crotty:2002mv}, yielding a depletion of the emission
spectrum by a factor $\sim e^{-E_\nu/150~{\rm GeV}}.$ Finally, the
$\nu_\mu$ flux reaching the Earth is,
\begin{equation}
\phi^{\nu_\mu}_\odot = \frac{1}{4 \pi d^2}
\left. \frac{dF^{\nu_\mu}_\odot}{dE_\nu}\right|_{\rm source} \,\,,
\end{equation}
where $d \approx 1.5 \times 10^8~{\rm km}$ is the Earth-Sun distance.

By replacing this flux in Eq.~(\ref{eq:numuevents}) it is easily seen
that the CDMS bound implies that the event rate at IceCube for
spin-independent interactions is at the level of the atmospheric
neutrino background~\cite{Halzen:2005ar}. Thus, isolating a signal
from WIMP's which scatter with nucleons mostly spin-independently in
the IceCube data sample would be a challenge. On the other hand, a
WIMP with a largely spin-dependent scattering cross section with
protons may be capable of generating large event rates at IceCube.
For example, a 300~GeV WIMP with a cross section near the experimental
limit, leads to rates as high as $\sim 10^6$ per
year~\cite{Halzen:2005ar}.  The elastic scattering and annihilation
cross sections of a neutralino depend on its various couplings and on
the mass spectrum of the Higgs bosons and superpartners. The
neutralino couplings depend on its composition.  Generally, the
lightest neutralino can be any mixture of bino, wino and the two
CP-even higgsinos, although in most models a largely bino-like
neutralino is lightest. Spin dependent, axial-vector, scattering of
neutralinos with quarks within a nucleon is made possible through the
t-channel exchange of a $Z$, or the s-channel exchange of a squark.
Spin independent scattering occurs at the tree level through s-channel
squark exchange and t-channel Higgs exchange, and at the one-loop
level through diagrams involving a loop of quarks and/or squarks. For
generality, here we do not assume any particular SUSY breaking
scenario or unification scheme, and scan the entire MSSM parameter
space using the DarkSUSY program~\cite{Gondolo:2000ee}. We find that,
after appling the cuts to comply with all collider constraints and to
produce a thermal relic density that saturates the observational
limits~\cite{Ellis:2003cw}, very large spin-dependent cross sections
($\sigma_{\rm SD} \agt 10^{-3}$~pb) are possible even in models with
very small spin-independent scattering rates.

\begin{figure}[t]
\includegraphics[width=2.4in,angle=90]{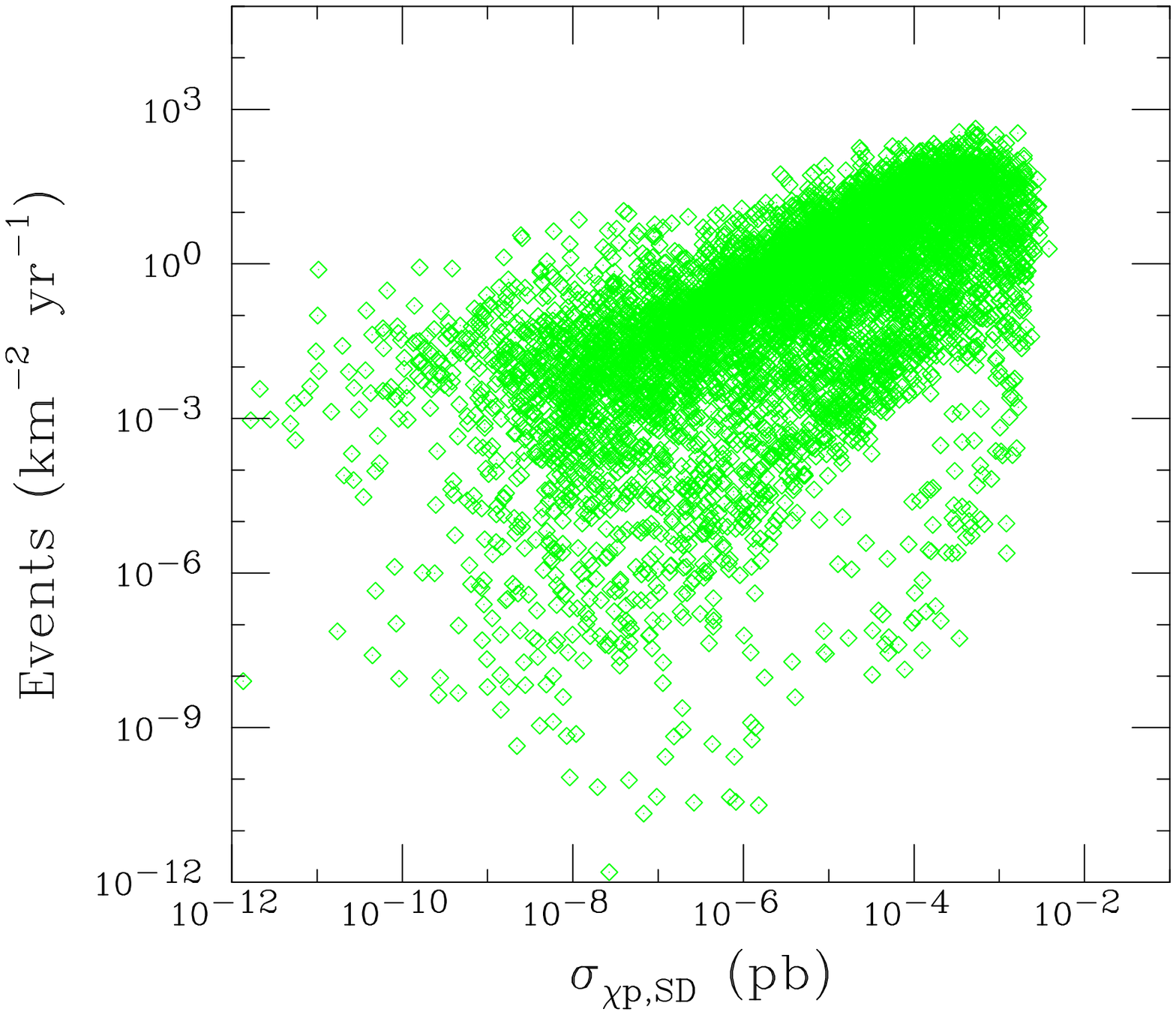}
\includegraphics[width=2.4in,angle=90]{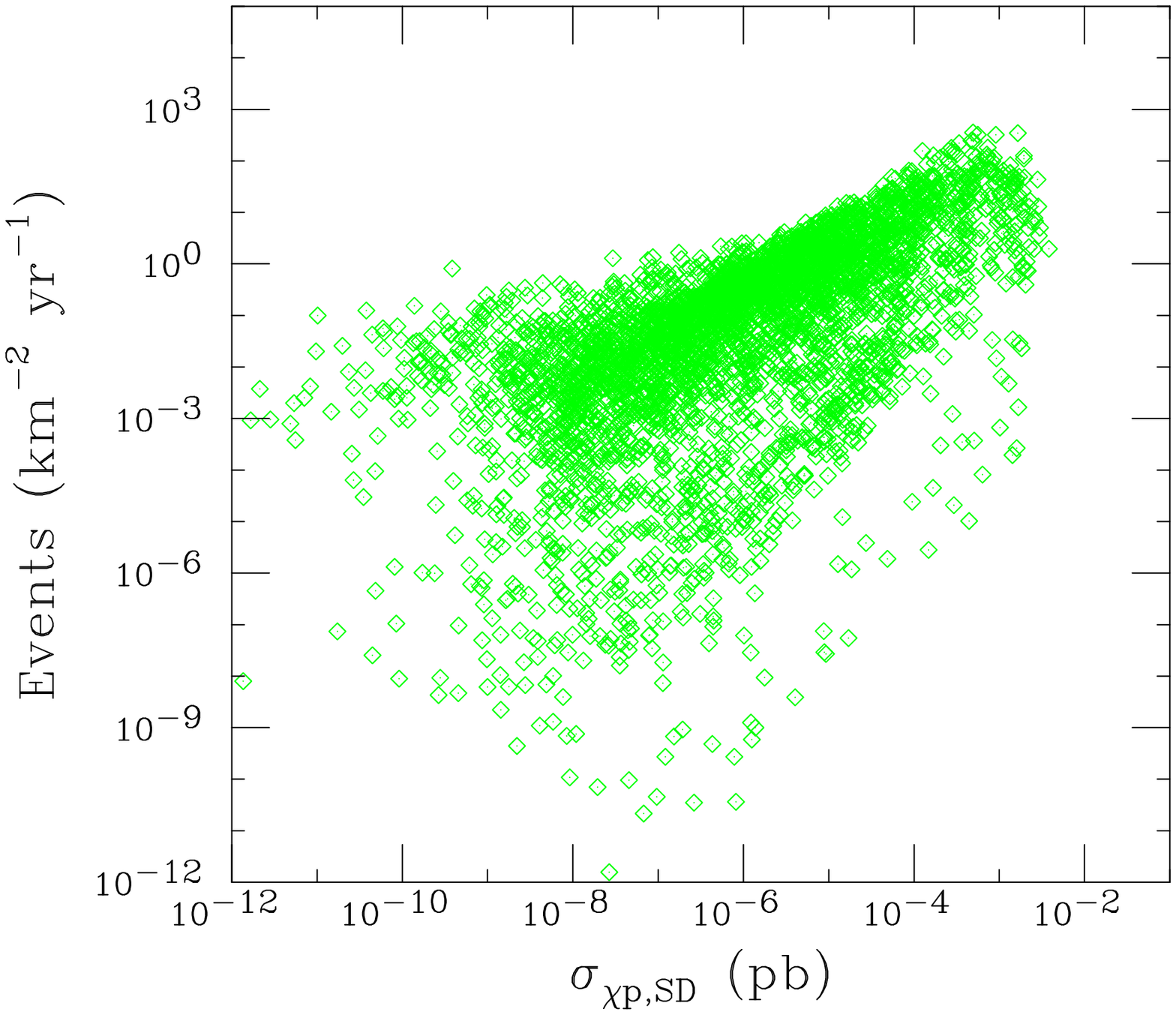}
\caption{IceCube event rate 
  from dark matter annihilations in the Sun, as a function of the
  WIMP's spin-dependent elastic scattering cross section. In the left
  frame, no points shown violate the current spin-independent
  scattering constraints of CDMS. In the right frame, no points would
  violate the a spin-independent bound 100 times stronger~\cite{Halzen:2005ar}.}
\label{ratesd}
\end{figure}

In Fig.~\ref{ratesd} we show the event rate at IceCube from WIMP
annihilation in the Sun versus the WIMP's spin-dependent cross section
with protons. In the left frame, all points evade current constraints
from CDMS. In the right frame we show the same result, but only
showing those points which evade a constraint {\it 100 times stronger}
than current CDMS bound. We can thus conclude that the next generation
of direct detection experiments will not be able to probe the MSSM
parameter space accessible to IceCube.

In summary, direct detection dark matter experiments, lead by the CDMS
Collaboration, have placed increasingly stronger constraints on the
cross section for elastic scattering of WIMP's on nucleons. Although
these constraints impact the prospects for indirect detection
experiemnts, the latest bounds placed on the scalar WIMP-nucleon cross
section by the CDMS experiment do not seriously limit the WIMP
parameter space to be probed by IceCube. This is largely due to the
fact that the spin-dependent scattering of WIMP's with protons is the
most efficient process for capture in the Sun for many particle dark
matter models. The spin-dependent scattering cross section of a WIMP
is not nearly as strongly constrained as the spin-independent quantity
and large values of spin-dependent cross sections are beyond the reach
of any planned direct detection experiments.  Thus, IceCube will play
an important as well as complementary role in the search for particle
dark matter.

\section{Topological Defects and Superheavy Relics}

A plethora of explanations have been proposed to address the
production mechanism of ultra-high energy cosmic
rays~\cite{Anchordoqui:2002hs}. In the absence of a single model which
is consistent with all data, the origin of these particles remains a
mystery. Clues to solve the mystery are not immediately forthcoming
from the data, particularly since various experiments report mutually
inconsistent results. 

In recent years, a somewhat confused picture {\it vis--{\`a}--vis} the
energy spectrum and arrival direction distribution has been emerging.
Since 1998, the AGASA Collaboration has consistently
reported~\cite{Takeda:1998ps} a continuation of the spectrum beyond
the GZK cutoff~\cite{Greisen:1966jv}.  In contrast, the most recent
results from HiRes~\cite{Abu-Zayyad:2002sf} describe a spectrum which
is consistent with the expected GZK feature. This situation exposes
the challenge posed by systematic errors (predominantly arising from uncertainties 
in hadronic interaction models~\cite{Anchordoqui:1998nq}) in these types of
measurements. Although there seems to be a remarkable agreement among
experiments on predictions about isotropy on large scale
structure, this is certainly
not the case when considering the two-point correlation function on a
small angular scale. The AGASA Collaboration reports observations of
event clusters which have a chance probability smaller than 1\% to
arise from a random distribution~\cite{Hayashida:bc}. Far from
confirming the AGASA results, the recent analysis reported by the
HiRes Collaboration showed that their data are consistent with no
clustering among the highest energy events~\cite{Abbasi:2004ib}.  The
discovery of such clusters would be a tremendous breakthrough for the
field, but the case for them is not yet proved.  Special care must be
taken when computing the statistical significance in such an analysis.
In particular, it is important to define the search procedure {\it a
  priori} in order to ensure one does not inadvertently perform
``trials'' by studying the data before deciding upon the cuts. Very
recently, with the aim of avoiding accidental bias on the number of
trials performed in selecting the angular bin, the original claim of
the AGASA Collaboration~\cite{Hayashida:bc} was re-examined
considering only those events observed after the original
claim~\cite{Finley:2003ur}. This study showed that the evidence for
clustering in the AGASA data set is weaker than was previously
supposed, and is consistent with the hypothesis of isotropically
distributed arrival directions.

Further confusing the issue, recent HiRes data have been interpreted
as a change in cosmic ray composition, from heavy nuclei to protons,
at $\sim 10^9$~GeV~\cite{Bergman:2004bk}. This is an order of
magnitude lower in energy than the previous crossover deduced from the
Fly's Eye data~\cite{Bird:1993yi}. The end-point of the galactic flux
is expected to be dominated by iron, as the large charge $Ze$ of heavy
nuclei reduces their Larmor radius (containment scales linearly with
$Z$) and facilitates their acceleration to highest energy (again
scaling linearly with $Z$).  The dominance of nuclei in the high
energy region of the Galactic flux carries the implication that any
changeover to protons represents the onset of dominance by an
extra-galactic component.  The inference from this new HiRes data is
therefore that the extra-galactic flux is beginning to dominate the
Galactic flux already at $\sim 10^9$~GeV. Significantly, this is well
below $E_{\rm GZK} \sim 10^{10.7}$~GeV~\cite{Greisen:1966jv}, and
so samples sources even at large redshift.

The Pierre Auger Observatory is confronting the low statistics problem
at the highest energies by instrumenting a huge collection area
covering 3000 square kilometers on an elevated plane in Western
Argentina~\cite{Abraham:2004dt}. The instrumentation consists of 1600
water \v{C}erenkov detectors spaced 1.5~km apart.  Showers occurring
on clear moonless nights (about 10\% of the operational time) are also
viewed by four fluorescence detectors, allowing powerful
reconstruction and cross-calibration techniques.  Simultaneous
observations of showers using two distinct detector methods will help
to control the systematic errors that have plagued cosmic ray
experiments to date.  While no breakthroughs were
reported~\cite{Sommers:2005vs}, preliminary data forecast rapid
progress and imminent results in deciding event rates at the high end
of the spectrum.

The difficulties so far encountered in modeling the production of
ultra-high energy cosmic rays arise from the need to identify a source
capable of launching particles to extreme energy~\cite{Torres:2004hk}. 
In contrast to the
``bottom-up'' acceleration of charged particles, the ``top-down''
scenario avoids the acceleration problem by assuming that charged and
neutral primaries simply arise in the quantum mechanical decay of
supermassive elementary $X$ particles~\cite{Bhattacharjee:1998qc}. 
To maintain an appreciable decay rate
today, it is necessary to tune the $X$ lifetime to be longer (but not
too much longer) than the age of the universe, or else ``store''
short-lived $X$ particles in topological vestiges of early universe
phase transitions.

According to current unified models of high energy interactions, the
Universe may have experienced several spontaneous symmetry breakings,
where some scalar field, generally referred to as the Higgs field,
acquired a non-vanishing expectation value in the new vacuum (ground)
state. Quanta associated with these fields are typically of the order
of the GUT symmetry-breaking scale. During a phase transition,
non-causal regions may evolve towards different states, so that in
different domain borders the Higgs field may keep a null expectation
value. Energy is then stored in a topological defect whose
characteristics depend on the topology of the manifold where the Higgs
potential reaches its minimum~\cite{Kibble:1976sj}.  The relic defects
such as magnetic monopoles~\cite{Hill:1982iq}, cosmic
strings~\cite{Bhattacharjee:vu}, superconducting cosmic
strings~\cite{Hill:1986mn}, and cosmic
necklaces~\cite{Berezinsky:1997td} (a possible hybrid topological
defect consisting of a closed loop of cosmic string with monopole
``beads'' on it~\cite{Hindmarsh:xc}) are all relatively topologically
stable, but can release part of their energy (through radiation,
annihilation, or collapse) in the form of $X$ particles that typically
decay to quarks and leptons.

The highest energy cosmic rays may also be produced from the decay of
some metastable superheavy relic particle with mass $m_X \agt
10^{12}$~GeV and lifetime exceeding the age of the
Universe~\cite{Berezinsky:1997hy,Kuzmin:1997cm}.  Discrete gauged
symmetries~\cite{Hamaguchi:1998wm} or hidden
sectors~\cite{Ellis:1990iu} are generally introduced to stabilize the
$X$ particles. Higher dimensional operators, wormholes, and instantons
are then invoked to break the new symmetry super-softly to maintain
the long
lifetime~\cite{Berezinsky:1997hy,Kuzmin:1997cm,Hamaguchi:1998wm,Ellis:1990iu}
(collisional annihilation has been considered as
well~\cite{Blasi:2001hr}).  Arguably, these metastable super-heavy
relics may constitute (a fraction of) the dark matter in galactic
haloes~\cite{Berezinsky:1997hy,Sarkar:2001se}.  If this were the case,
due to the non-central position of the Sun in our Galaxy the flux of
ultra-high energy cosmic rays from $X$ particle decay would exhibit a
dipole anisotropy~\cite{Dubovsky:1998pu}.

The cascade decay to cosmic ray particles is driven by the ratio of
the volume density of the $X$-particle ($n_X = \rho_c
\,\Omega_X\,/m_X$) to its decay time ($\tau_{_X}$).  This is very
model dependent, as neither the cosmic average mass density
contributed by the relics $(\Omega_X),$ nor $\tau_{_X}$ is known with
any degree of confidence ($\rho_c \approx 1.05 \times 10^{-4}\,
h^2$~GeV cm$^{-3}$). Moreover, the internal mechanisms of the decay
and the detailed dynamics of the first secondaries do depend on the
exact nature of the particles.  Consequently, no firm prediction on
the expected flux of cosmic rays can be made. However, if there are no
new mass scales between $\Lambda_{\rm SUSY} \sim 1$~TeV and
$m_X,$\footnote{This hypothesis, known as the desert ``hypothesis'',
  is well motivated by the fact that the existence of new physics
  between the GUT scale and the SUSY breaking scale would destroy the
  very impresive feature of ``natural'' unification of the gauge
  couplings at $M_{\rm GUT} \sim 10^{16}~{\rm GeV}$ occuring in the
  MSSM.}  the squark and sleptons would behave like their
corresponding supersymmetric partners, enabling one to infer from the
``known'' evolution of quarks and leptons the gross features of the
$X$-particle cascade: the quarks hadronize producing jets of hadrons
containing mainly pions together with a 3\% admixture of
nucleons~\cite{Sarkar:2001se,Coriano:2001rt}.  This implies
that the injection spectrum is a rather hard fragmentation-type shape
(with an upper limit usually fixed by the GUT scale) and dominated by
$\gamma$-rays and neutrinos produced via pion decay.  Therefore, the
photon/proton ratio~\cite{Aharonian:1992qf} and the ultra-high energy
neutrino flux (which would be beyond the WB-flux of transparent
sources)~\cite{Gondolo:1991rn} can be used as diagnostic tools to 
probe top-down models.

In light of the mounting evidence that ultra-high energy cosmic rays
are not $\gamma$-rays~\cite{Ave:2000nd}, one may try to force a proton
dominance at ultra-high energies by postulating efficient absorption
of the dominant ultra-high energy photon flux on the universal and/or
galactic radio background. According to current estimates of the
strengths of the magnetic field and of the radio wave background in
our own galaxy, most ultra-high energy photons produced in the halo
are expected to reach the Earth. In what follows we assume that the
photon interaction length in the galaxy has been greatly
over-estimated~\cite{Sarkar:2003sp}, and explore this assumption for
neutrino signals.\footnote{It is also possible that the energy
  measurements in air shower arrays are biased due to increasingly
  photon dominance at super-GZK energies, such that the superheavy
  dark matter model remains consistent with observational 
  limits~\cite{Busca:2006tr}.}

If a top down scenario is to explain the origin of ultra-high energy
cosmic rays, the injection spectrum should be normalized to account
for the super-GZK events without violating any observational flux
measurements or limits at higher or lower energies~\cite{Sigl:1998vz}.
In particular, neutrino and $\gamma$-ray fluxes depend on the energy
released integrated over redshift, and thus on the specific top down
model. Note that the electromagnetic energy injected into the Universe
above the pair production threshold on the cosmic microwave background
is recycled into a generic cascade spectrum below this threshold on a
short time scale compared with the Hubble time. Therefore, it can have
several potential observable effects, such as modified light element
abundances due to $^4$He photodisintegration, or induce spectral
distortions of the cosmic microwave background~\cite{Sigl:1995kk}.
Additionally, measurements of the diffuse GeV $\gamma$-ray
flux~\cite{Sreekumar:1997un}, to which the generic cascade spectrum
would contribute directly, limit significantly the parameter space in
which cosmologically distant $X$'s can generate the flux of the
ultra-high energy cosmic rays~\cite{Sigl:1996im}. Such bounds are thus
relevant only for topological defects, not for superheavy dark matter which is
clustered locally~\cite{Sarkar:2003sp}.

A point worth noting at this juncture: as we discussed in the previous
section, the cold dark matter hunt traditionally concentrates on
WIMP's that were once in thermal equilibrium in the early Universe.
Its present abundance is determined by the self-annihilation cross
section.  Since the largest annihilation cross section at early times
is expected to be $\propto m_{\rm WIMP}^{-2},$ heavy WIMP's would have
such a small cross section that their present abundance would be too
large.  Consequently, the mass of a thermal WIMP is found to be less
than about 500~TeV. Then, in associating metastable superheavy relics
with cold dark matter we assume that the $X$-particles have never
experienced thermal equilibrium.  Moreover, since the density of
thermal WIMP's saturates the WMAP limit~\cite{Ellis:2003cw}, the
contribution from superheavy relics to the cold dark matter density
must be much less than that of the relic neutralinos.  As noted above
the cosmic ray flux resulting from $X$-particle decays depends on the
dimensionless parameter $r_X\equiv \xi_Xt_0/\tau_{_X}$, where $\xi_X =
\Omega_X/\Omega_{\rm CDM},$ and $t_0$ is the age of the universe.
Scenarios which include in the rate normalization photon flux from
$X$-particles clustered in the halo lead to a value $r_X \sim 5\times
10^{-11}$~\cite{Berezinsky:1997hy}. Omission of the photon channel in the
normalization increases $r_X$ by a factor of about $2 - 10.$ Since models
of $X$-production and decay typically lead to exponential dependence
of both $\xi_X$ and $\tau_{_X}$ on a reheating temperature $T_R$ and a
quantum mechanical tunneling action, respectively; there is no
impediment on accommodating this change in $r_X$ while maintaining
$\Omega_X\ll\Omega_{\rm CDM}.$ For example, assumuing a relic
overdensity of $10^5$ in the vicinity of our galaxy, as expected for
$X$ particles that move freely under the influence of
gravity~\cite{Berezinsky:1997hy}, an Earthly proton dominated spectrum
requires $\xi_X \propto e^{-2m_X/T_R} \sim 10^{-4} -
10^{-8}$~\cite{Anchordoqui:2004qh}.

\begin{figure}[tbp]
\begin{minipage}[t]{0.49\textwidth}
\postscript{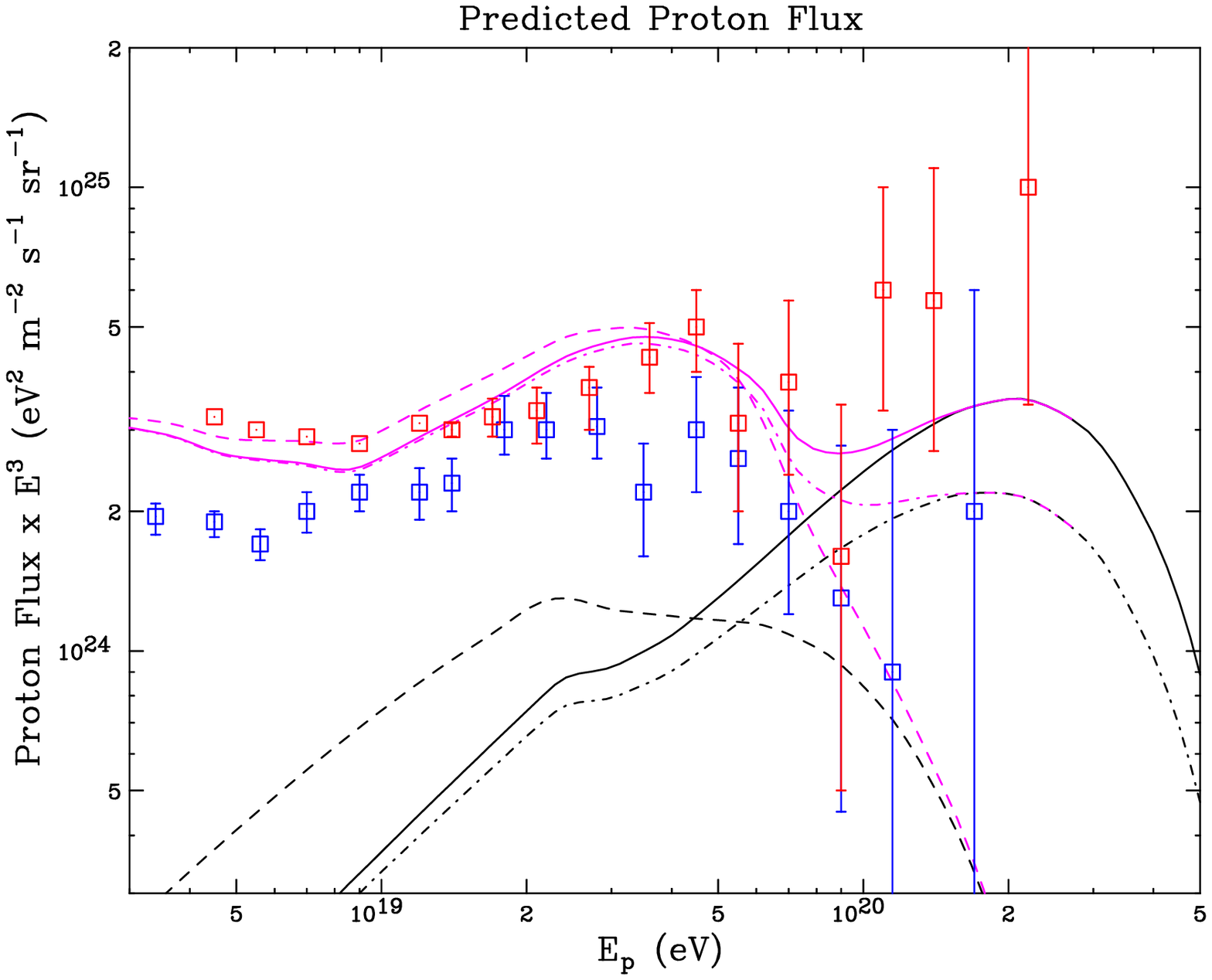}{0.99}
\end{minipage}
\hfill
\begin{minipage}[t]{0.49\textwidth}
\postscript{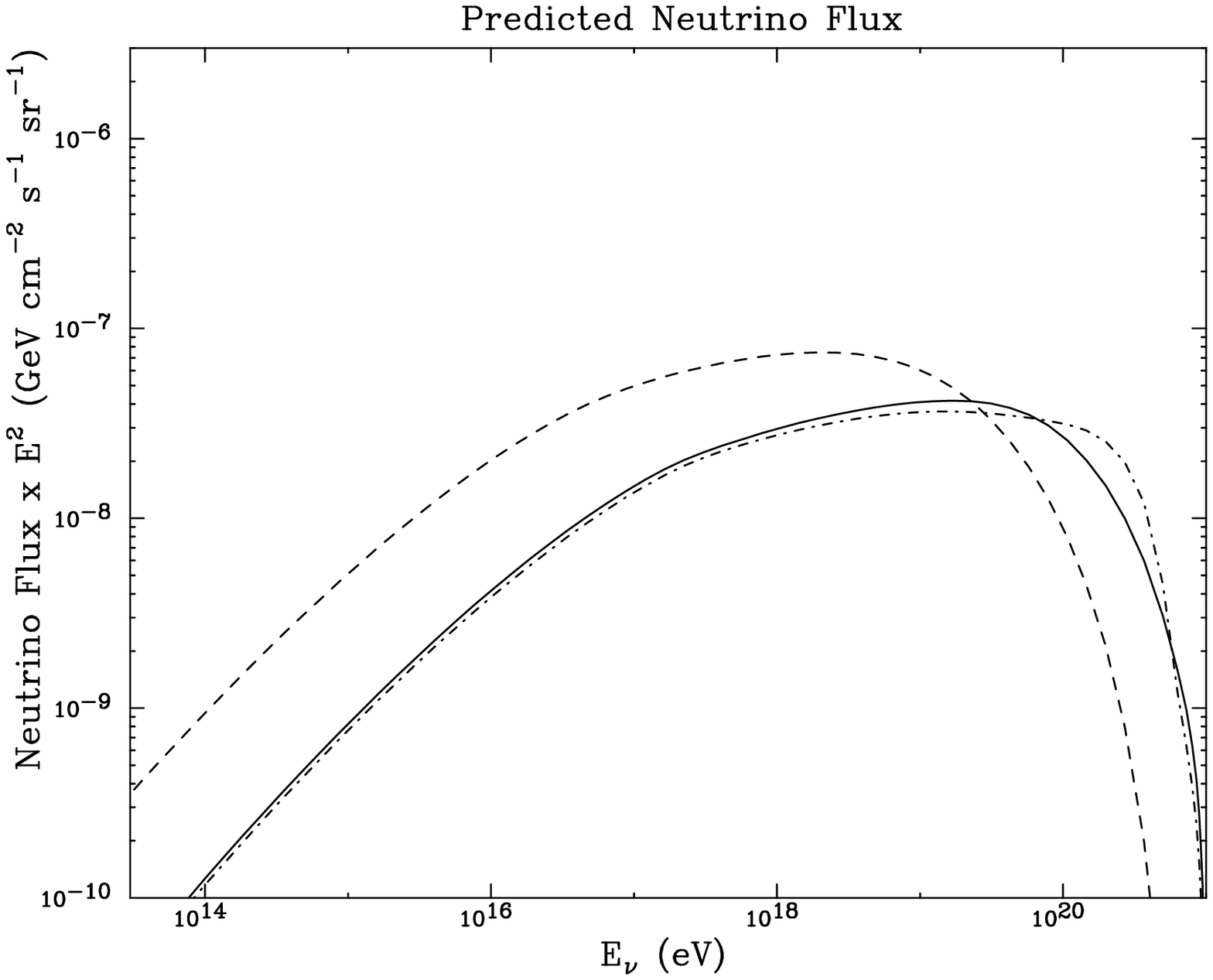}{0.99}
\end{minipage}
\caption{Right: The ultra-high energy cosmic ray flux predicted for
  the decay of superheavy particles with mass $m_X = 2 \times
  10^{12}$~GeV is compared to the HiReS and AGASA cosmic ray data. The
  distribution of jets used includes an overdensity factor of $10^{5}$
  within 20~kpc of the galaxy. Spectra are shown for quark+antiquark
  (solid), quark+squark (dot-dash) and 5 quarks + 5 squarks (dashes)
  initial states. Dark lines are from top-down origin alone whereas
  lighter lines are top-down plus contributions from an homogeneous
  population of astrophysical sources.  Left: The neutrino plus
  antineutrino flux corresponding to the cosmic ray spectra from decay
  of superheavy particles~\cite{Barbot:2002kh}.}
\label{X}
\end{figure}

In Fig.~\ref{X} we show the neutrino spectra produced in top-down
models for a variety of $X$-decay channels. We have assumed that $X$
particles have an overdensity of $10^5$ in the vicinity of our galaxy.
The exact profile of the halo of $X$ particles does not affect the
results as long as most ultra-high energy cosmic ray events originate
at distances well below one GZK interaction length. The spectra of
stable particles (protons, photons, neutralino, electrons and
neutrinos of the three species) have been computed using the SH-decay
program~\cite{Barbot:2003cj}. At the energies under consideration, it
is necessary to take into account all the gauge couplings of the MSSM;
indeed, at the scale of unification, they are all of the same
strength, so that electroweak (and some Yukawa) interactions can be as
relevant as the QCD ones. The perturbative part of the shower was
computed by solving numerically the complete set of evolution
equations for the relevant fragmentation functions of the MSSM. We
modelled carefully the decays of unstable particles with mass near
$\Lambda_{\rm SUSY} \sim 1$ TeV, as well as the hadronization process
at the GeV scale for light quarks and gluons.\footnote{The primary
  10--body decay $X \rightarrow 5q 5 \tilde q$ has been modeled using
  phase space only, i.e. ignoring any possible dependence of the
  matrix element on external momenta.}  The normalization is
determined by matching the $X$-particle baryon flux to the difference
between the observed cosmic ray spectrum at $E\sim 10^{11}$~GeV and
contributions from a homogeneous population of astrophysical sources.
All fluxes shown in Fig.~\ref{X} are consistent with current
measurements of the diffuse GeV $\gamma$-ray background.

\begin{table}
\begin{tabular}{|c|c|c|}
\hline
\hline
 ~~~~$X$-decay channel~~~~ & ~~~~AMANDA-II~~~~ &  ~~~~IceCube~~~~ \\
\hline
$q \overline q$ & 0.39~yr$^{-1}$ & 12.2~yr$^{-1}$ \\
$q \tilde q$ & 0.36~yr$^{-1}$ & 11.4~yr$^{-1}$ \\
$5 \times q \tilde q$ & 1.40~yr$^{-1}$ & 44.6~yr$^{-1}$ \\
\hline \hline
\end{tabular}
\caption{Neutrino event rates at AMANDA-II and IceCube for various 
$X$-decay channels.}
\label{nevents2}
\end{table}

The event rates at AMANDA-II and IceCube are given in
Table~\ref{nevents2}~\cite{Barbot:2002kh}.  They are estimated using
Eqs.~(\ref{eq:numuevents}) and (\ref{FTjpg}) with a threshold energy
$10^5$~GeV for showers and muons. This energy cut effectively removes
any background from atmospheric neutrino events.  For simplicity, the
muon effective area and shower effective volume of the AMANDA-II
detector are taken to be uniform: $A_{\rm eff}^0 \sim
0.05$~km$^2$~\cite{Ahrens:2003pv} and ${\cal V}_{\rm eff} \sim
0.008~{\rm km}^3$~\cite{Ackermann:2004zw}, respectively.

In summary, IceCube can test the viability of top-down models.  If a
signal is found, future high statistics experiments should be able to
map out the neutrino spectrum, therby allowing us experimental access
to physics at energy scales many order of magnitude beyond the scope
of any conceivable particle collider on Earth.

On a separate track, the cubic kilometer of ice provides a large
detector area for direct search of fast magnetic monopoles. This
topological defect appears in phase transitions which leave an
unbroken $U(1)$ symmetry group.  A symmetry-breaking temperature
$T_{\rm c} \sim \langle H \rangle$ at which the order parameter
$\langle H \rangle$ turns on, leads to a monopole mass $m_{X_{\rm m}}
\sim T_{\rm c}/\alpha,$ where $\alpha$ is the fine structure constant
at scale $T_{\rm c}$.  To avoid violations of Standard Model physics the
value of $\langle H \rangle$ should be at or above the electroweak
scale, yielding a lower limit on the monopole mass: $m_{X_{\rm m}}
\agt 40~{\rm TeV}$.

The number density and therefore the flux of monopoles emerging from
a phase transition are determined by the Kibble mechanism~\cite{Kibble:1976sj}.
At the time of the phase transition, roughly one monopole
or antimonopole is produced per correlated volume.
The resulting monopole number density today is
\begin{equation}
n_{X_{\rm m}} \sim 10^{-19}\, (T_{\rm c}/10^{11}{\rm GeV})^3 (l_H/\xi_{\rm c})^3\,{\rm cm}^{-3},
\label{density}
\end{equation}
where $\xi_{\rm c}$ is the phase transition correlation length,
bounded from above by the horizon size $l_H$ at the time when the
system relaxes to the true broken--symmetry vacuum.
In a second order or weakly first order phase transition,
the correlation length is comparable to the horizon size.
In a strongly first order transition,
the correlation length is considerably smaller than the horizon size.

These monopoles easily pick up energy from the magnetic fields
permeating the Universe and can traverse unscathed through the
primeval radiation.  The kinetic energy imparted to a magnetic
monopole on traversing a magnetic field along a particular path
is~\cite{Kephart:1995bi}
\begin{equation}
E = g \, \int_{\rm path} \vec B . d\vec l \sim g\,\, B\,\, \xi\,\, \sqrt{n} \,\,,
\end{equation}
where $g = e/2\alpha = 3.3 \times 10^{-8}~{\rm dynes/G}$ is the 
magnetic charge according to the Dirac quantization~\cite{Dirac:1931kp},
$B$ is the magnetic field strength, $\xi$ specifies the
field's coherence length, and $\sqrt{n}$ is
a factor to approximate the random--walk through the $n$ domains
of coherent fields traversed by the path.
From Eq.~(\ref{density}) then,
the general expression for the relativistic monopole
flux may be written as
\begin{equation}
\phi_{X_{\rm m}} = c\: n_{X_{\rm m}}/4\pi
 \sim 2\times 10^{-4}\, \left(\frac{m_{X_{\rm m}}}{{10^{15}{\rm GeV}}}\right)^3
\left(\frac{l_H}{\xi_c}\right)^3\,
{\rm cm}^{-2} \;{\rm sec}^{-1}\;{\rm sr}^{-1}\,.
\label{flux}
\end{equation}
Phenomenologically, the monopole flux is constrained by cosmology and
by astrophysics.  Cosmology requires that the monopole energy density
$\Omega_{X_{\rm m}}$ should be less than the cold dark matter density of
the Universe. From Eq.~(\ref{density}) comes a constraint
$\Omega_{X_{\rm m}} \sim 0.1\, (m_{X_{\rm m}}/10^{13} {\rm GeV})^4
(l_H/\xi_c)^3,$ which implies that monopoles with $m_{X_{\rm m}} \alt
10^{13} (\xi_c/l_H)^{3/4}$~GeV do not ``over-close'' the Universe.
Astrophysics requires that there not be so many monopoles around as to
effectively ``short out'' the Galactic magnetic field, $\phi_{X_{\rm
    m}} < 10^{-15} \ {\rm cm}^{-2} \ {\rm s}^{-1}\ {\rm
  sr}^{-1}$~\cite{Turner:ag}. To satisfy this constraint, the Kibble
flux in Eq.~(\ref{flux}) requires $m_{X_{\rm m}} \alt 10^{11}
(\xi_c/l_H)\, \rm{GeV}\,.$ Note that with the usual GUT scale, the
fractional monopole mass density $\Omega_{X_{\rm m}} \sim 10^{16}$
overcloses the Universe by sixteen orders of magnitude.  However, to
dilute the monopole density inflation can be invoked after the phase
transition~\cite{Guth:1980zm}, or else the $U(1)$ group can be broken
temporarily so as to create cosmic string defects which connect
monopoles to anti-monopoles pairwise, which then
annihilate~\cite{Langacker:1980kd}.

A magnetic monopole with velocity $v \to c$ would emit \v{C}erenkov
radiation along its path. The total power 
emitted per unit frequency $\nu$ and per unit length $l$ 
\begin{equation}
\frac{d^{2}W}{d\nu \,dl} = \frac{\pi \, n_{\rm r}^2  \nu}{4 \alpha}
\left[ 1 - \frac{c^2}{v^{2}\,n_{\rm r}^{2}} \right],
\label{eq:cherenkov}
\end{equation}
exceeds that of a bare relativistic muon by $n_{\rm r}^2/4\alpha$,
where $n_{\rm r}$ is the refractive index of the
medium~\cite{Tompkins}.  This corresponds to an enhancement factor of
4700 for monopoles interacting in vacuum and 8300 for monopole
interactions in ice.  Clearly, the large light output of a monopole
track would be a rather unique signal at IceCube~\cite{Wick:2000yc}.
The non-observation of monopole signals at the large \v{C}erenkov
detector in the lake Baikal leads to a 90\% CL upper limit
$\phi_{X_{\rm m}}^{\rm max} = 0.5 - 0.7 - 1.9 \times 10^{-16}~{\rm
  cm}^{-2} \, {\rm s}^{-1} \, {\rm sr}^{-1}$ for $\beta = 1.0\  -
0.9\  - 0.8\,$ respectively 
(where $\beta \equiv v/c$)~\cite{Aynutdinov:2005sg}.  The AMANDA
(data taken during 1997)~\cite{Niessen:2001ci} and
MACRO~\cite{Ambrosio:2002qq} collaborations reported slightly weaker
flux limits. Since the muon effective area of IceCube is about a
factor of 20 larger than that of AMANDA, in 5 years of operation the
kilometer scale telescope will reach a sensitivity nearly two orders
of magnitude below existing limits.

\section{Outlook}

Today, precision data from man-made accelerators can, without exception,
be accommodated by the Standard Model of particle physics. Whenever the
experimental precision increases, the higher precision measurements
invariably collapse into Standard Model values. Nevertheless, since all the
big questions remain unanswered, there is no feeling that we are now
dotting the i's and crossing the t's of a mature theory. Worse, the theory
has its own demise built into its radiative corrections.

Led by a string of fundamental experimental measurements that unmasked the
leptonic flavor mixing, neutrino physics has wounded the Standard Model. The
hope is that at this point the glass is half full and that high-precision
high-statistics IceCube data will continue the process and pierce the Standard
Model's resistant armor.

\acknowledgments{\noindent We have benefited from discussions with
  Felix Aharonian, Markus Ahlers, Jaime Alvarez Mu\~niz, George
  Alverson, Cyrille Barbot, Vernon Barger, John Beacom, Tere Dova,
  Manuel Drees, Luis Epele, Jonathan Feng, Tom Gaisser, Haim Goldberg,
  Concha Gonzalez Garcia, Tao Han, Dan Hooper, Chung Kao, Mitja
  Khangulyan, Michele Maltoni, Alan Martin, Tom McCauley, Teresa
  Montaruli, Pran Nath, Charly Nu\~nez, Tom Paul, Steve Reucroft,
  Andreas Ringwald, David Saltzberg, Subir Sarkar, Sergio Sciutto, Al
  Shapere, Guenter Sigl, Todor Stanev, Floyd Stecker, Ed Stoeffhaas,
  John Swain, Lucas Taylor, Diego Torres, Huitzu Tu, Ricardo Vazquez,
  Alan Watson, Tom Weiler, Enrique Zas, and the AMANDA and IceCube
  collaborations.  This work has been partially supported by the US
  National Science Foundation (NSF Grants Nos.  PHY-0457004 and OPP-
  0236449), the US Department of Energy (DoE Grant No.
  DE-FG02-95ER40896), and the University of Wisconsin Research
  Committee with funds granted by the Wisconsin Alumni Research
  Foundation.}

\end{document}